\documentclass[12pt,headinclude,bibtotoc]{scrbook}
\areaset{465pt}{640pt}
\usepackage[english,ngerman]{babel}
\usepackage{bm}
\usepackage[intlimits]{amsmath} 
\usepackage{amssymb}
\usepackage{amsfonts}
\usepackage{sublabel}
\usepackage{mathbbol}
\usepackage{mathrsfs}
\usepackage{exscale}              
\usepackage{hyperref}
\hypersetup{
    pdftitle={A Class of Almost Equilibrium States in Robertson-Walker Spacetime},
    pdfauthor={Muharrem Küskü},
    pdfsubject={Dissertation},
    pdfkeywords={Physics, Quantum field theory in Curved Spacetime, Cosmology}
}

\usepackage{theorem}                           
{\theoremstyle{plain}\theorembodyfont{\upshape}\newtheorem{Definition}{Definition}[chapter]}
{\theoremstyle{plain}\newtheorem{Proposition}[Definition]{Proposition}}
{\theoremstyle{plain}{\newtheorem{Theorem}[Definition]{Theorem}}}
{\theoremstyle{plain}{\newtheorem{Lemma}[Definition]{Lemma}}}
{\theoremstyle{plain}}

\newcommand{\ket}[1]{|{#1}\rangle}
\newcommand{\sproduct}[2]{\langle{#1},{#2}\rangle}
\providecommand{\abs}[1]{\lvert{#1}\rvert}
\providecommand{\norm}[1]{\lVert{#1}\rVert}
\newcommand{\ren}[1]{:{#1}:}
\renewcommand{\d}{\mathrm d}
\newcommand{\m}{\mathrm m}
\newcommand{\transpose}[1]{{#1}^T}
\newcommand{\angles}[1]{\left\langle #1 \right\rangle}
\newcommand{\C}{\mathrm C}

\renewcommand{\Bar}[1]{\overline{#1}}

\newcommand{\F}{\bm F}
\newcommand{\G}{\bm G}
\renewcommand{\H}{\bm H}
\newcommand{\K}{\bm K}
\newcommand{\T}{\bm T}

\newcommand{\DD}{\mathscr D}
\newcommand{\EE}{\mathscr E}
\newcommand{\FF}{\mathscr F}
\newcommand{\HH}{\mathscr H}
\newcommand{\KK}{\mathscr K}
\newcommand{\MM}{\mathscr M}
\newcommand{\MMa}{\hat{\mathscr M}}
\renewcommand{\SS}{\mathscr S}
\newcommand{\TT}{\mathscr T}
\newcommand{\VV}{\mathscr V}

\renewcommand{\vec}{\mathbf}
\newcommand{\borS}{\mathcal S}

\newcommand{\D}{\mathrm D}
\renewcommand{\L}{\mathrm L}
\renewcommand{\k}{\mathrm k}
\renewcommand{\a}{\mathrm a}
\renewcommand{\gg}{\mathrm g}

\renewcommand{\i}{\mathrm i}
\newcommand{\e}{\mathrm e}
\newcommand{\g}{\mathfrak g}
\newcommand{\h}{\mathfrak h}
\newcommand{\s}{\mathfrak s}
\renewcommand{\det}[1]{\left[{#1}\right]}
 
\DeclareMathOperator{\supp}{supp} 

\DeclareMathOperator{\tr}{Tr} 
\DeclareMathOperator{\WF}{WF}

\DeclareMathOperator{\arcoth}{Arcoth}
\newcommand{\RS}{S^{\Re}_{01}}
\newcommand{\aes}{{\beta,f}}
\newcommand{\sole}{{\infty,f}}
\newcommand{\sigmaS}{{\sigma_\SS}}
\newcommand{\sigmaM}{{\sigma_\MM}}
\newcommand{\sigmaMa}{{\hat\sigma}}
\renewcommand{\transpose}{\intercal}

\newcommand{\proof}{\textit{Proof. }}
\newcommand{\qed}{\hspace{\stretch{1}}$\square$\par}

\usepackage{fancyhdr}
\setkomafont{pagenumber}{\sffamily} 
\begin{document}
\sloppy
\selectlanguage{english}

\pagestyle{empty}
\begin{center}
\vspace*{\stretch{1}}
\textbf{\sffamily
\LARGE A Class of Almost Equilibrium States\\[0.5em]
 in Robertson-Walker Spacetimes}

\vspace*{\stretch{1}}
{\LARGE
Dissertation\\[0.25em]
zur Erlangung des Doktorgrades\\[0.25em]
des Department Physik\\[0.75em]
der Universit\"at Hamburg}

\vspace*{\stretch{1}}
{\large
vorgelegt von\\
Muharrem K\"usk\"u \\[0.25em]
aus Istanbul
}

\vspace*{\stretch{1}}
{\large
Hamburg\\
2008
}
\end{center}

\newpage
 \vspace*{\stretch{1}}
 \begin{tabular}[t]{ll}
 Gutachter der Dissertation                 & Prof. Dr. K. Fredenhagen\\
                                            & Prof. Dr. W. Buchm\"uller\\
 Gutachter der Disputation:                 & Prof. Dr. K. Fredenhagen\\
                                            & Prof. Dr. V. Schomerus\\
 Datum der Disputation:                     & 29.07.2008\\
 Vorsitzender des Pr\"ufungsausschusses:    & Prof. Dr. J. Bartels\\
 Vorsitzender des Promotionsausschusses:    & Prof. Dr. J. Bartels\\
 Dekan der Fakult\"at f\"ur
Mathematik, \\Informatik und Naturwissenschaften:   & Prof. Dr.
A. Fr\"uhwald\\
 \end{tabular}

\cleardoublepage
\section*{Abstract} In quantum field theory in curved spacetimes
the construction of the algebra of observables of linear fields is today well
understood. However, it remains a non-trivial task to construct physically
meaningful states on the algebra. For instance, we are in the unsatisfactory
situation that there exist no examples of states suited to describe local thermal
equilibrium in a non-stationary spacetime. In this thesis, we construct a class
of states for the Klein-Gordon field in Robertson-Walker spacetimes, which seem
to provide the first example of thermal states in a spacetime without time
translation symmetry. More precisely, in the setting of real, linear, scalar
fields in Robertson-Walker spacetimes we define on the set of homogeneous,
isotropic, quasi-free states a free energy functional that is based on the
averaged energy density measured by an isotropic observer along his worldline.
This functional is well defined and lower bounded by a suitable quantum energy
inequality. Subsequently, we minimize this functional and obtain states that we
interpret as 'almost equilibrium states'. It turns out that the states of low
energy, which were recently introduced in \cite{Olbermann:2007i}, are the ground states
of the almost equilibrium states. Finally, we prove that the almost equilibrium
states satisfy the Hadamard condition, which qualifies them as physically
meaningful states. 
\selectlanguage{ngerman}
\section*{Zusammenfassung} In der Quantenfeldtheorie in gekr\"ummter Raumzeit ist
die Konstruktion der Algebra der Observablen linearer Felder heutzutage gut
verstanden. Es ist jedoch eine nicht-triviale Aufgabe physikalische
Zust\"ande auf der Algebra zu konstruieren. Zum Beispiel sind wir in der
unzufriedenstellenden Situation, dass keine Zust\"ande bekannt sind, die es erlauben ein lokales thermales Gleichgewicht in einer
nicht-station\"aren Raumzeit zu beschreiben. In dieser Arbeit konstruieren wir
ein Klasse von Zust\"anden f\"ur das Klein-Gordon Feld in Robertson-Walker
Raumzeiten, die das erste Beispiel thermaler Zust\"ande in einer Raumzeit ohne
Translationssymmetrie bez\"uglich der Zeit darstellen d\"urften. Genauer gesagt
definieren wir f\"ur das reelle lineare skalare Feld in Robertson-Walker
Raumzeiten ein Funktional f\"ur die freie Energie auf der Menge der homogenen
isotropen quasifreien Zust\"ande, welches auf der gemittelten Energiedichte
basiert, die ein isotroper Beobachter auf seiner Weltlinie misst. Dieses
Funktional ist wohldefiniert und von unten beschr\"ankt, dank einer geeigneten
Quanten-Energie-Ungleichung. Danach minimieren wir dieses Funktional und
erhalten so Zust\"ande, welche wir als Fast-Gleichgewichts-Zust\"ande (``almost
equilibrium states'') interpretieren. Es stellt sich heraus, dass die Zust\"ande
niedriger Energie (``states of low energy''), welche k\"urzlich in
\cite{Olbermann:2007i} definiert wurden, die nat\"urlichen Grundzust\"ande der
Fast-Gleichgewichts-Zust\"ande sind. Schlie{\ss}lich beweisen wir, dass die
Fast-Gleichgewichts-Zust\"ande die Hadamard-Bedingung erf\"ullen, was sie als
physikalisch sinnvolle Zust\"ande auszeichnet.
\selectlanguage{english}

\cleardoublepage
\vspace*{\stretch{1}}
\begin{center}
\textit{Dedicated to the memory of Bernd Kuckert.} 
\end{center}
\vspace*{\stretch{4}}

\setcounter{tocdepth}{2}
\tableofcontents
\addtocontents{toc}{\protect\thispagestyle{empty}}
\addtocontents{toc}{\protect\vspace*{-2\baselineskip}}
\addtocontents{toc}{\protect\enlargethispage{3\baselineskip}}
\cleardoublepage

\pagestyle{fancy}
\chapter{Introduction}
There are plenty of reasons to investigate quantum fields in curved spacetimes.
Accepting for the time being that we don't know how to formulate a quantum
theory of gravity, we may still have the desire to investigate the physics in the
intermediate regime, where the quantum nature of fields is already apparent and
spacetime is curved, but the quantum nature of gravitation plays no crucial
role. That such a theory can produce interesting results is undoubted since the
discovery of the Hawking effect, which establishes a remarkable connection
between two seemingly distinct branches of physics, namely, gravitational (black
hole) physics and thermodynamics. The latter field is also of major importance in
modern cosmology, where the cosmic microwave background (CMB), its
isotropy on large scales, and its anisotropy at small scales remains a startling
puzzle. It is also remarkable that cosmological observations, e.g., the ones leading to the
postulation of dark matter and dark energy, nowadays become a major trigger for
particle physics research. For example, the Large Hadron Collider (LHC) at CERN,
scheduled to start operating this year, is hoped to provide clues on the constituents of dark matter. Surely,
the interplay between cosmology and particle physics will increase in the years
to come.

The formulation of quantum field theory in curved spacetime is hampered by the
fact that the traditional formulation relies heavily on Poincar\'e symmetry,
which is lacking in a general curved spacetime. Consequently, notions like
'particles' and 'vacuum', which depend on Poincar\'e symmetry, are not
well defined in this setting. Furthermore, quantum field theory
becomes ambiguous because of the appearance of unitarily inequivalent Hilbert space representations of the algebra
of observables. An expanding, homogeneous, isotropic universe is described by a
Robertson-Walker metric, which is non-stationary and comes in three types, of
which two have curved spacelike sections. Owing to the present symmetry, the
curvature of the spacelike sections does not pose a serious problem for the formulation of quantum field theory. On has analogues
of the Fourier transform, and the quantum field can be written, as usual, as an
integral over modes, albeit the coefficients in the mode decomposition, which in
stationary spacetimes are interpreted as creation and annihilation operators,
loose their meaning. However, non-stationarity, i.e., the absence of time
translation symmetry, brings about, among others, the problems mentioned above. In
particular, the lack of a timelike Killing field results in a failure of energy
conservation and, consequently, the lack of a Hamiltonian, which generates
time-evolution in the non-stationary case.

A thorough analysis of these problems is viable within the algebraic approach to
quantum field theory, which can be rigorously formulated for quantum fields on
curved backgrounds. In this approach, one first constructs the algebra of
observables for the quantum system, satisfying, for instance, the canonical
commutation relations. In a second step, one constructs states with desired
properties as linear functionals on the algebra. This course of action
disentangles, effectively, problems tied to the algebra of observables from
problems arising at the level of representations and states. While the first step
is, at least for free fields, to a large degree understood, the second one is,
put mildly, less explored. One has to admit, that the number of known examples
for physically meaningful states in curved spacetimes is small. For example, to
date, there are no explicitly known thermal states with respect to global time
evolution in a non-stationary spacetime.

A different problem -- which is already present in flat spacetime --, namely,
the lower unboundedness of point-like energy densities in quantum field theory
has led to the so-called quantum inequalities. These uncertainty-type inequalities
give lower bounds on the averaged energy densities measurable in (a class of)
quantum states, where the averaging procedure involves a sampling function with
suitable properties. Quantum inequalities were established in flat spacetimes as
well as in arbitrary globally hyperbolic spacetimes. They are important in
general relativity since they reestablish the macroscopic energy conditions,
e.g., the weak energy condition, utilized for proving singularity theorems.
Furthermore, it has been argued that they may serve as stability conditions
for quantum systems on a mesoscopic scale between the Hadamard condition
(microscopic scale) and passivity (macroscopic scale) \cite{FewsterVerch}.

The initial motivation for the project presented in this thesis can be
stated as follows. Since there are, by quantum inequalities, locally
meaningful energy-quantities even in curved spacetimes, how do the ground states and (thermal) equilibrium
states of these 'energies' look like?  Regarding ground states, the analysis
carried out by Olbermann in \cite{Olbermann:2007i} in the setting of 
real, linear, scalar fields in Robertson-Walker spacetimes resulted in the
definition of states of low energy. These are pure, homogeneous, isotropic,
quasi-free states that minimize the averaged energy density measured by an
observer along a timelike curve. Moreover, it could be proven, by comparison with
adiabatic vacuum states, that the two-point distributions of the states of low
energy satisfy the Hadamard condition, which is a condition on the short distance
behaviour of the two-point distribution to be satisfied by physically reasonable
states. Encouraged by this result, we construct, in the same setting, a
class of (thermal) almost equilibrium states with respect to the timelike averaged
energy density.

The basic idea of our construction is the following. For a finite quantum
statistical system in contact with a much larger reservoir at a constant positive
temperature $\frac{1}{\beta}$, the free energy $\mathcal F=\mathcal
E-\frac{1}{\beta}\mathcal S$, where $\mathcal E$ is the inner energy and
$\mathcal S$ denotes the entropy, is the maximum amount of work that the system
can perform. It turns out that the equilibrium state of such system is
characterized uniquely by the property that it minimizes the free energy.
Following this general principle, we define on the set of homogeneous, isotropic,
quasi-free states of the Klein-Gordon field in Robertson-Walker spacetimes a free
energy functional, where a timelike averaged energy density takes the role of
$\mathcal E$. We show that, owing to the symmetry of the spacetimes, this
energy quantity has the form of an integral over the modes of the quantum
field. In view of this, we associate to each mode an entropy functional and
require the mode to evolve according to the KMS condition, i.e., as a system in
thermal equilibrium at inverse temperature $\beta$. By a certain quantum
inequality proved in \cite{Fewster:2000}, our free energy functional is bounded
below within the class of Hadamard states. Indeed, we succeed in calculating the
two-point distribution of the unique state that minimizes our free energy
functional for a given sampling function. We call the resulting state an almost
equilibrium state a inverse temperature $\beta$ associated to the sampling
function used for averaging the energy density. We finally prove that the almost
equilibrium states satisfy the Hadamard condition. The last step is crucial
since, as noted above, the quantum inequality on which we base our definition is
valid only within the class of Hadamard states. At the same time, the Hadamard
property proves the almost equilibrium states to be physically meaningful states.

\paragraph*{Outline} The outline of this thesis is as follows. In section 2 we
give an account of the basics of quantum field theory in curved spacetimes. We
consider both the quantization scheme via construction of a Hilbert space as well
as the algebraic method. We introduce the notion of quasi-free states and
Hadamard states. The main sources for this chapter are \cite{WaldQFT} and
\cite{KayWald} but the reader, who is not well acquainted with algebraic quantum
field theory will also benefit from \cite{Haag} and \cite{Emch}. In section 3,
we specialize to Robertson-Walker spacetimes and recall relevant facts about
homogeneous, isotropic, quasi-free states of the Klein-Gordon field in this
setting. In particular, we introduce the two-point distribution of such states.
Most of the the material presented here can be found in the seminal paper
\cite{LR}. We add to the known facts a characterization of the two-point
distribution in terms of the mode solutions of the Klein-Gordon equation. Chapter
4 is devoted to quantum inequalities. A brief survey of the topic is followed by
a quotation of the quantum inequality on which our construction is
based \cite{Fewster:2000}. Then, we calculate the specific expression for the averaged
stress-energy tensor, since this will be part of the free energy that we will
minimize. Section 5 is the main part of this thesis, in which we explain and
accomplish the construction of the almost equilibrium states. As a necessary step
for our construction, but maybe also interesting in its own right, we calculate
the generator of KMS states on the Weyl algebra of a system with one degree of
freedom. Then, we define the almost equilibrium states and prove that they are
indeed Hadamard states. In section 6, a summary of the construction is given and
supplemented by a few ideas on possible future projects.

\chapter{Quantum field theory in curved spacetime}\label{ChapterQFTCST}
The overall setting of this thesis is quantum field theory in curved
spacetime. This is a generalization of ordinary quantum field theory in the sense
that the fields propagate on a curved Lorentzian manifold rather than flat
Minkowski space. It may be seen as a semiclassical approach to 'quantum
gravity' since the fields are quantized but the background spacetime is taken to be classical and
fixed. Any backreaction of the field on the spacetime is, in principle, encoded
in the semiclassical Einstein equations
\begin{equation}
G_{\mu\nu}+\Lambda \g_{\mu\nu}=8\pi\angles{\T_{\mu\nu}}\ , 
\end{equation}
though, to date, few is known about backreaction effects, due to some serious
difficulties in calculating $\angles{\T_{\mu\nu}}$. See the discussion in
\cite{WaldQFT} on this point.

The absence of symmetries is a big obstacle for the traditional formulation of
quantum field theory on a general spacetime. However, the algebraic quantum field
theory, as defined in \cite{HaagKastler}, has a well defined generalization to
curved spacetimes \cite{Dimock:1980}. The fundamental object in the algebraic
formulation is a net of local algebras on a spacetime. Such a net associates to
each open region $\mathcal O$ in spacetime an algebra $\mathfrak A(\mathcal O)$
of observables that are measurable in the region. The latest enhancement of the
algebraic approach is the so-called locally covariant quantum field theory
\cite{BFV}. In this approach a locally covariant quantum field theory is defined
as a covariant functor between the category of globally hyperbolic Lorentzian
spacetimes with admissible (isometric, orientation and time-orientation preserving, causally convex) embeddings as morphisms and the
category of unital C*-algebras with certain homomorphisms as morphisms.

In the first section of this chapter, we introduce the necessary facts about
globally hyperbolic spacetimes and the Cauchy problem thereon. Then, we present
the quantization scheme in curved spacetime, where we follow the
presentation found in the monograph \cite{WaldQFT} and the
review \cite{KayWald}. At last, we introduce the notions of quasi-free
states and Hadamard states. Our presentation of the general theory is strongly
biased by our needs in the later chapters.

\section{Classical preliminaries}
\subsection{Spacetime structure}
We adopt the common viewpoint that spacetime is modelled by a smooth (in the
sense that it is $\C^\infty$, Hausdorff, paracompact, and connected)
four-dimensional manifold $\mathcal M$ with Lorentzian metric $\g_{\mu\nu}$. We
also assume that the spacetime under consideration is orientable and
time-orientable. Actually, we will restrict ourselves to the class of
Robertson-Walker spacetimes, which fulfill these hypotheses. Following the
loosely established tradition of field theorists, we choose the signature of the
metric $\g_{\mu\nu}$ to be $(+,-,-,-)$. According to this convention, a vector
$v^\mu$ is called timelike if its norm $\g_{\mu\nu}v^\mu v^\nu$ is
positive, spacelike if the norm is negative, and null if it is
zero. A curve $\gamma(t):\mathbb R\rightarrow\mathcal M$ with tangent vectors
$\dot\gamma^\mu(t)$ that are timelike everywhere is called a timelike
curve. It is called a causal curve if the tangent vectors  are timelike
or null everywhere. For $\mathcal K\subset \mathcal M$ the causal future
$J^+(\mathcal K)$ and the causal past $J^-(\mathcal K)$ are the sets of all
points that can be reached from $\mathcal K$ by a future/past directed causal
curve.  A point $x\in\mathcal M$ is said to be the future endpoint of a
causal curve if for every neighborhood $\mathcal O$ of $x$ there is a $t_0$ such
that $\gamma(t)\in \mathcal O$ for every $t\geq t_0$. A causal curve is said to
be future inextendible if it has no future endpoint. Analogously, we can define
past inextendible curves. A causal curve is called inextendible if it is
future and past inextendible \cite{HawkingEllis}. Less precisely, one can say
that an inextendible curve can end only at infinity or at some initial or final
singularity. The future(+)/past(-)  domain of dependence $D^\pm(\Sigma)$ of
$\Sigma$ is the set of all $x\in\mathcal M$ such that every past/future inextendible
causal curve through $x$ intersects $\Sigma$ \cite{WaldGR}.

A Cauchy surface for a Lorentzian manifold $\mathcal M$ is a spacelike
hypersurface which is intersected exactly once by every inextendible causal curve
in $\mathcal M$ \cite{Fulling}. Equivalently, the domain of dependence
$D(\Sigma)=D^+(\Sigma)\cup D^-(\Sigma)$ of a Cauchy surface is the entire
spacetime. A spacetime $(\mathcal M,\g_{\mu\nu})$ is said to be stationary
if there exists a one-parameter group of isometries $\Xi_t:\mathcal
M\rightarrow\mathcal M$, $\Xi_t^*\g_{\mu\nu}=\g_{\mu\nu}$, $t\in\mathbb R$ whose
orbits are timelike curves. Equivalently, there exists a global timelike
Killing vector field $\xi^\mu$ satisfying Killing's equation
$\nabla_\mu\xi_\nu+\nabla_\nu\xi_\mu=0$, namely, the generator of $\Xi_t$. The
spacetime is said to be static if it is stationary and if, in addition, there exists a spacelike hypersurface $\Sigma$ that is orthogonal to the orbits
of the isometry. In a local coordinate system $x^i, i=1,2,3$, the metric of a
static spacetime may be written as $ds^2=N(\vec x)^2 dt^2-\h_{ij}(\vec x)dx^i
dx^j$, where $\h_{ij}$ is the induced metric on the Cauchy surfaces, and
$N\in\C^\infty(\Sigma)$ is the lapse function. In this case, the timelike
Killing vector field satisfies $\xi_{[\mu}\nabla_{\nu}\xi_{\rho]}=0$. For
completeness, we note that a spacetime is called ultrastatic if it
possesses a timelike Killing vector field $\xi^\mu$ that is orthogonal to the
spacelike hypersurfaces and obeys $\g_{\mu\nu}\xi^\mu\xi^\nu=1$. In this case, the lapse
function is $N=1$.

In order to guarantee the well posedness of the classical time evolution of a
field, one usually assumes globally hyperbolicity of the underlying spacetime.
There are different equivalent definitions of global hyperbolicity. For example,
a spacetime $(\mathcal M,\g_{\mu\nu})$ is called globally hyperbolic if there are
no closed causal curves in $\mathcal M$ and the collection of all causal curves
joining two arbitrary points $x_1,x_2\in\mathcal M$ is compact (in a suitable
topology). Due to a theorem by Geroch \cite{Geroch:1970} an $n$-dimensional
globally hyperbolic spacetime $(\mathcal M,\g_{\mu\nu})$ can be foliated with a
family of Cauchy surfaces that are diffeomorphic to an $(n-1)$-dimensional
manifold $\Sigma$. This means that a globally hyperbolic spacetime has the
topology $\mathcal M=\mathbb R\times\Sigma$. Global hyperbolicity is a condition
on the geometry of spacetime that ensures the existence and uniqueness of global
solutions to hyperbolic differential equations. In a broader sense, global
hyperbolicity of Lorentzian spacetimes serves as an analog for the notion of
completeness of Riemannian manifolds \cite{Baer}. Since quantum dynamics is
usually modelled around classical dynamics, global hyperbolicity of the
underlying spacetime is a reasonable assumption in a semiclassical approach like
quantum field theory in curved spacetime. We refer the interested reader to
\cite{HawkingEllis,WaldGR,Baer} for further elaborations on this topic.

\subsection{The Cauchy problem}
A thorough and general treatment of the Cauchy problem on globally hyperbolic
spacetimes can be found in the monograph \cite{Baer}. The results there are valid
for arbitrary (complex) vector bundles on general globally hyperbolic manifolds.
However, in order to keep in touch with the physical literature and to avoid
unnecessary complexity, we will use the traditional formulation found in
\cite{Dimock:1980,WaldQFT}, keeping in mind the more general results. 

In this thesis, we are concerned with the special case of a real, linear, scalar
field $\phi$ on a globally hyperbolic spacetime $(\mathcal M,\g_{\mu\nu})$
satisfying the Klein-Gordon equation,
\begin{equation}\label{ScalarFieldEquation}
(\Box_\g+\m^2)\phi=0\ , 
\end{equation}
where $\Box_\g=\g^{\mu\nu}\nabla_\mu\nabla_\nu$ is the wave operator associated
to $\g_{\mu\nu}$ and $\m\geq 0$ is the mass parameter. By the global hyperbolicity
of $\mathcal M$, the Klein-Gordon equation \eqref{ScalarFieldEquation} has a well-posed initial value formulation.
This means that there exist unique continuous linear operators
\begin{equation}
E^\pm:\mathscr D(\mathcal M)\rightarrow\mathscr E(\mathcal M)\ ,
\end{equation}
called the advanced fundamental solution, $E^+$, and retarded fundamental
solution, $E^-$, with the following properties:
\begin{gather}
(\Box_\g+\m^2) E^{\pm}f= f= E^{\pm}(\Box_\g+\m^2)f\ ,\\
\supp(E^+ f)\subset J^+(\supp f)\ ,\\
\supp(E^- f)\subset J^-(\supp f)
\end{gather}
for $f\in\mathscr D(\mathcal M)$. We use the notation $\DD(\mathcal
M):=\C^\infty_0(\mathcal M)$ and $\EE(\mathcal M):=\C^\infty(\mathcal M)$,
denoting the set of (complex-valued) smooth functions (with compact support in
the case of $\DD$) on the manifold $\mathcal M$. The first of these properties
says that $E^+$ and $E^-$ are solutions to the inhomogeneous Klein-Gordon equation. The other properties are sensible support properties. Combining the advanced and retarded fundamental solutions, one constructs the fundamental solution
\begin{equation}\label{FundamentalSolution}
E:=E^+-E^-:\mathscr D(\mathcal M)\rightarrow \mathscr E(\mathcal M)
\end{equation}
of the Cauchy problem, which has the properties
\begin{gather}
(\Box_\g+\m^2) Ef = 0 = E(\Box_\g+\m^2)f\ ,\\
\supp(Ef) \subset \big( J^+(\supp f) \cup J^-(\supp f)\big)
\end{gather}
for $f\in\mathscr D(\mathcal M)$. 

There are different schemes available for the quantization of a classical fields,
some of which start from complex solutions and some of which use real-valued
functions. In this chapter we will be concerned with the latter spaces. Denote by
$\SS\subset\mathscr E(\mathcal M,\mathbb R)$ the space of classical, real-valued,
smooth solutions to the Klein-Gordon equation that have compact support on Cauchy
surfaces, and denote by $\mathscr T:=\mathscr D(\mathcal M,\mathbb R)$ the test
function space of real-valued, smooth functions of the fundamental solution. with
compact support. Then the following lemma states some important properties
\begin{Lemma}[\cite{WaldQFT}]\label{FundamentalSolutionLemma}
The map $E:\mathscr T \rightarrow \SS $ satisfies the following three properties
\begin{enumerate}
\item $E$ is onto, i.e., every $\phi\in\SS$ can be expressed as $\phi=Ef$ for some
$f\in\mathscr T$. \item $Ef=0$ if and only if $f=(\Box_\g+\m^2)g$ for
some $g\in\mathscr T$. \item For all $\phi\in\SS$ and all
$f\in\mathscr T$, we have
\begin{equation}
\phi(f):=\int  d^4x\ \phi(x) f(x) = \sigma(Ef,\phi)\ .
\end{equation}
\end{enumerate}
\end{Lemma}
The fundamental solution plays an important role in the quantization procedures
to be described later. For example, it determines the values of the commutators
of the quantum fields. Even more important for us, it can be used to calculate
the four-smeared two-point distribution from the three-smeared two-point
distribution of the quantum fields, as we will see in section
\ref{TwoPointDistribution}.

The fundamental solution $E$ maps test functions on a globally hyperbolic
spacetime to solutions that arise from initial data on some Cauchy surface. More
precisely, this correspondence can be formulated as follows \cite{Dimock:1980}.
Define for a given Cauchy surface $\Sigma$ the restriction operator $\rho_0$ and
the forward normal derivative $\rho_1$ by
\begin{align}
\rho_0:\mathscr E(\mathcal M,\mathbb R) &\rightarrow \mathscr E(\Sigma,\mathbb R) & \rho_1:\mathscr E(\mathcal M,\mathbb R)&\rightarrow \mathscr E(\Sigma,\mathbb R)\\
\phi &\mapsto \phi\vert_{\Sigma}\ ,   & \phi &\mapsto
(n^\mu\nabla_\mu\phi)\vert_{\Sigma}\ ,
\end{align}
where $n^\mu$ denotes the unit forward normal to $\Sigma$ and $n^\mu\nabla_\mu$
is the Lie derivative in this direction. Now, the following holds. Let $\Sigma$
be any Cauchy surface and let $u,p\in \mathscr D(\Sigma,\mathbb R)$ be any pair of smooth
functions with compact support on the Cauchy surface $\Sigma$. Then, there
exists a unique solution $\phi\in\SS$ defined on all of $\mathcal M$
to the Klein Gordon equation \eqref{ScalarFieldEquation} that is related to its
initial values on $\Sigma$ by $\rho_0(\phi)=u$ and $\rho_1(\phi)=p$.
Furthermore, 
\begin{equation}\textstyle{
\supp \phi\subset\left(\bigcup_\pm J^\pm(\supp u)\right)\cup\left(\bigcup_\pm
J^\pm(\supp p)\right)\ ,}
\end{equation}
i.e., for any closed subset of $\Sigma$ the solution $\phi$
restricted to the corresponding domain of dependence depends only on the initial data in that subset. 

\subsection{Phase spaces}\label{PhaseSpaces}
The quantum theory of linear systems is usually modelled after the classical
theory as it becomes apparent in the 'Poisson bracket goes to commutator'
rule. A sophisticated version of this rule is used for linear fields
in curved spacetimes. Now, we introduce the classical part of this
correspondence, namely, the classical phase space.

A classical phase space is a symplectic vector space, i.e., a pair
$(\VV,\sigma)$, where $\VV$ is a vector space and $\sigma$ is a bilinear form
$\sigma:\VV\times\VV\rightarrow\mathbb R$ that is symplectic,
$\sigma(f,g)=-\sigma(g,f)$, and non-degenerate, which means if $\sigma(f,g)=0$
for all $f\in\VV$ then $g=0$. The space of solutions $\SS$ to the Klein-Gordon
admits a natural symplectic form defined by
\begin{equation}\label{SymplecticSolutions}
\sigmaS(\phi_1,\phi_2):=\int_{\Sigma} d^3 x \sqrt{\abs \h}\
n^\mu(\phi_1\nabla_\mu\phi_2-\phi_2\nabla_\mu\phi_1)\ .
\end{equation}
The integral is evaluated on a Cauchy surface $\Sigma$, but $\sigma$ is
independent of the particular choice of $\Sigma$. This is due to the
conservation, $\nabla_\mu j^\mu=0$, of the current
$j^\mu:=\phi\nabla^\mu\phi'-\phi'\nabla^\mu\phi$ as can be shown by the
application of Stokes' theorem to a timelike cylindrical spacetime volume bounded
by portions of Cauchy surfaces. The space $(\SS,\sigmaS)$ may be called the
covariant phase space of the theory. 

Alternatively, one can regard the canonical phase space. This is the space
\begin{equation}
\MM:=\{(u,p),u,p\in\mathscr D(\Sigma_t,\mathbb R)\} 
\end{equation}
of initial values on $\Sigma_t$ of the Klein-Gordon equation equipped with the symplectic form
\begin{equation}\label{SymplecticCauchy}
\begin{split}
\sigmaM(F_1,F_2)&:=-\int_{\Sigma_t} d^3x\ (u_1 p_2-u_2 p_1)\\
\end{split}
\end{equation}
for $F_i:=(u_i,p_i)\in\MM$, $i=1,2$, where $p_i:=\sqrt{\abs\h}(n^\mu\nabla_\mu
u_i)$ is the canonical conjugate to the configuration variable $u_i$. 

The relation between the distinct phase spaces can be summarized as follows: By
the unique correspondence between the solutions to the field equation and the
initial values on a given Cauchy surface $\Sigma_t$, the spaces $\SS$ and $\MM$
are isomorphic, i.e., there exists an isomorphism $\mathcal I_t:\MM\rightarrow\SS$. This
isomorphism induces a symplectic map $\sigmaM=\mathcal I_t^* \sigmaS$, where
$\mathcal I_t^*$ denotes the pulled back isomorphism. Consequently, both
phases spaces are equally well suited for quantization \cite{TorreVaradarajan}.

\section{Quantization}\label{SectionQuantization}
Let us introduce some basic notions regarding algebras
\cite{BaumgaertelWollenberg}. Let $\mathfrak A$ be an algebra over $\mathbb C$ with a map  $^*: \mathfrak A\rightarrow \mathfrak A$ such that for all
$A,B \in\mathfrak A$ and $\alpha,\beta\in\mathbb C$ we have
$(\alpha A + \beta B)^*=\Bar\alpha A^* + \Bar\beta
B^*$, $(AB)^*=B^*A^*$, and  $(A^*)^*=A$.
Then, $^*$ is called an involution and $\mathfrak A$ is called an
involutive algebra or a *-algebra (star-algebra). If $\mathfrak
A$ contains a unit element $\mathbb 1$ such that $\mathbb 1 A=A\mathbb 1$ for all $A\in\mathfrak
A$ then it is called a unital *-algebra. If the *-algebra $\mathfrak A$
is also a Banach space where the norm satisfies $\norm{A B}\leq
\norm{A}\norm{B}$ for all $A,B\in\mathfrak A$ then $\mathfrak A$ is a
Banach *-algebra. If, in addition, $\norm{A}^2=\norm{A^*A}$ then
$\mathfrak A$ is a C*-algebra. A *-subalgebra $\mathfrak I$ of
$\mathfrak A$ is a *-ideal if $AB,BA\in \mathfrak I$ for all
$A\in\mathfrak A$ and $B\in\mathfrak I$. A C*-algebra  $\mathfrak A$ is called
simple if it contains no non-trivial, i.e., different from $0$ and $\mathfrak
A$, closed *-ideals.

The observables of the quantized theory are represented
by the self-adjoint elements of a suitable algebra, e.g., a (unital) *-algebra
or, if a stronger structure is desired, a C*-algebra. So, owing to the different necessities,
there exist several formulations of the algebra of observables, which are
substantially equivalent, albeit technically inequivalent. In this section, we introduce the
formulation in terms of a Weyl algebra, which is a C*-algebra. For a
thorough treatment of the quantization scheme in curved spacetimes see \cite{WaldQFT,KayWald}. All relevant
facts regarding Weyl algebras and their representations can be found in
\cite{BratteliRobinson2}.

Based on each of the symplectic spaces defined in section \ref{PhaseSpaces},
we can define an abstract C*-algebra that obeys the canonical commutation relations
(CCR) via the Weyl construction. Consider a real symplectic vector space
$(\VV,\sigma)$.
\begin{Definition}\label{WeylAlgebra}
A Weyl algebra $\mathfrak W(\VV,\sigma)$ is a simple C*-algebra
with unit generated by objects $W(f)$ that are labeled by functions $f\in\VV$ and
that satisfy the relations
\begin{enumerate}
\item $W(0)=\mathbb 1$,
\item $W(f)^*=W(-f)$,
\item $W(f_1)W(f_2)=\e^{-\frac{\i}{2}\sigma(f_1,f_2)}W(f_1+f_2)$
\end{enumerate}
for all $f,f_1,f_2\in\VV$.
\end{Definition}
Condition (iii) is the Weyl form of the canonical commutation relations (CCR);
Thus, a Weyl algebra is often called a CCR algebra. The elements of $\mathfrak W$
represent the basic observables of the quantum theory. They  are bounded
operators, which avoids possible domain problems, and they correspond, formally,
to exponentiated field operators $W(f)=\e^{-\i\Phi(f)}$. This interpretation is
mathematically well defined in regular representations (see section \ref{AlgebraicConstruction}). Note that,
since $\VV$ is not given a topology, the elements $W(f)$ need not be continuous. Provided that
$\sigma(f,g)$ is non-degenerate, the Weyl algebra is unique in the sense that
given two Weyl algebras $\mathfrak W_1$ and $\mathfrak W_2$ there exists a unique
*-isomorphism $\alpha:\mathfrak W_1\rightarrow\mathfrak W_2$ such that for any
$W_1\in\mathfrak W_1$ and $W_2\in\mathfrak W_2$, we have $\alpha\cdot W_1=W_2$.

\subsubsection{States}
A state $\omega$ on the algebra of observables $\mathfrak A$ is a
positive, $\omega(A^*A)\geq 0$ , normalized, $\omega(\mathbb 1)=1$, linear functional $\omega:\mathfrak A\rightarrow\mathbb C$
for all $A\in\mathfrak A$. The set of all states is a convex set, i.e., a mixture
$\omega:=\lambda_1\omega_1+\dots+\lambda_n\omega_n$ of states
$\omega_1,\dots,\omega_n$ with $\lambda_i\geq 0$, $\sum \lambda_i=1$ is again a
state. A pure state is extremal in this set in the sense that it cannot be
expressed as the sum of two other states with positive coefficients $\lambda_i$.

The general definition of states on an algebra is, on the one hand, clear and
concise, but on the other hand, far to general for concrete applications. The
space of states satisfying these conditions is enormous and requires further
criteria that single out subspaces of states that are appropriate for a given
physical situation. On a Weyl algebra one typically restricts attention to the
class of regular states, which allows to introduce the quantum fields $\Phi(f)$
as self-adjoint generators of the Weyl elements $W(f)$. A further condition is to
require the states to be quasi-free, i.e., to be completely specified by their
two-point distribution, which makes the set of states tractable without removing
most of the physically interesting states. Further conditions that we will use
are homogeneity and isotropy, and the Hadamard condition. All these
requirements will be introduced in due place.

The algebraic approach makes contact with  the traditional Hilbert space
formulation of quantum theory via the GNS theorem, which says that every state
$\omega$ on a C*-algebra $\mathfrak A$ gives rise to a representation of
$\mathfrak A$ on some Hilbert space. 
\begin{Theorem}[GNS construction] Let
$\omega$ be a state on a C*-algebra $\mathfrak A$ with unit-element. Then,
there exists a complex Hilbert space $\HH_\omega$, a unit-preserving representation $\pi_\omega$ in
terms of linear operators on $\HH$ , and a vector $\Omega_\omega\in\HH_\omega$
such that
\begin{equation}
\omega(A)=\angles{\Omega_\omega,\pi_\omega(A)\Omega_\omega}_{\HH_\omega}
\end{equation}
for all $A\in\mathfrak A$. The vector $\Omega_\omega$ is cyclic, i.e.,
$\pi_\omega(\mathfrak A)\Omega_\omega$ is dense in $\HH_\omega$. The
representation $\pi_\omega$ is unique up to unitary equivalence.
\end{Theorem}
The triple $(\HH_\omega,\pi_\omega,\Omega_\omega)$ is called the GNS
 triple and the representation is called the GNS representation.

The  folium of a state $\omega$ is the set of all states that can be
represented as density matrices $\rho$ in the GNS representation of
$\omega$. So the folium consist of all states of the form
\begin{equation}
\omega(A)=\tr \rho \pi(A)\ .
\end{equation}
with a positive, trace class, i.e., $\tr \rho < \infty$, operator $\rho$.       
An important theorem due to Fell \cite{Fell} states that the folium of a
faithful representation of a C*-algebra is weakly dense in the set of all states. Since every physical
experiment consists of a finite number of measurements and, furthermore, these
measurements have limited accuracy, it is impossible to determine more than a
weak neighborhood of a state. Thus, by Fell's theorem, it is impossible to find
out in which folium the state lies. Note that, since all Weyl algebras are
simple \cite{Simon:1972}, all their representations are faithful. For a
discussion of further implications of Fell's theorem see \cite{Haag,WaldQFT}.

The GNS representations of different states need not be unitarily equivalent. In
fact, the Stone-von~Neumann uniqueness theorem fails for systems with infinitely
many degrees of freedom and it is known that infinitely many inequivalent,
irreducible Hilbert space representations of the Weyl algebra exist (see, e.g.,
\cite{WaldQFT}). Consequently, the folium of a single state does
not encompass all possible algebraic states. This is in contrast to the finite case,
where all irreducible regular representations, in particular, the Schr\"odinger
and the Heisenberg representation, are unitarily equivalent.

A representation $\pi$ on $\HH_\pi$ of the Weyl algebra $\mathfrak W$ is called
regular if the unitary groups $\lambda\mapsto \pi(W(\lambda f))$,
$\lambda\in\mathbb R$ are strongly continuous for all $f$. If $\pi$ is
regular, one can, by Stone's theorem, introduce self-adjoint infinitesimal generators $\Phi_\pi(f)$ of the
Weyl elements, which act on $\HH_\pi$. These operators may then be
used to define annihilation and creation operators (see theorem \ref{GNSFockRepresentation}).
A state $\omega$ on the Weyl algebra $\mathfrak W$ is said to be regular if its
GNS representation is regular.

 An automorphism $\alpha$ on a *-algebra $\mathfrak A$ is a one-to-one linear
 mapping of the algebra onto itself that is compatible with the algebraic
 structure, i.e., it satisfies
$\alpha(A\cdot B) = \alpha(A)\cdot\alpha(B)$ and $\alpha(A^*) = \alpha(A)^*$ for
all $A,B\in\mathfrak A$. A classical symplectic transformation on $(\VV,\sigma)
$ is a map that leaves the symplectic form invariant, i.e.,  a symplectic
transformation is given by an operator $\mathcal T:\VV\rightarrow\VV$ such that
$\sigma(\mathcal Tf_1,\mathcal Tf_2)=\sigma(f_1,f_2)$ for all $f_1,f_2\in\VV$.
A symplectic transformation on a classical symplectic vector space corresponds
directly to an automorphism on the associated Weyl algebra. For example, the time
translation on stationary spacetime is implemented on the classical phase
$(\SS,\sigmaS)$ by a one-parameter group of symplectic transformations $\mathcal
T_t:\SS\rightarrow\SS$, which gives rise to a one-parameter group of automorphisms
\begin{align}
\alpha_t:\mathfrak W\rightarrow \mathfrak W\ ,\quad  
\alpha_t(W(\phi)) :=W(\mathcal T_t \phi))
\end{align}
for all $\phi\in\SS$. We note that a pair $(\mathfrak A,\alpha_t)$ of a C*-algebra
$\mathfrak A$ and a strongly continuous automorphism group
$\{\alpha_t\}_{t\in\mathbb R}$ acting on $\mathfrak A$ is called a 
C*-dynamical system. This kind of system provides the basis for
the definition of KMS states (see section
\ref{KMSStates}) and passive states (section \ref{StabilitityConditions}).

Given a Hilbert space representation $\pi$ of $\mathfrak A$ on some Hilbert space
$\HH$, we say that the symplectic transformation $\mathcal T$ is unitarily
implementable if there exists a unitary transformation $U:\HH\rightarrow\HH$ such
that
\begin{equation}\label{UnitaryEquivalence}
U\pi(A)U^{-1}=\pi(\alpha\cdot A)
\end{equation}
for all $A\in\mathfrak A$. While there is no problem with the unitary
implementation of time-translations in stationary spacetimes, the situation
changes significantly for non-stationary spacetimes. The two-parameter family
of symplectic transformations $\mathcal T_{t_2,t_1}$, describing time-evolution in
that case, gives rise to a family of automorphisms $\alpha_{t_2,t_1}$ on the
algebra. However, these automorphisms are no longer implementable as unitary
operators on a Fock space, as it has been shown in \cite{TorreVaradarajan} for
the Klein-Gordon field on the torus $\mathbb T^2$ with non-flat Cauchy surfaces.
One may actually conjecture that only transformations defined by the isometry
group of a spacetime can be represented as unitary transformations on a Hilbert
space.

\subsection{Hilbert space quantization}
In this section, we review the quantization of a linear, scalar field in a
formalism that is close to the traditional Hilbert space quantization and
directly applicable to curved spacetimes. The formalism starts with a
real, symplectic vector space - in our case, the vector space $(\SS,\sigmaS)$ of
solutions to the Klein-Gordon equation. Then, on $(\SS,\sigmaS)$ an inner
product $\mu$ with suitable properties is chosen, which gives rise to a map $\K$ from
$\SS$ to a (complex) Hilbert space $\HH$. The quantum field theory is then constructed on the symmetric
Fock-space $\FF_s(\HH)$ associated to the one-particle space $\HH$. We do not
give all details of the construction, just the general procedure. The
authoritative references caring for all contingencies are \cite{KayWald,WaldQFT}.

First, we need to construct an inner product structure on the
real, symplectic vector space$(\SS,\sigmaS)$. So, choose any positive,
symmetric, bilinear map $\mu:\SS\times\SS\rightarrow \mathbb R$ such that
\begin{equation}\label{NonUniqueScalarProduct}
\frac{1}{4}{\sigmaS(\phi_1,\phi_2)}^2\leq\mu(\phi_1,\phi_1)\mu(\phi_2,\phi_2)
\end{equation}
for all $\phi_1,\phi_2\in\SS$. Since $\sigmaS$ is non-degenerate, the map $\mu$
defines a real inner product on the vector space $\SS$.

One can show that there always exists a $\mu$ satisfying
\eqref{NonUniqueScalarProduct}, but in a general curved spacetime there is no way
to select a preferred one. While in the case of theories with finitely many
degrees of freedom different choices of $\mu$ lead to unitarily equivalent
theories, in the non-finite case the theories turn out to be, in general,
unitarily inequivalent. In a stationary spacetime an operator $\K$
associated to $\mu$ can be defined that projects solutions in $\SS$ onto the subspace of
`positive frequency solutions'. These solutions have positive frequency in a
generalized sense, namely, with respect to the timelike Killing vector field
$\xi^\mu$ present in a stationary spacetime
\cite{Ashtekar:1975zn,Kay:1978yp,WaldQFT}.

For non-stationary spacetimes there is
no unique subspace of positive frequency solutions on which $\K$ could project.
Nevertheless, one can proceed with a non-unique decomposition of $\SS$ by the
following results due to Kay and Wald.
\begin{Theorem}[\cite{KayWald}]\label{KayWald}
Let $\SS$ be a real vector space on which are defined both a
bilinear symplectic form $\sigmaS$ and a bilinear positive symmetric
form $\mu$ satisfying \eqref{NonUniqueScalarProduct}. Then, one can always find
a complex Hilbert space $\HH$ together with a real-linear map $\K:\SS\rightarrow \HH$ such that
\begin{enumerate}
\item the complexified range of $\K$, i.e., $\K\SS+\i \K\SS$,
is dense in $\HH$,
\item $\mu(\phi_1,\phi_2)=\Re\angles{\K\phi_1,\K\phi_2}_\HH$ for all
$\phi_1,\phi_2\in\SS$,
\item $\sigmaS(\phi_1,\phi_2)=2\Im\angles{\K\phi_1,\K\phi_2}_\HH$ for all
$\phi_1,\phi_2 \in\SS$.
\end{enumerate}
Moreover, the pair $(\K,\HH)$ is uniquely determined up to
equivalence, where we say $(\K',\HH')$ is equivalent to
$(\K,\HH)$ if there exists an isomorphism $V:\HH\rightarrow \HH'$ such that $V
\K=\K'$.
\end{Theorem}
So, to every triple $(\SS,\sigmaS,\mu)$ there corresponds a pair
$(\HH,\K)$. The pair $(\HH,\K)$ is called a
one-particle Hilbert space structure.
The equations given in theorem \ref{KayWald} are often stated in the form
\begin{equation}
\angles{\K\phi_1,\K\phi_2}_\HH=\mu(\phi_1,\phi_2)+\frac{\i}{2}\sigmaS(\phi_1,\phi_2)\
.
\end{equation}  
A corresponding operator $\overline \K:\SS\rightarrow \overline{\HH}$ can be
defined which projects into the subspace of `negative frequency solutions', where
$\overline{\HH}$ is the complex conjugate Hilbert space to $\HH$. To remind the
reader, the complex conjugate space $\Bar\HH$ differs from $\HH$ by the scalar
multiplication: $c\odot f=\overline c\cdot f$, $f \in\HH$, $c\in\mathbb C$,
where the bar denotes complex conjugation. One may as well say that an antilinear
isometry $\Gamma$ satisfying $\Gamma^2=1$ makes the transition between the spaces
$\HH$ and $\overline{\HH}$. It follows immediately that $\K+\overline \K=\mathbb
1$.

Once having defined a one-particle structure $\K$, the remaining quantization
procedure is straightforward. Define by 
\begin{equation}
\FF_s(\HH):=\mathbb C\oplus\HH\oplus(\HH
\otimes_s \HH)\oplus\dots
\end{equation}
the symmetric Fock space over the one particle Hilbert
space $\HH$. To each solution in $\SS$ a corresponding operator
$\sigmaS(\Phi,\cdot)$ on the Fock space $\FF_s(\HH)$ is defined by
\begin{equation}
\label{SymplecticallySmearedOperator}
\sigmaS(\Phi,\phi):=\i a({\K\phi})-\i a^*(\K\phi)\ ,
\end{equation}
where the standard creation and annihilation operators $a^*,a$ on $\FF_s(\HH)$
satisfy
\begin{align}
[a(\psi),a^*(\psi')]&=\angles{\psi,\psi'}_\HH\ ,\\
a(\psi)\Omega&=0
\end{align}
for all $\psi,\psi'\in\HH$ and the vacuum state
$\Omega:=\mathbb 1\oplus \mathbb 0\oplus \mathbb 0\oplus\dots$. This
concludes the construction of the quantum field theory.

The quantization scheme presented here, leads directly to a Fock space
representation of the algebra. However, as already noted, these
representations need not be unitarily equivalent. If $\mu_1\neq \mu_2$ the
resulting representations $\{\FF_s(\HH_1),\sigmaS_{\!,1}(\Phi,\phi)\}$ and
$\{\FF_s(\HH_2),\sigmaS_{\!,2}(\Phi,\phi)\}$ may be unitarily inequivalent. It
follows immediately that different notions of 'particles' arise by different
choices of $\mu$. In a spacetime without time translation symmetry no
preferred choice for $\mu$ exists. On stationary
spacetimes, owing to the existence of a timelike Killing field, a satisfactory definition of
a preferred $\mu$, and thus a meaningful notion of 'particles' can be given
\cite{Ashtekar:1975zn,Kay:1978yp}.

The viewpoint taken in algebraic quantum field theory is that unitary
equivalence on the level of representations is not fundamental to the quantum theory. Rather,
the algebraic relations between the collection of operators
$\{\sigmaS(\Phi,\phi)\}$ on $\FF_s(\HH)$ are important. Hence, one postulates
that a net of local algebras satisfying these relations completely determines the
quantum theory and the fields play a role similar to coordinates in differential
geometry: useful tools for daily work, but dispensable for the essential
assertions.

Before we come to the algebraic formulation, let us note that there is an
alternative procedure for the selection of a representation. Rather than
specifying the bilinear form $\mu$, one may define a complex structure $J$, i.e.,
a bounded linear map  satisfying $J^2=-\mathbb 1$, on $\SS$ for which
$-\sigma(\phi,J\phi)$ is a positive definite inner product. This procedure is,
e.g., used in the geometric quantization programme \cite{Woodhouse:1992de}. Its
application and importance for quantum field theory in curved spacetimes is
illustrated best in a series of papers by Ashtekar and Magnon-Ashtekar
\cite{Ashtekar:1975zn,Ashtekar:1980ya,Ashtekar:1980yw}. Both schemes, the choice
of an inner product $\mu$ and the choice of a complex structure $J$ are roughly
equivalent. See, once more, \cite{WaldQFT} for details.

\subsection{The algebra of observables}\label{AlgebraicConstruction}
\subsubsection{The Weyl algebra}
We have seen how a Fock space representation $\{\FF_s(\HH),\sigmaS(\Phi,\phi)\}$
of the algebra of observables can be obtained via the choice of a bilinear form
$\mu(\phi,\phi)$. Now, a Weyl algebra (see definition \ref{WeylAlgebra}) can be
defined through the unitary operators
\begin{equation}\label{WeylOperatorSymplectic}
W(\phi):=\e^{- \i \sigmaS(\Phi,\phi)}\ .
\end{equation}
These operators satisfy the Weyl relations
\begin{align}
W(\phi)^*&=W(-\phi)\ ,\\
W(\phi_1)W(\phi_2)&=\e^{-\frac{\i}{2}\sigmaS(\phi_1,\phi_2)}W(\phi_1+\phi_2)\ .
\end{align}
The C*-completion, which is known to exist, of the space generated by all
$W(\phi)$ via formal finite sums $\sum_i \lambda_i W(\phi_i)$ comprises a Weyl
algebra, which can be seen as the minimal algebra of observables of a
quantum field theory in curved spacetime. This makes sense, since the Weyl algebras
arising from different choices of $\mu$ are always isomorphic even if the
corresponding Fock space representations are unitarily inequivalent
\cite{Slawny}.

A second Weyl algebra, which is isomorphic to the former via lemma
\ref{FundamentalSolutionLemma}, can be constructed as follows. Define using the
third property of lemma \ref{FundamentalSolutionLemma} for
each $f\in\TT$ a smeared operator on $\FF_s(\HH)$ by
\begin{equation}
\Phi(f):=\i a({\K(E f)})-\i a^*(\K(E f))\ ,
\end{equation}
and to consider the Fock space  $\{\FF_s(\HH),\Phi(f)\}$. The operators 
\begin{equation}\label{WeylOperatorInitial}
W(f):=\e^{-\i \Phi(f)}\ ,
\end{equation}
where we set $W(f'):=W(f)$ if $Ef=Ef'$ again define a Weyl algebra, 
\begin{align}\label{WeylRelationsTestFunctions}
W(f)^*&=W(-f)\ ,\\
W(f_1)W(f_2)&=\e^{-\frac{\i}{2} E(f,g)}W(f_1+f_2)\ , 
\end{align}
where  
\begin{equation}
E(f_1,f_2):=\int d^4x f_1 E f_2=\sigmaS(E f_1,E f_2) \ .
\end{equation}

\paragraph{The field algebra}
An abstract definition of a *-algebra $\mathfrak A$ of field operators for a
real, linear, scalar field can be obtained as follows. Take the unit $\mathbb
1$ and the formal smeared field operators $\Phi(f)$, where $f\in\DD(\mathcal M)$ and demand that
\begin{enumerate}
\item $\Phi(f)$ is linear,
\item $\Phi(f)$ is hermitian: $\Phi(f)=\Phi(f)^*$, 
\item $\Phi(f)$ satisfies the field equation:  $\Phi([\Box_\g+\m^2]f)=0$,
\item $\Phi(f)$ satisfies the commutation relations.
$[\Phi(f_1),\Phi(f_2)]=\i E(f_1,f_2)\mathbb 1$.
\end{enumerate}
To be precise, one takes the free algebra over the field of
complex numbers generated by the symbols $\Phi(f),\Phi(f)^*$, and $\mathbb 1$
and divides by the *-ideals generated by the properties stated above.

The formal correspondence between the field algebra and the Weyl algebra becomes
mathematically well defined in regular GNS representations. For those
representations, as already noted, the map
$\lambda\mapsto\pi(W(\lambda f)=\pi(\e^{-\i\Phi(\lambda f)})$,
$\lambda\in\mathbb R$ defines a strongly continuous unitary group for every $f$.
Thus, by Stone's theorem, the operators $\Phi(f)$ are self-adjoint generators of these groups. These generators satisfy the requirements for a *-algebra of
fields by lemma \ref{FundamentalSolutionLemma}. The collection of all such
operators may, equally well as their Weyl counterparts, be interpreted as the
collection of fundamental observables of the theory. The same is valid for the
collection of $\sigmaS(\Phi,\phi)$ in a regular representation.

\subsection{Quasi-free states} 
There is a distinct class of regular algebraic states, namely, the quasi-free
states, also known as Gaussian states, which have GNS Hilbert spaces that look
like the familiar Fock spaces built on a one-particle Hilbert space (see theorem
\ref{GNSFockRepresentation} below). The class of quasi-free states contains the
usual vacuum states in stationary spacetimes  as well as other vacua obtained by
mode decomposition of the field operators. In quantum statistical mechanics,
quasi-free states represent the general form of equilibrium states for
free bosonic systems \cite{BratteliRobinson2,HoneggerRieckers}. Besides this,
quasi-free states are well suited for calculations as they are exclusively determined by
their two-point distribution. 

A quasi-free state can be defined as an abstract linear functional of a special
kind on the algebra of observables, or in a regular GNS representation by the
requirement that it is completely fixed by the two-point distribution.  Let us
investigate the first possibility. To define a state on a Weyl algebra $\mathfrak
W$, it suffices to specify its expectation values on the Weyl operators
$W\in\mathfrak W$. As before, let $\mu$ be a real scalar product on $\SS$ and
consider the Weyl algebra $\mathfrak W(\SS,\sigmaS)$. Define a functional
$\omega_\mu:\mathfrak W\rightarrow \mathbb C$ by
\begin{equation}\label{QuasifreeState}
\omega_\mu(W(\phi))=\e^{-\frac{1}{2}\mu(\phi,\phi)}
\end{equation}
for all $\phi\in\SS$. Its action is extended to the whole algebra by linearity
and continuity. Now, if $\mu$ satisfies \eqref{NonUniqueScalarProduct} then
$\omega_\mu$ is positive on the whole algebra and thus a state. Any
quasi-free state $\omega_\mu$ can be realized in a representation as the vacuum
state in a Fock space by the following theorem.
\begin{Theorem}[\cite{KayWald}]\label{GNSFockRepresentation}
Let $(\K,\HH)$ be the one-particle Hilbert space structure obtained
from $\omega_\mu$ by Theorem \ref{KayWald}.
The GNS-triple $(\HH_{\omega_\mu},\pi_{\omega_\mu},\Omega_{\omega_\mu})$ of the
state $\omega_\mu$ is equivalent to a triple
$(\FF(\HH),\pi^\FF,\Omega^\FF)$ with the following properties.
\begin{enumerate}
\item The GNS space $\FF(\HH):=\FF_s(\HH)$ is the symmetric Fock space built
on $\HH$.
\item The representation $\pi^\FF$ is specified by 
$\pi^\FF(W(\phi))= \e^{-\Bar{\left[ a({\K\phi}) -
 a^*(\K\phi)\right]}}$\ ,
where the bar denotes the closure.
\item The state $\Omega^\FF:=\mathbb 1\oplus \mathbb 0\oplus \mathbb
0\oplus\dots$ is the (cyclic) Fock vacuum in $\FF_s(\HH)$.
\end{enumerate} 
\end{Theorem}
The purity of $\omega_\mu$ is equivalent to the irreducibility of the
representation $\pi^\FF$. Moreover this is equivalent to the property that
$\K\SS$ alone, rather than $\K\SS+\i\K\SS$, is dense in $\HH$ \cite{KayWald}.

The two-point distribution of a quantum field in the state $\omega$
on the Weyl algebra $\mathfrak W(\SS,\sigmaS)$ is defined by
\begin{equation}
\angles{\sigmaS(\Phi,\phi_1)\sigmaS(\Phi,\phi_2)}_\omega:=-\left.\frac{\partial^2}{\partial
s\partial t}\left(\omega(W(s\phi_1+t\phi_2))\e^{-\i st\sigma(\phi_1,\phi_2)/2}\right)\right\vert_{s=t=0} \end{equation} provided that the
right hand side exists.  For quasi-free states, as defined by
\eqref{QuasifreeState} the two-point distribution always exists and is given by
\begin{equation}
\omega^{(2)}(\phi_1,\phi_2)=\mu(\phi_1,\phi_2)+\frac{\i}{2}\sigmaS(\phi_1,\phi_2)\
.
\end{equation}  

If one works with the field algebra of operators $\Phi(f)$, a quasi-free state
$\omega$ can be defined, by the requirement that all odd $n$-point
distributions vanish and the even ones are determined by the two-point
distribution via the the recursion formula
\begin{equation}
\omega^{(2j)}(\Phi(f_1),\dots,\Phi(f_{2j})):=
\sum_\sigma\prod_{i=1}^j
\omega^{(2)}(f_{\sigma(i)},f_{\sigma(i+j)})
\end{equation} for $j\in\mathbb N$, where the sum
is taken over all permutations $\sigma$ of $\{1,2,\dots,2j\}$ with
$\sigma(1)<\sigma(2)<\dots<\sigma(j)$ and $\sigma(i)<\sigma(i+j), i=1,\dots,j$.

According to theorem \ref{GNSFockRepresentation}, the two-point
distribution $S(f_1,f_2)$ of a quasi-free state can be calculated in the GNS
representation by
\begin{align}
S(f_1,f_2)&:=\angles{\Omega^{\FF},\Phi(f_1)\Phi(f_2)\Omega^{\FF}}\\
&=\angles{\K E f_1,\K E f_2}_{\HH}\\
&=\mu(E f_1,E f_2)+\frac{\i}{2}\sigma(E f_1,E f_2)
\end{align}
Thus $S(f,f)=\mu(Ef,Ef)=\angles{\K E f,\K E f}_{\HH}$ for $f\in\mathscr
D(\mathcal M)$ and we see that we can define the generator
\eqref{QuasifreeState} of a quasi-free state either by the use of $\mu$, $S$, or
$\angles{\ ,\ }_\mathscr H$. 

Saturating the inequality \eqref{NonUniqueScalarProduct} turns out to be equivalent to the irreducibility of the
GNS representation arising from $\omega_\mu$. Thus, (remember the comments after
theorem \ref{GNSFockRepresentation}) such a $\mu$ results in pure quasi-free
states. However, already in stationary spacetimes, thermal equilibrium states at
finite temperature are represented by quasi-free states which fail to satisfy the
saturated version of \eqref{NonUniqueScalarProduct}, and in a spacetime with a
non-compact Cauchy surface, states in the folium of a thermal equilibrium state
do not, in general, lie in the folium of any pure quasi-free state \cite{WaldQFT}.
Thus, in order to incorporate thermal states, we cannot restrict ourselves to the
case of pure states. Note that the common notion of a `vacuum state' or, more
generally, ground state corresponds to a pure, quasi-free state in the algebraic
approach. We note that all pure, quasi-free state are Fock states and
thus related through a Bogoliubov transformation \cite{ManuceauVerbeure}.

\subsection{Hadamard states}\label{HadamardStates}
Even in the restricted class of quasi-free states there exist many states that
cannot be considered physical; this should not be misunderstood
as the statement that all physical states are quasi-free. A further condition
that is believed to reasonably narrow down the class of states is the Hadamard
condition. It has several formulations and was used already for a long time (see, e.g.,
 \cite{DeWittBrehme}) before Kay and Wald \cite{KayWald} put it on
 a sound mathematical foundation. The Hadamard condition essentially restricts
 the singularity structure of the two-point distribution at coinciding points
 such that it comes close the Minkowski vacuum state.  On a heuristic level, one
 could say that in a Hadamard state the high frequency modes of the field are
 `close' to their ground state.

The principle motivation for the Hadamard condition comes from the
point-splitting renormalization scheme. The absence of a preferred vacuum state
makes the normal ordering procedure of standard quantum field theory nonviable in
a general curved spacetime. The point-splitting prescription is a replacement for
normal ordering in the sense that it defines sensibly differences of
stress-energy expectation values even in curved spacetime. 

The basic idea is as
follows. The calculation of the stress-energy tensor involves the calculation of
 objects like $\angles{\phi(x)^2}$. These objects are, in general, ill
 defined as they involve the calculation of products of two distributions at
 a point. Such products are only well defined in special
 cases, where the wavefront sets (see below) of the involved distributions 'fit'
 to each other \cite{Hoermander}. The point-split object
 $\angles{\phi(x)\phi(x')}$, however, makes sense as a bi-distribution on $\mathcal M$. Now, one demands that for physically reasonable states the singularity structure of $\angles{\phi(x)\phi(x')}$ as $x'$ approaches $x$ should be the same as for $\angles{\Omega,\phi(x)\phi(x')\Omega}$. This provided, the difference $\angles{\phi(x)\phi(x')}-\angles{\Omega,\phi(x)\phi(x')\Omega}$ is a
smooth function, which allows taking the coincidence limit $x'\rightarrow x$.

States satisfying the Hadamard condition yield a
renormalized stress-energy tensor $\T_{\mu\nu}$ that satisfies Wald's axioms
\cite{WaldQFT}, which are believed to be reasonable physical assumptions on
$\T_{\mu\nu}$. In brief, the assumptions are as follows. For states
$\omega,\omega_1,\omega_2$:
\begin{enumerate}
\item 
$\angles{\T_{\mu\nu}}_{\omega_1}-\angles{\T_{\mu\nu}}_{\omega_2}$ should be well
defined by the point splitting procedure.
\item $\T_{\mu\nu}$ should be locally covariant.
\item $\angles{\T_{\mu\nu}}_\omega$ should be conserved,
$\nabla^\mu\angles{\T_{\mu\nu}}_\omega=0$, for all states $\omega$.
\item In Minkowski spacetime, $\angles{\Omega,\T_{\mu\nu}\Omega}_\omega=0$.
\end{enumerate}
These assumptions fix the expectation value $\angles{\T_{\mu\nu}}_\omega$
uniquely up to a conserved local curvature term that is independent of $\omega$.
Actually, it is possible to locally construct a bi-distribution $\H(x,x')$ such
that if
\begin{equation}
\F_\omega(x,x'):=\angles{\phi(x)\phi(x')}_\omega-\H(x,x')
\end{equation}
is a smooth function for a state $\omega$ then $\angles{\T_{\mu\nu}}_\omega$ is
well defined. We refer the reader to \cite{WaldQFT} for subtleties of the
construction.

It has not only been proved that there exist many Hadamard states for the
linear, scalar field in a globally hyperbolic spacetime; One also knows that the
canonical ground state and the thermal equilibrium states in stationary, globally
hyperbolic spacetimes are quasi-free Hadamard states \cite{SahlmannVerch}.
Regarding interacting fields, the construction of renormalized perturbative
quantum field theory in a generic spacetime is possible for Hadamard states
\cite{BrunettiFredenhagen2000,HollandsWald}. On the other hand, only few
Hadamard states on curved spacetimes have been explicitly constructed.
Among these are the ground states on ultrastatic spacetimes
\cite{FNW,Junker:1996} and KMS states on ultrastatic spacetimes with compact
Cauchy surfaces; Furthermore, we mention the adiabatic vacuum states of
infinite order \cite{Junker:1996,Junker:2002}. None of these examples deals
with thermal states on a non-stationary spacetime.

\subsubsection{The formulation by Kay and Wald}
Let $t(x)$ be a any global time function that increases towards the future and
let $t(x,x'):=t(x)-t(x')$. Furthermore, let $\sigma(x,x')$ be the squared
geodesic distance, i.e.,
\begin{equation}
\sigma(x,x')=\pm\left( \int_\gamma \left\vert
\g_{\mu\nu}(y(\tau))\frac{dy^\mu(\tau)}{d\tau} \frac{dy^\nu(\tau)}{d\tau}
\right\vert^{\frac{1}{2}} \d\tau \right)^2\ ,
\end{equation}
where $\gamma$ is the unique geodesic connecting $x$ and $x'$ with
parametrization $y(\cdot)$ and the sign is plus for spacelike
$y(\cdot)$ and minus for timelike $y(\cdot)$. The geodesic distance
is well defined and smooth in the set $\mathcal O$ in
$\mathcal M\times\mathcal M$, which is taken to be a neighborhood of
the set of causally related points $(x,y)$ such that $J^+(x)\cap
J^-(y)$ and $J^+(y)\cap J^-(x)$ are contained within a convex normal
neighborhood. A convex normal neighborhood of a point $x$ is a
neighborhood $\mathcal U$ of $x$ such that there exists a unique geodesic
connecting each pair of points in $\mathcal U$ and staying entirely within
$\mathcal U$. A manifold can always be covered by such neighborhoods
\cite{HawkingEllis}.

For each $n\in\mathbb N$ and $\epsilon\in\mathbb R, \epsilon>0$ define a
complex-valued function in $\mathcal O$ by
\begin{equation}
G_\epsilon^{t,n}(x,x')=\frac{1}{(2\pi)^2}\left(\frac{u(x,x')}{\sigma^t_\epsilon(x,x')}+v^{(n)}(x,x')
\ln \sigma^T_\epsilon(x,x') \right)\ ,
\end{equation}
where
\begin{equation}
\sigma^t_\epsilon(x,x')=\sigma(x,x')+2\i\epsilon t(x,x')+\epsilon^2\ .
\end{equation}
The smooth function $u(x,x')$ is the square root of the van~Vleck-Morette
determinant, and $v^{(n)}(x,x')\in \EE(\mathcal O)$ is a real
function defined by the power series
\begin{equation}
v^{(n)}(x,x'):=\sum_{m=0}^n v_m(x,x') \sigma(x,x')^m\ ,
\end{equation}
where the $v_m(x,x')$ are determined by the Hadamard recursion
relations. The branch cut for the logarithm is taken to lie
along the negative real axis. The field equations and commutation
relations require that $u$ and $v$ are uniquely determined by the
local geometry, i.e., by the metric $\g_{\mu\nu}$ and its derivatives.
Of course, $G_\epsilon^{t,n}(x,x')$ is singular for coinciding points $x$ and
$x'$, but it is also singular for points connected by a null geodesic.

Let $\Sigma$ be a Cauchy surface and let $\mathcal N$  be a causal normal
neighborhood of $\Sigma$ \cite{KayWald}. Denote by $\mathcal O'$ an open
neighborhood in $\mathcal N\times\mathcal N$ of the set of causally related points  such that the
closure of $\mathcal O'$ in $\mathcal N\times\mathcal N$ is contained in
$\mathcal O$. Now, define a function $\chi(x,x')\in\mathscr E(\mathcal
N\times\mathcal N)$ with the property
\begin{align}
\chi(x,x')=
\begin{cases}
0\ , & \text{ for } (x,x')\not\in\mathcal O\\
1\ , & \text{ for } (x,x')\in\mathcal O' \ . 
\end{cases}
\end{align}
A state $\omega$ is said to be a Hadamard state if its two-point
distribution satisfies the following requirement: for each $n\in\mathbb N$ there exists a
$\C^n$-function $w^n(x,x')$ on $\mathcal N\times\mathcal N$ such that
for all $f_1,f_2\in \mathscr D(\mathcal N)$ we have
\begin{equation}
S(f_1,f_2)=\lim_{\epsilon\rightarrow 0}\int_{\mathcal
  N\times\mathcal N} d\mu(x)\ d\mu(x')  f_1(x)
f_2(x')\ \Lambda^{t,n}_\epsilon(x,x')\ ,
\end{equation}
where
\begin{equation}
\Lambda^{t,n}_\epsilon(x,x')= \chi(x,x')G_\epsilon^{t,n}(x,x')+ w^n(x,x')
\end{equation}
and the measures are the induced measures $d\mu(x):=d^4x \sqrt{\abs\g}$. The
functions $w^{n}(x,x')$ depend on the individual state in the folium. Note that
the function $\chi(x,x')$ introduces a localization of the singular object
$G_\epsilon^{t,n}(x,x')$ in spacetime, which has the consequence that the Hadamard condition only
cares for the ultraviolet modes of a field.

\subsubsection{The microlocal formulation}
For linear quantum fields the microlocal spectrum condition, which we will state
below, is equivalent to the Hadamard condition. This has been shown by
Radzikowski \cite{Radzikowski,Radzikowski:1996}. The microlocal spectrum
condition has been investigated and extended to curved spacetimes and Wick
powers of scalar fields in \cite{BFK}. In this sense, it is more general than the
original Hadamard condition as it also can be generalized to non-linear fields.
Microlocal analysis shifts the study of singularities of distributions from the
base space to the cotangent bundle. The 'micro-localization' property makes this
formulation well suited for curved spacetimes.  

We define the notion of wavefront sets by first introducing the set of regular
directed points. Let $u\in\mathscr D'(\mathbb R^n)$ be a distribution. A point
$(x,\xi)$ in the cotangent bundle of $\mathbb R^n$ is called a regular
directed point of $u$ if there exists a smooth function $\phi\in\mathscr D(X)$,
$X\subset\mathbb R^n$ which does not vanish at $x$ and such that for any
$m\in\mathbb N$ there exists a constant $B_{m}$ satisfying
\begin{equation}
\abs{\widetilde{\phi u}(\xi')}\leq  B_m (1+\abs{\xi'})^{-m}
\end{equation}
for all $\xi'$ in a conical neighborhood $U\subset\mathbb R^n\backslash\{0\}$ of
$\xi$. A neighborhood $U$ is called conical if $\xi\in U\  \Rightarrow\ t\xi\in
U, t\in \mathbb R^+$. Now the wavefront set $\WF(u)$ of a distribution
$u\in\mathscr D'(X)$ is the complement in $X\times\mathbb R^n\backslash\{0\}$ of
the space of all regular directed points of $u$.
\begin{Theorem}[Microlocal spectrum
condition]\label{MicrolocalSpectrumCondition} A quasi-free state $\omega$ of
the Klein-Gordon field on globally hyperbolic spacetime $\mathcal M$ is a (global) Hadamard state if and only if its two-point distribution has the wave-front set
\begin{equation}
\WF(\omega^{(2)}(x_1,x_2))=C^+\ ,
\end{equation}
where
\begin{equation}
C^+=\left\{ [(x_1,\xi_1),(x_2,\xi_2)]\in (T^*(\mathcal M)\backslash
0)\times(T^*(\mathcal M)\backslash 0); (x_1,\xi_1)\sim(x_2,-\xi_2); \xi_1^0\geq
0 \right\}\ .
\end{equation}
The notation $(x_1,\xi_1)\sim(x_2,\xi_2)$ means that $x_1$ and $x_2$
can be connected by a null geodesic such that $\xi_1^\mu$ is
tangential to $\gamma$ at $x_1$ and $\xi_2^\mu$ is the parallel
transport of $\xi_1^\mu$ along $\gamma$ at $x_2$.
\end{Theorem}
So, singularities in the two-point distribution $\omega^{(2)}(x,x')$ only occur
if $x$ and $x'$ are lightlike connected and the singularities have positive
frequencies. 

Wavefront sets have the property that for two distributions $u,v$ it
holds that
\begin{equation}
\WF(u+v)\subseteq \WF(u) \cup \WF(v)\ .
\end{equation}
Thus, $\WF(u)\subseteq \WF(u-v)\cup \WF(v)$ and $\WF(u)=\WF(v)$ if
$\WF(u-v)=\emptyset$, i.e., the distributions $u$ and $v$ have the same wavefront
set, and hence the same singularity structure if the difference of $u$ and $v$
is a smooth function.

\chapter{Scalar fields in Robertson-Walker spacetimes}\label{QFTinRW}
Robertson-Walker spacetimes are a particularly important class of spacetimes to
investigate. They are homogeneous and isotropic solutions of the Einstein
equations. These assumptions on the geometry of spacetime determine the solutions
up to three discrete types of spatial geometry -- flat, closed, and open geometry
-- and an arbitrary positive function $\a(t)$, which determines the expansion of
the spacelike sections. Although these spacetimes do not possess a time evolution
symmetry, they possess a preferred class of observers, namely, the ones who are
orthogonal to the homogeneous surfaces. These are also called isotropic (or
comoving) observers. Owing to the present symmetries, the field can be written,
in the usual way, as an integral (or sum, for the closed geometry) over modes. The spatial dependence of
the modes is known explicitly, but the time dependent part does not acquire the
usual form $\e^{\pm i\omega t}$. Rather, it satisfies a certain differential
equation with time-dependent coefficients. A determination of solutions to this
equation is possible by a method motivated by a WKB type approximation. This
led Parker to define the adiabatic vacuum states \cite{Parker:1969au}. The
adiabatic vacuum states are defined such that the particle creation is minimized
in an expanding universe. Later, the adiabatic vacuum states were redefined by
L\"uders and Roberts \cite{LR} in a more rigorous setting of
 quantum field theory in curved spacetime. 

In this chapter we introduce, first, the quantum theory of the scalar field in
the formulation used in \cite{LR}. We also give a brief account of adiabatic
vacuum states. Then, we quote the important theorem on the general form of the
two-point distribution of a homogeneous, isotropic, quasi-free state for the
scalar field in Robertson-Walker spacetime. Finally, we calculate, as a
preparation for the construction the almost equilibrium states, a
'four-smeared' version of the two-point distribution.

\section{The algebra and the states}
We have seen in chapter \ref{ChapterQFTCST} how the Weyl algebra or the field
algebra for a quantum field is constructed in a general, curved spacetime. There
is an alternative way to obtain the algebra of observables, which goes back
to Araki, namely, by constructing a self-dual algebra
\cite{Araki,ArakiShiraishi,ArakiYamagami}. The self-dual algebra is a
different route to the construction of the CCR algebra, distinguished by the
fact that one chooses a different set of generators that act on a complex
vector space $\KK$. This approach is used in \cite{LR}, where the
Weyl algebra associated to a self-dual algebra is is taken as the algebra of
observables.

\subsubsection{The self-dual algebra}
A self-dual algebra is based on a phase space triple $(\mathscr
K,\Gamma,\boldsymbol\gamma)$, where $\KK$ is a complex linear space, $\Gamma$
is an antilinear involution of $\mathscr K$, and $\boldsymbol\gamma$ is a Hermitian form on $\KK$ satisfying
\begin{equation}\label{ArakiGamma}
\boldsymbol\gamma(\Gamma f,\Gamma g)=-\boldsymbol\gamma(f,g)^*\ .
\end{equation} 
The space $\KK$ serves as the test function space for creation and annihilation
operators. The indefinite inner product $\boldsymbol\gamma(f,g)$ gives rise to
the canonical commutation relations and the involution $\Gamma$ combines
complex conjugation with the interchange of test functions for creation and
annihilation operators \cite{ArakiYamagami}.

For the Klein-Gordon field on a globally hyperbolic spacetime $(\mathcal
M,\g_{\mu\nu})$ one defines
\begin{gather}
\KK:=\mathscr D(\mathcal M)/[(\Box_\g+\mathrm m^2)\mathscr D(\mathcal M)]\ ,
\label{ArakiK} \\ \boldsymbol\gamma(f,g):= \i G(\bar f,g)\ , \label{ArakiSigma}
\end{gather}    
where $G(f,g):=\int d^4x\ fEg$, $f,g\in\KK$, and $\Gamma$ is defined
by the requirement
\begin{equation}
\Gamma f=\bar f\ . \label{ArakiI}
\end{equation}
Note that $\mathscr D(\mathcal M):=\C^\infty_0(\mathcal M)$ is a space of complex
functions, as opposed to the space of real functions used in the construction in
chapter \ref{ChapterQFTCST}, and the fundamental solution $E$ is defined in
equation \eqref{FundamentalSolution}. Equation \eqref{ArakiK} ensures that the
corresponding field satisfies the Klein-Gordon equation. The involution $\Gamma$
satisfies \eqref{ArakiGamma} by the properties of the Green's function $G(f,g)$.

On the phase space $(\mathscr K,\Gamma,\boldsymbol\gamma)$ one constructs the
self-dual CCR algebra $\mathfrak A(\KK,\Gamma,\boldsymbol\gamma)$ as follows.
First, consider the complex, free, *-algebra over $\KK$ generated by the
symbols $\phi(f)$, their conjugates $\phi(f)^*$, and the identity $\mathbb 1$, where
$f\in\KK$. Then, take the quotient of it by the two-sided *-ideal
that is generated by the relations
\begin{align}
&\text{$\phi(f)$ is complex linear in $f$}\ ,\\
&\phi(f)^* \phi(g)- \phi(g) \phi(f)^*=  \boldsymbol\gamma(f,g)\mathbb 1\ ,\\
&\phi(\Gamma f)^*=\phi(f)\ .
\end{align}
for $f,g\in\KK$. In order to circumvent problems with unbounded operators,
one passes, as usual, from $\mathfrak A(\KK,\Gamma,\boldsymbol\gamma)$ to an
associated Weyl algebra. The Weyl algebra $\mathfrak
W(\KK_\Re,\i\boldsymbol\gamma)$ is based on the real linear space
$\KK_\Re:=\{f\in\KK: \Gamma f=f\}$ equipped with the real symplectic form $\i\boldsymbol\gamma(f,g)$:
\begin{equation}
W(f)W(g)=\e^{-\frac{1}{2}\boldsymbol\gamma(f,g)}W(f+g)\ .
\end{equation}
Note that this is formally equal to \eqref{WeylRelationsTestFunctions}
if $\KK_\Re$ coincides with  $\TT$. A net of C*-algebras is obtained by taking
$\mathfrak A(\mathcal O)$ to be the C*-subalgebra generated by the Weyl elements $W(f)$, $f\in\DD(\mathcal M)$ with $\supp f\subset\mathcal O$.

\subsubsection{Quasi-free states} 
A quasi-free state for a real, linear, scalar field is specified by its
two-point distribution
\begin{equation}
S(f,g)=\omega_S(\phi(f)^*\phi(g))\ 
\end{equation}
for $f,g\in\KK$. This two-point distribution may be seen as a complex scalar product on the space
$\KK$. It specifies the corresponding quasi-free state in terms of Weyl operators
by
\begin{equation}
\omega_{S}(W(f)):=e^{-\frac{1}{2}S(f,f)}\ .
\end{equation}
The two-point distribution of a quasi-free state is a polarization on the
phase space $(\KK,\Gamma,\boldsymbol\gamma)$. This means that  $S(\cdot,\cdot)$
is a positive Hermitian form on $\KK$ such that
\begin{equation}\label{Polarization}
S(f,g)-S(\Gamma g,\Gamma f)=\boldsymbol\gamma(f,g)\ .
\end{equation}
Given $S(\cdot,\cdot)$ one defines a scalar product $(\cdot,\cdot)_S$ on $\KK$
by 
\begin{equation}
(f,g)_S=S(f,g)+S(\Gamma g,\Gamma f)
\end{equation}
so that by the Schwarz inequality and the triangle inequality one obtains \cite{ArakiShiraishi}
\begin{equation}
\abs{\gamma(f,g)}^2\leq (f,f)_S\  (g,g)_S\ ,
\end{equation}
which is the complex version of \eqref{NonUniqueScalarProduct}.

It is remarkably easy to characterize pure, quasi-free
states in this scheme. Namely, denote by $\KK_S$ the Hilbert space completion of $\KK$ by
$(\cdot,\cdot)_S$. On $\KK_S$ we have an operator $\hat S$ satisfying
$(f,\hat S g)_S=S(f,g)$. The state $\omega_S$ is a pure quasi-free state if
and only if $\hat S$ is a basis projection on $\KK_S$
\cite{ArakiShiraishi}.

\section{The spacetime and the field}
As already noted, Robertson-Walker spacetimes are homogeneous and isotropic
solutions to the Einstein equations. They are globally hyperbolic Lorentzian
manifolds with topology $\mathcal M^\varepsilon=\mathbb
R\times\Sigma^\varepsilon$, where $\varepsilon=0,+1,-1$ discriminates three
types of spacelike hypersurfaces. The Cauchy surfaces
$\Sigma^\varepsilon$ are homogeneous Riemannian manifolds with constant curvature of sign $\varepsilon$. The
homogeneous and isotropic spacetimes can be endowed with the
Robertson-Walker metrics
\begin{align}\label{RobertsonWalkerMetric}
ds^2=dt^2-\a(t)^2 \left[ \frac{dr^2}{1-\varepsilon r^2} + r^2 \left(d\theta^2 + \sin^2\theta\ d\varphi^2 \right) \right]\ , 
\end{align}
where the coordinates cover the ranges $r\in[0,\infty)\ ,\ \theta\in[0,\pi]\
,\ \varphi\in[0,2\pi]$ for $\varepsilon=0,-1$ and $r\in[0,1]$ for $\varepsilon=+1$.
The function $\a(t)$ is a strictly positive, smooth function describing the
expansion of the universe, and $H(t)=\frac{\dot \a(t)}{\a(t)}$ is
the Hubble parameter.

The induced metric on the Cauchy surfaces can be written as 
\begin{align}
ds^2&=dt^2-\a(t)^2 \s_{ij}^\varepsilon dx^i dx^j\ ,
\end{align}
where we wrote the induced metric on the Cauchy surface  as
$\h_{ij}^\varepsilon(t)=\a(t)^{2}\s_{ij}^\varepsilon$. Note that
the metric $\h_{ij}$ is time-dependent while $\s_{ij}$ is not. We
use $\Sigma$ to denote the manifold with metric $\s_{ij}$, while $\Sigma_t$ is
endowed with the metric $\h_{ij}$. The future directed normal fields of all the
hypersurfaces $\Sigma^\varepsilon$ are given by $n^\mu=(1,0,0,0)$. These fields
are geodesic, i.e., $n^\mu\nabla_\mu n^\nu=0$. 

It is convenient to regard the Cauchy surfaces $\Sigma^\varepsilon$ as being
embedded in $\mathbb R^4$ by
\begin{align}
\Sigma^{0}\,&=\Bigg\{ x\in\mathbb R^4: x^0=0  \Bigg\}\ ,\\
\Sigma^{+}&=\left\{ x\in\mathbb R^4: (x^0)^2+\sum_{i=1}^3 (x^i)^2=1   \right\}\ ,\\
\Sigma^{-}&=\left\{ x\in\mathbb R^4: (x^0)^2-\sum_{i=1}^3 (x^i)^2=1\ , x^0>0 \right\}\ .
\end{align}
Generally speaking, one calls the spacetime $\mathcal M^+$ a closed universe since
$\Sigma^+$ is compact. It is also customary to call $\mathcal M^-$ and $\mathcal
M^0$ open and flat universes, respectively. Each of the Cauchy surfaces
$\Sigma^\varepsilon$ is a homogeneous surface for a different symmetry
group $\G^\varepsilon$. In detail, these are the groups: $\G^0=\boldsymbol E(3)$, the Euclidean group;
$\G^+=\boldsymbol{SO}(4)$, the rotation Group; and $\G^-=\mathcal
L^\uparrow_+(4)$, the Lorentz group.

\subsubsection{The field equation}
In Robertson-Walker spacetimes the Klein-Gordon equation
\eqref{ScalarFieldEquation} becomes
\begin{equation}
\frac{\partial^2}{\partial t^2}\phi(t,\vec x)+ 3H(t)
\frac{\partial}{\partial t}\phi(t,\vec x)+(-\Delta_\h+m^2)\phi(t,\vec x)=0\ ,
\end{equation}
where $\Delta_\h$ denotes the Laplacian on $\Sigma^\varepsilon_t$. If one assumes
that the field modes $u_\vec k(x)=\Bar T_{\k}(t) Y_{\vec k}(\vec x)$ and their
complex conjugates form a complete orthonormal basis, the general
solution of the Klein-Gordon equation can be written as
\begin{align}\label{GeneralHermitianSolution}
\phi(t,\vec x)
&=\int\d\vec k\left[ \Bar T_{\k}(t) Y_{\vec k}(\vec x)a(\vec k)+T_{\k}(t)\Bar
Y_{\vec k}(\vec x) a(\vec k)^*\right]\ .
\end{align}
We think of $a(\vec k),a(\vec k)^*$ as arbitrary complex coefficients with no
reference to particles. This decomposition of the field is valid on all types of
Robertson-Walker spacetimes if one defines the measure $\d\vec k$ accordingly. Our notation is as follows:
\begin{subequations}\label{Measures}
\begin{align}
\label{Measure2}\varepsilon& = 0\phantom{+} : & \int\d\vec
k&:=\int_{\mathbb R^3}d^3\vec k\ , & \vec k&:=(k_1,k_2,k_3)\in\mathbb R^3,\ 
\k:=\abs{\vec k}\ , \\
\label{Measure1}\varepsilon& =+1: & \int\d\vec k&:=\sum_{\k=0}^\infty
\sum_{l=0}^\k\sum_{m=-l}^l  , & \vec k&:=(\k,l,m)\in\mathbb N\times\mathbb
N\times\mathbb Z\ ,\\
 \label{Measure3}\varepsilon& =-1: & \int\d\vec
k&:=\int_{\mathbb R^3}d^3\vec k\ , & \vec k&:=(k_1,k_2,k_3)\in\mathbb R^3,\  \k:=\abs{\vec k}\ .
\end{align}
\end{subequations}
Note the subtle difference in notation between the absolute value of the
three-momentum, $\k:=\abs{\vec k}$, and the four-momentum
$k:=(k_0,k_1,k_2,k_3)=(k_0,\vec k)$. The functions $T_\k(t)$ depend on $\k$ and
the functions $Y_{\vec k}(\vec x)$ depend on $\vec k$.

\subsubsection{Homogeneous, isotropic, quasi-free states}
The groups $\G^\varepsilon$, act as isometry groups on the manifolds $\mathcal
M^\varepsilon:=\mathbb R\times\Sigma^\varepsilon$ by $\gg(t,\vec x)=(t,\gg\vec
x)$, $\gg\in\G^\varepsilon$. It follows that they must commute with the
fundamental solution $E$ and thus act as a group of transformations on
the phase space. In turn, this defines a group of (Bogoliubov) automorphisms
$\{\alpha_\gg\}$ on the Weyl algebra via
\begin{align}
\alpha_\gg(W(f)):=W(\gg f)
\end{align}
for $\gg\in\G^\varepsilon$. Now, we say that a state $\omega$ is
homogeneous and isotropic if $\omega\circ\alpha_\gg=\omega$,
$\gg\in\G^\varepsilon$. Since a quasi-free state $\omega_S$ is, by definition, uniquely determined by its two-point
distribution $S(f_1,f_2)$, this translates to the necessary and sufficient
condition
\begin{equation}
S(\gg f_1,\gg f_2)=S(f_1,f_2)
\end{equation}
for all $\gg\in\G^\varepsilon$. The $\G$-invariance of the two-point distribution
is analyzed in \cite{LR} by passing to the phase space $(\MMa,\sigmaMa)$ of
initial data for $\phi(x)$ at time $t$, where 
\begin{equation}
\MMa:=\{(u,\a(t)^3p),u,p\in\mathscr
D(\Sigma,\mathbb R)\}
\end{equation}
 and the symplectic form is a variant of \eqref{SymplecticCauchy}:
\begin{align}
\sigmaMa(F_1,F_2)&:=\a(t)^3\int_\Sigma \d\vec x \left(p_1 u_2 - u_1
p_2 \right)
\end{align}
for $F_i:=(u_i,\a(t)^3p_i)\in\MMa$, $i=1,2$, where $\d\vec x=d^3x\sqrt{\abs\s}$.
Please note that $\sigmaMa$ is defined using the measure derived from $\s_{ij}$
and not from $\h_{ij}$. 

By introducing suitable topologies on $\MMa$, one finds
that it is sufficient to compute the commutants of the unitary representations
$U^\varepsilon$ of $\G^\varepsilon$ on $\L^2(\Sigma)$ given by $U^\varepsilon_\gg
f:=f\circ \gg^{-1}$, $f\in \DD(\Sigma)$ \cite{LR}. The representations
$U^\varepsilon$ are decomposed into irreducible representations by
(generalized analogues to) Fourier transforms
\begin{equation}\label{FourierTransform}
\tilde f(\vec k):=\int \d\vec x\ \Bar Y_\vec k(\vec x)f(\vec x)
\end{equation}
for $f\in \L^2(\Sigma)$. 
In each case, the Fourier transform is a unitary operator from
$\L^2(\Sigma^\varepsilon)$ to $\L^2(\widetilde\Sigma^\varepsilon)$, where
$\widetilde\Sigma^\varepsilon$ denotes the momentum space associated to
$\Sigma^\varepsilon$. And, again in each case, a bounded operator on
$\L^2(\Sigma^\varepsilon)$ commuting with $U^\varepsilon$ corresponds on
$\L^2(\Sigma^\varepsilon)$ to a multiplication by a bounded function of $\k$.

The functions $Y_\vec k(\vec x)$ constitute an orthonormal basis of
eigenfunctions of the Laplacian $\Delta_\s$ in $\L^2(\Sigma)$ (see below for the
precise forms). The inverse Fourier transform is given by
\begin{equation}
f(\vec x)=\int\d\vec k\ Y_{\vec k}(\vec x)\tilde f(\vec k)
\end{equation}
and one has the usual completeness relations
\begin{align}
\int \d\vec k\ Y_\vec k(\vec x) \Bar Y_{\vec k}(\vec x')&=\delta(\vec x,\vec x') \label{FourierK}\ ,\\
\int \d\vec x\ Y_\vec k(\vec x) \Bar Y_{\vec k'}(\vec x)&=\delta(\vec k,\vec
k') \label{FourierX}\ .
\end{align}
The $\delta(\vec k, \vec k')$-distribution is
to be taken with respect to the measures $\d\vec k$ defined in \eqref{Measures}:
\begin{equation}
\int \d\vec k' f(\vec k')\delta(\vec k,\vec k')=f(\vec k)\ .
\end{equation}

\subsubsection{Spatial mode functions} 
We give a brief account of the functions $Y_\vec k(\vec x)$ in each of the three
types of Robertson-Walker spaces. What is needed is a direct sum or
direct integral decomposition of the space $\L^2(\Sigma^\varepsilon)$ in terms of
eigenfunctions of the Laplacian $\Delta_\s$, i.e.,
\begin{equation}
\Delta_\s Y_\vec k(\vec
x)=-E(\k)Y_\vec k(\vec x)\ .
\end{equation}
In each case, the decomposition is obtained by
different means, but it exists and allows to treat, to a great extent, the
solutions in the different types of Robertson-Walker spacetimes on an equal
footing \cite{LR}.

\paragraph{[$\varepsilon=0$]:} In the flat case we have
$\Delta_\h=\frac{1}{\a^2}\Delta_\s$. Thus, it follows that $\Delta_\s Y_{\vec
k}(\vec x)=-\k^2 Y_{\vec k}(\vec x)$ and that the generalized eigenvectors
$Y_\vec k(\vec x)$ are independent of $t$. The eigenvectors and their
negative eigenvalues are given by
\begin{align}\label{SpacelikeFlatSolution}
Y_\vec k(\vec x)&=\frac{1}{(2\pi)^{\frac{3}{2}}}e^{\i\vec k \vec x}\ , &
E(\k)&:=\k^2 \ .
\end{align}
The direct integral decomposition amounts to the ordinary Fourier
transform. Note the useful relations $\overline{Y_{\mathbf k}(\vec x)}=Y_{-\mathbf k}(\vec
x)$ and $\abs{Y_{\mathbf k}(\vec x)}^2=\frac{1}{(2\pi)^3}$.

\paragraph{[$\varepsilon=+1$]:} For the closed universe, the solutions
$Y_\vec k(\vec x)$ are the spherical harmonics
\begin{align}
Y_\vec k(\psi,\vartheta,\varphi)&=A_{\k l}\ \Pi^+_{\k l}(\psi)\
Y_{l,m}(\vartheta,\varphi)\ , & E(\k)&:=\k(\k+2)\ ,
\end{align}
($\k=0,1,\dots;\ l=0,1,\dots,\k;\ m=-l,-l+1,\dots,l$), where the $Y_{l,m}$ are the
harmonics on the 2-sphere, the $\Pi^+_{\k l}$ are real polynomials in
$\sin{\psi}$ and $\cos{\psi}$, and the $A_{\k l}$ are real normalization
constants. One has a direct sum
decomposition $L^2(\Sigma^+)=\bigoplus_{\k=0}^\infty\mathcal H_\k$, where
$\mathcal H_\k$ denotes the span of $Y_{\k,l,m}$ as $l$ and $m$ vary. 

\paragraph{[$\varepsilon=-1$]:} In the open universe the Cauchy surface
$\Sigma^-$ is regarded as being embedded in $\mathbb M$, $\xi=(1,\boldsymbol\xi)\in\mathbb
M$ and $x\cdot\xi$ is the Minkowski scalar product. A set of generalized
eigenvectors are
\begin{align}\label{SpacelikeSolutionOpen}
Y_{\vec k}(x)&=\frac{1}{(2\pi)^{\frac{3}{2}}}(x\cdot \xi)^{-1+\i\k}\ , &
E(\k)&:=\k^2+1\ ,
\end{align}
where $\vec k=\k\boldsymbol \xi\in\mathbb R^3$. Here, the Fourier transform is a
map with values in $\mathrm L^2(\mathrm S^2,d\Omega)$, where $\mathrm S^2$ is the
two-sphere embedded in $\mathbb R^3$, i.e., a function on the set of horospheres,
in the language of \cite{GelfandVilenkin5}.

\subsubsection{Time-dependence}
The time-dependent function $T_\k(t)$, which appears in
\eqref{GeneralHermitianSolution}, is required to be a solution to the
differential equation
\begin{align}\label{KGRWTimeDependentPart}
\ddot T_\k(t)+3H(t)\dot T_\k(t)+\omega_\k^2(t)T_\k(t)&=0\ ,
\end{align}
where the frequencies $\omega_\k(t)$ are given by
\begin{equation}\label{Frequencies}
\omega_\k^2(t):=\frac{E(\k)}{\a(t)^2}+\m^2 
\end{equation}
and, additionally, to satisfy the condition
\begin{equation}\label{Wronskian}
\overline T_\k(t)\dot T_\k(t)- T_\k(t)\dot{\overline T}_\k(t)
=\frac{\i}{\a(t)^3}\ .
\end{equation}
The explicit form of $E(\k)$ in each type of Robertson-Walker spacetime has been
given in the last section. The condition \eqref{Wronskian}, which amounts to the
Wronskian, determines the commutation relations of the $a(\vec k),a(\vec k)^*$
\cite{HNS,ParkerFulling:1974}.

Equation \eqref{KGRWTimeDependentPart} is an ordinary,
homogeneous, linear differential equation with variable coefficients. 
In case of a static spacetime it has the explicit solutions
\begin{equation}\label{StaticSolution}
T_\k(t)=\frac{1}{\sqrt{2\a^3\omega_\k}}\e^{-\i\omega_\k t}\ .
\end{equation}
In the general case it has a fundamental system  $T_\k(t),\Bar T_\k(t)$ of
solutions, which cannot be calculated explicitly.

\subsubsection{Adiabatic vacuum states} 
The only freedom one has in the construction of pure, homogeneous, isotropic,
quasi-free states is the choice of initial data for the function $T_\k(t)$. This
choice can be made on physical grounds. For example, one can chose initial
values such that the resulting states, in a certain sense, minimize particle creation in
an expanding universe and reduce to the known particle notion in the static
case. Then, one obtains the so-called adiabatic vacuum states \cite{Parker:1969au}.

In \cite{LR} the former definition of adiabatic vacuum states was put on a firm
basis. The adiabatic vacuum states were redefined by fixing the large $\k$
behaviour of  $T_\k(t)$ and $\dot T_\k(t)$ such that the resulting states adhered
the principle of local definiteness \cite{HNS}, which, roughly, requires
that the the set of expectation values measurable in a bounded region $\mathcal
O$ of the spacetime in a GNS representation of a state $\omega$  should not
depend on the state $\omega$. This rules out inequivalent representations of the
local algebra of bounded observables $\mathfrak A(\mathcal O)$ arising from
different global states, which would be considered a pathology. For a precise
definition and discussion of the principle of local definiteness and other
related notions see \cite{Verch:1994eg}.

One can obtain iterative solutions $T^n_\k(t)$ of \eqref{KGRWTimeDependentPart}
and hereby the adiabatic vacuum states \cite{Parker:1969au,LR}, by
using a WKB-type ansatz
\begin{align}
T_\k(t)=\frac{1}{\left(2a(t)^3\Omega_\k(t)\right)^{1/2}}\exp\left(\i\int_{t_0}^{t}dt\
\Omega_\k(t)\right)
\end{align}
with yet undetermined positive functions $\Omega_\k(t)$. This ansatz satisfies
the normalization, and it satisfies \eqref{KGRWTimeDependentPart} if
\begin{align}
\Omega_\k^2=\omega_\k^2-\frac{3}{4}\left(\frac{\dot\a}{\a}\right)^2-\frac{3}{2}\frac{\ddot\a}{\a}+\frac{3}{4}\left(\frac{\dot\Omega_\k}{\Omega_\k}\right)^2-\frac{1}{2}\frac{\ddot\Omega_\k}{\Omega_\k}\
.
\end{align}
Now start with an iteration
\begin{align}
\left(\Omega_\k^{(0)}\right)^2&=\omega_\k^2\ ,\\
\left(\Omega_\k^{(n+1)}\right)^2&=\omega_\k^2-\frac{3}{4}\left(\frac{\dot\a}{\a}\right)^2-\frac{3}{2}\frac{\ddot\a}{\a}+\frac{3}{4}\left(\frac{\dot\Omega_\k^{(n)}}{\Omega_\k^{(n)}}\right)^2-\frac{1}{2}\frac{\ddot\Omega_\k^{(n)}}{\Omega_\k^{(n)}}\
.
\end{align}
The iteration procedure breaks down when we obtain a negative value for
$\left(\Omega_k^{(n+1)}\right)^2$. This does not happen if one restricts to a
time interval, $t\in I$, where $I\in\mathbb R$ is finite, and, in
addition, $\k$ is chosen sufficiently large, $\k\geq\k_{\mathrm{min}}$
\cite{LR}.
Then, one defines for $t_0,t\in I$ the function 
\begin{align}
W_\k^{(n)}(t)=\frac{1}{\left(2a(t)^3\Omega_\k^{(n)}(t)\right)^{1/2}}\exp\left(i\int_{t_0}^{t}dt\
\Omega_k^{(n)}(t)\right)\ .
\end{align}
An adiabatic vacuum state of order $n$ is the pure, quasi-free state
obtained as the exact solution of \eqref{KGRWTimeDependentPart} with the initial conditions
\begin{align}\label{AdiabaticInitialValues}
T_\k(t)&=W_\k^{(n)}(t)\ , & \dot T_\k(t)&=\dot W_\k^{(n)}(t)\ .
\end{align}
The adiabatic vacuum states depend on several quantities involved in their
definition. First, they depend on the initial time $t$ used  for the initial
values in \eqref{AdiabaticInitialValues}. This has no effect on the adiabatic
vacuum state, as it amounts to common phase change of the initial conditions.
Second, they depend on the extrapolation of $\Omega^{(n)}$ to small $\k$, which
is always possible in a continuous manner, and amounts to some Bogoliubov
transformation on the state, not affecting the large $\k$ behaviour. Of
course, they depend also on the order of iteration, $n$.

It has been shown in \cite{Junker:1996,Junker:2002} that adiabatic vacuum states
of infinite order are Hadamard states. This fact is of indirect importance for
our construction as is is used in \cite{Olbermann:2007ii} to prove the Hadamard 
property of the states of low energy, and we base our proof of Hadamard property
on the latter fact. The notion of adiabatic vacuum states was extended to more
general spacetimes by the usage of Sobolev wavefront sets in
\cite{JunkerSchrohe}.

\section{The two-point distribution}
The two-point distribution of a homogeneous, isotropic, quasi-free state can be
characterized by different means. For example, in \cite{LR} it is given in a form
that takes data on the surface $\Sigma$. So, let $F_i:=(u_i,\a(t)^3 p_i)\in\MMa$
be a pair of initial values on the phase space associated to $\Sigma$.
\begin{Theorem}[\cite{LR}]\label{LR21}
The homogeneous and isotropic states on Robertson-Walker spacetimes have
two-point distributions of the form
\begin{equation}\label{HomogeneousIsotropicTwoPoint}
\omega^{(2)}(F,F'):=\int \d\vec k \angles{{\tilde F(\vec k)},S(\k)\tilde
F'(\vec k)}\ ,
\end{equation}
where
\begin{equation}
\begin{split}
\angles{{\tilde F(\vec k)},S(\k)\tilde F'(\vec k)}&:=\sum_{i,j=0}^1 \bar{\tilde
F}_i(\vec k)S_{ij}(\k)\tilde F_j'(\vec k)\ .
\end{split}
\end{equation}
Here $\k\mapsto S_{ij}(\k)$ is measurable and polynomially bounded. For the
matrix $S(\k)$ it holds almost everywhere in $\k$ that
\begin{subequations} 
\begin{align}
S_{01}(\k)-S_{10}(\k)&=\i\label{CommutatorSa}\\
S_{01}(\k)&=\Bar S_{10}(\k)\label{CommutatorSb}\\
S_{00}(\k)&\geq 0\label{CommutatorSc}\\
S_{00}(\k)S_{11}(\k)&\geq \abs{S_{01}(\k)}^2\label{CommutatorSd}\ .
\end{align}
\end{subequations} 
\end{Theorem}
Obviously equations \eqref{CommutatorSc} implement \eqref{CommutatorSc}
the positivity of the state (compare to \eqref{NonUniqueScalarProduct}). We
remark that the two-point distributions depend only on the magnitude $\k$ of the
three-momentum $\vec k$ because of the symmetry of the states. Exploiting these
relations, it is possible to write $S_{01}(\k)=\RS(\k)+\frac{\i}{2}$. So, with
$\mathbb J=\left(\begin{smallmatrix}0 & \mathbb 1 \\ -\mathbb 1 &
0\end{smallmatrix}\right)$ the matrix $S(\k)$ can be written as
\begin{equation}
S(\k)=S+\frac{\i}{2}\mathbb J\ ,
\end{equation}
where we introduced
\begin{equation}\label{RealS}
S:=
\begin{pmatrix}
S_{00} & \RS \\
\RS & S_{11}
\end{pmatrix}\ .
\end{equation}
We distinguish the original matrix $S(\k)$ and its real part $S$ by
omitting the $\k$-dependence in the latter. This will cause no trouble since,
from now on, we will almost exclusively deal with $S$. We remark that
equation \eqref{CommutatorSd} becomes
\begin{align}\label{GeneralizedUncertainty}
[S] &\geq \frac{1}{4}\ ,
\end{align}
where $[S]=\operatorname{det} S$. Equation \eqref{GeneralizedUncertainty}
resembles a generalized uncertainty relation for the
covariance matrix of a state in quantum statistical mechanics. States of minimum
uncertainty are pure states, which satisfy $\det S=\frac{1}{4}$.

The two-point distribution of pure, quasi-free states is given by
\eqref{HomogeneousIsotropicTwoPoint} with
\begin{equation}
S(\k):=\begin{pmatrix}\abs{p(\k)}^2 & -q(\k)\Bar{p(\k)}\\  -\Bar{q(\k)} p(\k)&
\abs{q(\k)}^2  \end{pmatrix}\ ,
\end{equation}
where $p(\k)$ and $q(\k)$ are (essentially polynomially bounded and measurable)
complex-valued functions satisfying
\begin{equation}
\Bar{q(\k)} p(\k) - q(\k)\Bar{p(\k)}=-\i\ .
\end{equation}
The functions $p(\k)$ and $q(\k)$ are proportional to the initial data of
$T_\k(t)$ \cite{Junker:1996}:
\begin{align*}
S(\k)=
\begin{pmatrix}
\a^6 \dot{\Bar{T}}_\k \dot T_\k  & -\a^3 \dot{\Bar{T}}_\k  T_\k   \\
-\a^3 \Bar{T}_\k \dot T_\k  &   \Bar{T}_\k T_\k 
\end{pmatrix}\ .
\end{align*}

\subsubsection{The fundamental solution} 
For the calculation of the explicit four-smeared two-point distribution of
a quasi-free state we will need the following characterization of the
fundamental solution $E$ (see \cite{LR}). First, define the generalized function $G(x,y)$ by
\begin{gather}
G(x,y):=\int\d\vec k\ G_\k(x^0,y^0)Y_\vec k(\vec x)\Bar{Y}_\vec k(\vec y)\ ,\\
G_\k(x^0,y^0):=\i(T_\k(x^0)\Bar{T}_\k(y^0)-\Bar{T}_\k(x^0){T_\k(y^0)})\ .\label{TheKernelG}
\end{gather}
Then the fundamental solution $E:\mathscr D(\mathcal M)\rightarrow \mathscr
E(\mathcal M)$ can be written as
\begin{align}
(Ef)(x^0,\vec x):=-G(x,f):=\int dy^0\int\d\vec k\ Y_\vec k(\vec x)
G_\k(x^0,y^0) \check{f}(y^0,\vec k)\ ,
\end{align}
where
\begin{equation}
\check f(y^0,\vec k):=\a(y^0)^3\int d^3y \sqrt{\abs\s}\ \Bar{Y}_{\vec k}(\vec
y)f(y^0,\vec y)\ .
\end{equation}
The kernel $G_\k(\cdot,y^0)$ satisfies
for each fixed $y^0$ equation \eqref{KGRWTimeDependentPart} (by linearity) with initial conditions
\begin{align}
G_\k(y^0,y^0)&=\i(T_\k(y^0)\Bar{T}_\k(y^0)-\Bar{T}_\k(y^0){T_\k(y^0)})=0\ ,\\
\dot G_\k(y^0,y^0)&=\i(\dot
T_\k(y^0)\Bar{T}_\k(y^0)-\dot{\Bar{T}}_\k(y^0){T_\k(y^0)})=-\frac{1}{\a(y^0)^3}\ . 
\end{align}
This shows that $G_\k(x^0,y^0)$ is independent of the
particular solution $T_\k(t)$ used in equation \eqref{TheKernelG}.

\subsection{Four-smeared two-point distribution}\label{TwoPointDistribution}
As a first step to our construction, we need the explicit form of the two-point distribution $S(f,g)$ of a homogeneous, isotropic,
quasi-free state in terms of the real matrix $S$ from \eqref{RealS}  and the
solutions to equation \eqref{KGRWTimeDependentPart}. We prove the following
lemma.
\begin{Lemma}\label{LemmaTwoPoint}
Let $\omega$ be a quasi-free, homogeneous, isotropic state of the Klein-Gordon
field in a Robertson-Walker spacetime $\mathcal M$. Then the two-point
distribution of $\omega$ is given by
\begin{align}
S(f,g)
&=\int d\mu(x) \int d\mu(x')\ \bar{f}(x)g(x')\ \omega^{(2)}(x,x')
\end{align}
with $d\mu(x):=\sqrt{\det \g}d^4x$\ , where
\begin{align}\label{QuasifreeOmegaTwo}
&{\omega}^{(2)}(x,x')
:=\int \d\vec k \ Y_{\vec k}(\vec x)\Bar{Y}_{\vec k}(\vec x')\\
&\phantom{=}{\nonumber}\times\bigg[
T_\k(t) \Bar T_{\k}(t') \left( b_1 - \frac{1}{2} \right)+T_\k(t) T_{\k}(t')\cdot b_2 
+ \Bar T_\k(t) \Bar T_{\k}(t') \cdot \bar b_2 +\Bar T_\k(t) T_{\k}(t')  \left(
b_1+\frac{1}{2}\right) \bigg]
\end{align}
and
\begin{subequations}
\begin{align}\label{TheBees}
b_1&:=S_{00} \abs{T_\k(t_0)}^2+\a(t_0)^6S_{11}\abs{\dot T_\k(t_0)}^2+\left(\Bar
T_\k(t_0){\dot T_\k(t_0)}+{T}_\k(t_0)\dot{\Bar T}_\k(t_0)\right) \a(t_0)^3\RS
\ ,\\ b_2&:=-S_{00} \Bar{T}^2_\k(t_0)-\a(t_0)^6S_{11}\dot{\Bar
T}^2_\k(t_0)-2\Bar{T}_\k(t_0)\dot{\Bar T}_\k(t_0) \a(t_0)^3\RS \ .
\end{align}
\end{subequations}
\end{Lemma}
\proof In order obtain this formula for the two-point distribution, we need to
calculate the two-point distribution of a homogeneous,
isotropic, quasi-free state \eqref{HomogeneousIsotropicTwoPoint}, using the
initial values $F,H\in\MM$ defined by
\begin{align}\label{PhaseSpaceF}
F&:=\begin{pmatrix} \rho_0Ef \\ \a(t_0)^3\rho_1Ef\end{pmatrix}\ ,        &
H&:=\begin{pmatrix} \rho_0Eh \\ \a(t_0)^3\rho_1Eh\end{pmatrix}
\end{align}
on a Cauchy surface at time $t_0$. First, we need some Fourier transforms.
Using equations \eqref{FourierTransform}, \eqref{FourierX} and the abbreviated notation
\begin{align}
\rho_0 G_{\k}(x^0,y^0)&:=G_{\k}(t_0,y^0)\ ,\\
\rho_1 G_{\k}(x^0,y^0)&:=\dot G_{\k}(t_0,y^0)
\end{align}
one obtains
\begin{align}
{\widetilde{\rho_0Ef}}(\vec k)&=\int dy^0\ G_{\k}(t_0,y^0) \check{f}(y^0,\vec
k)\ ,\\
\Bar{\widetilde{\rho_0Ef}}(\vec k) &=\int dy^0\ \Bar G_{\k}(t_0,y^0)
\bar{\check f}(y^0,\vec y)\ ,\\
\widetilde{\rho_1Ef}(\vec k)&=\int dy^0\ \dot G_{\k}(t_0,y^0)
\check{f}(y^0,\vec k)\ ,\\ \Bar{\widetilde{\rho_1Ef}}(\vec k)&=\int dy^0\
\dot{\Bar G}_{\k}(t_0,y^0)\bar{\check f}(y^0,\vec y)\ .
\end{align}
The Fourier transform of  $\a(t_0)^3{\rho_1Ef}$ is simply
$\a(t_0)^3\widetilde{\rho_1Ef}$. Using this and \eqref{PhaseSpaceF} we can write
the integrand of the two-point distribution \eqref{HomogeneousIsotropicTwoPoint} as
\begin{align}
&\angles{{\tilde F(\vec k)},S(\k)\tilde H(\vec k)}=\sum_{i,j=0}^1 \bar{\tilde
F}_i(\vec k) S_{ij}(\k)\tilde H_j(\vec k)\\ &=\Bar{\widetilde{\rho_0Ef}}(\vec
k) S_{00}(\k)\widetilde{\rho_0Eh}(\vec k) +\a(t_0)^3 \Bar{\widetilde{\rho_0Ef}}(\vec k) S_{01}(\k)\widetilde{\rho_1Eh}(\vec k)\\ &\phantom{=}{\nonumber}+\a(t_0)^3 \Bar{\widetilde{\rho_1Ef}}(\vec k)
S_{10}(\k)\widetilde{\rho_0Eh}(\vec k) +\a(t_0)^6\Bar{\widetilde{\rho_1Ef}}(\vec k) S_{11}(\k)\widetilde{\rho_1Eh}(\vec k)\\ &= \int d\mu(y)\int d\mu(y') \bar f(y^0,\vec y) h(y'^{0},\vec y')  {Y}_{\vec k}(\vec y) \Bar{Y}_{\vec k}(\vec y') \times\\
&\phantom{=}{\nonumber} \Big[ \Bar G_{\k}(t_0,y^0) S_{00}(\k)
G_{\k}(t_0,y'^{0}) + \a(t_0)^3 \Bar G_{\k}(t_0,y^0)S_{01}(\k) \dot
G_{\k}(t_0,y'^0)\\ &\phantom{=}{\nonumber} +\a(t_0)^3 \dot{\Bar G}_{\k}(t_0,y^0) S_{10}(\k) G_{\k}(t_0,y'^{0})+\a(t_0)^6\dot{\Bar G}_{\k}(t_0,y^0) S_{11}(\k) \dot
G_{\k}(t_0,y'^0) \Big]\ ,
\end{align}
where $ d\mu(y):=\sqrt{\abs\g}d^4y=\a(y^0)^3\sqrt{\abs\s}dy^0 d^3y$. We can now
write the two-point distribution as
\begin{align}
S(f,h)&=\int\d\vec k \angles{{\tilde F(\vec k)},S(\k)\tilde H(\vec k)}\\
&=\int d\mu(y) \int d\mu(y')\ \bar{f}(y)h(y')\ \omega^{(2)}(y,y')
\end{align}
with
\begin{align}\label{TwoPointPreliminary}
{\omega}^{(2)}&(y,y')
:=\int\d\vec k \ Y_{\vec k}(\vec y)\Bar{Y}_{\vec k}(\vec y')\\
&\phantom{=}{\nonumber} \times\Big[ \Bar G_{\k}(t_0,y^0) G_{\k}(t_0,y'^{0})
S_{00}(\k) + \Bar G_{\k}(t_0,y^0) \dot G_{\k}(t_0,y'^0)
\a(t_0)^3S_{01}(\k)\\ &\phantom{=\times}{\nonumber} +\dot{\Bar G}_{\k}(t_0,y^0) G_{\k}(t_0,y'^{0}) 
\a(t_0)^3S_{10}(\k)+\dot{\Bar G}_{\k}(t_0,y^0) \dot
G_{\k}(t_0,y'^0)\a(t_0)^6 S_{11}(\k)\Big]\ .
\end{align}
We need to calculate the quantity in the square brackets. From the definition
\eqref{TheKernelG} of $G_\k(x^0,y^0)$ we have
\begin{align}
G_\k(t_0,y^0):=\i\left(T_\k(t_0)\Bar{T}_\k(y^0)-\Bar{T}_\k(t_0){T_\k(y^0)}\right)\
.
\end{align}
In the following we use the abbreviated notation
\begin{align}
G&:=G_\k(t_0,y^0)\ , & G'&:=G_\k(t_0,y'^0)\ ,\\
\dot G&:=\partial_{t_0}G_\k(t_0,y^0)\ , & \dot
G'&:=\partial_{t_0}G_\k(t_0,y'^0)\ ,
\end{align}
and
\begin{align}
T_0&:=T_\k(t_0)\ , & T&:=T_\k(y^0)\ , & T'&:=T_\k(y'^0)\ .  
\end{align}
Note that
\begin{align}
G&=\i({T_0}\Bar{T}-\Bar{T}_0{T})\ ,&  G'&=\i({T_0}\Bar{T'}-\Bar{T}_0{T'}) \
,\\ 
\dot G&=\i({\dot T_0}\Bar{T}-\dot{\Bar T}_0{T})\ , &
\dot G'&=\i({\dot T_0}\Bar{T'}-\dot{\Bar T}_0 {T'}) \ .
\end{align}
This gives the intermediate results 
\begin{align}
\Bar{G}G'&=
 \abs{T_0}^2 {T}\Bar{T'} 
-\Bar{T}_0^2 {T}{T'}
-{T}_0^2 \Bar{T}\Bar{T'} 
+\abs{T_0}^2 \Bar{T}{T'}\ ,\\
\dot{\Bar{G}}\dot G'&=
 \abs{\dot T_0}^2 {T}\Bar{T'} 
-\dot{\Bar T}_0^2 {T}{T'}obtain 
-{\dot T_0}^2 \Bar{T}\Bar{T'} 
+\abs{\dot T_0}^2 \Bar{T}{T'}\ ,\\
\dot{\Bar G}G'&=
{T_0}\dot{\Bar T}_0{T}\Bar{T'}
-\Bar{T}_0\dot{\Bar T}_0{T}{T'}
-{T}_0{\dot T}_0\Bar{T}\Bar{T'}
+\Bar{T}_0{\dot T}_0\Bar{T}{T'}\ ,\\ 
{\Bar G}\dot G'&=
 \dot{T}_0{\Bar T}_0{T}\Bar{T'}
-{\Bar{T}_0}\dot{\Bar T}_0{T}{T'}
-{T}_0\dot{T}_0\Bar{T}\Bar{T'}
+\dot{\Bar{T}_0}{T}_0\Bar{T}{T'}\ .
\end{align}
Making use of $S_{01}=\RS +\frac{\i}{2}$ and $S_{10}=\RS -\frac{\i}{2}$ we
finally arrive at
\begin{align}
[\dots]&={T}\Bar{T'}\left[ S_{00} \abs{T_0}^2+\a^6S_{11}\abs{\dot
T_0}^2+(\Bar{T}_0{\dot T_0}+{T}_0\dot{\Bar T}_0) \a^3\RS -\frac{1}{2}\right]\\
&\phantom{=}{\nonumber} +{T}{T'} \left[-S_{00} \Bar{T}_0^2-\a^6S_{11}\dot{\Bar
T}_0^2-2\Bar{T}_0\dot{\Bar T}_0 \a^3\RS \right]\\ &\phantom{=}{\nonumber}
+\Bar{T}\Bar{T'} \left[-S_{00} T_0^2-\a^6S_{11}{\dot T_0}^2-2{T}_0{\dot T_0}
\a^3\RS \right]\\ &\phantom{=}{\nonumber} +\Bar{T}{T'}\left[S_{00}
\abs{T_0}^2+\a^6S_{11}\abs{\dot T_0}^2 +(\Bar{T}_0{\dot T_0}+{T}_0\dot{\Bar
T}_0) \a^3\RS +\frac{1}{2}\right]\ ,
\end{align}
which gives the lemma. \qed

\chapter{Quantum energy inequalities}\label{ChapterQEI}
This chapter starts with a brief overview of quantum energy inequalities and
their interpretation as stability conditions. Then, we quote the worldline
quantum inequality of Fewster \cite{Fewster:2000}, which we will use as the
basis of our construction of almost equilibrium states. Afterwards, we give an
explicit expression for the worldline averaged stress-energy in Robertson-Walker
spacetimes in a form that is suitable for the minimization procedure to be
accomplished in chapter \ref{ChapterAES}.
\section{A brief review}
The geometry of spacetime is related to its matter content via Einstein's
equation
\begin{equation}
R_{\mu\nu}- \frac{1}{2} R\ \g_{\mu\nu}=-8\pi \T_{\mu\nu}\ .
\end{equation}
Without restrictions on the stress-energy tensor there would be no restrictions
on the metric, thus no clue on physically realizable solutions. Apart from
covariant conservation of energy, $\nabla_\mu\T^{\mu\nu}$, the stress-energy
tensor is believed to obey several more energy conditions, of which the most
fundamental one is the weak energy condition (WEC). It says that 
\begin{equation}
\T_{\mu\nu} u^\mu u^\nu\geq 0
\end{equation}
for timelike $u^\mu$ is satisfied by all forms of physically reasonable matter.
Interpreting $u^\mu$ as the four velocity of an observer, it guarantees that all
observers at all points in spacetime always measure nonnegative energy density.

The energy conditions are crucial ingredients in many important results
concerning the behaviour of solutions to Einstein's equations. For instance, a
result that requires energy conditions is the positivity of the asymptotic
gravitational mass of isolated objects
\cite{Schoen:1979,Ludvigsen:1981gf,Witten:1981}. This has implications on the
stability of Minkowski spacetime. Furthermore, energy conditions ensure that
entertaining phenomena like traversable wormholes and stargates in 'designer
spacetimes', time machines, i.e., spacetimes with closed timelike curves, and
warp drives are forbidden \cite{Friedman:1993ty, Hawking:1991nk}. The WEC is also
used in theorems, which prove that at a certain stage the formation of
singularities becomes inevitable.  These singularities are related to
gravitational collapse \cite{Penrose:1964} as well as to an initial big bang
\cite{HawkingEllis}.  Only recently, it was suggested that in inflationary
cosmology an initial singularity may exists even if the energy conditions are
violated \cite{Borde:2001nh}. Most important for us, energy conditions are
necessary for the second law of thermodynamics to hold (see the discussion
below).

As opposed to classical field theory, it is known that the weak energy condition
is violated in quantum field theory. In fact, in any Wightman field the
renormalized energy density may become arbitrarily negative at a spacetime point
\cite{Epstein:1965, Kuo:1997jd}. This may happen under very general assumptions
and for free as well as interacting local quantum fields. Simple examples of
states with negative energy density are given by certain superpositions of the 
vacuum state with a two-particle state \cite{PfenningPhD,Ford:2002kz}. Besides
this, there are also examples for negative energy densities that do not rely on the availability of particles, e.g., the Casimir effect or squeezed states of light
(see references in \cite{Flanagan:1997gn}).

Quantum inequalities were introduced originally in \cite{Ford:1978}, where it is
argued that unconstrained negative energy fluxes $F=E/t$ can violate the second
law of thermodynamics. However, all evidence tells us that the second law of
thermodynamics holds on a macroscopic level. Quantum inequalities provide a
mechanism that prevents the microscopical violations to survive on a macroscopic
level. The argument goes as follows. An absorber is a quantum system that has an
energy uncertainty $\Delta E\gtrsim  t^{-1}$, where $t$ is some timescale.
The magnitude of change in its energy due to absorption of negative flux is of
the order of $t\abs{F}$. Hence, no macroscopic effects remain if the magnitude and duration
of the negative energy flux is constrained by $\abs{F}\lesssim t^{-2}$.

To constrain the negative energy that observers can measure it was suggested
using the integral of the energy density over the worldline of a geodesic
observer \cite{Tipler:1978} -- a concept which led to the averaged weak energy
condition (AWEC) $\int_{-\infty}^{\infty}d\tau \angles{\T_{\mu\nu}u^\mu u^\nu} \
\geq 0$, where $u^\mu$ is the observers four-velocity and $\tau$ is his proper
time. This condition allows for the energy density to be pointwise negative as
long as there is enough positive energy elsewhere on the worldline to compensate
for this. A related condition is the averaged null energy condition (ANEC), which
follows from integration along null geodesics \cite{WaldYurtsever}.
Unfortunately, these conditions are violated easily for the vacuum state in
certain spacetimes. It was then discovered \cite{FordRoman} that introducing a
sampling function into the AWEC integral successfully constrains the magnitude
and duration of negative energy densities. The idea of averaging the renormalized
stress-energy tensor over regions or curves in spacetime has been very fruitful
since then and led to a variety of quantum weak energy inequalities of different
types in different settings -- see the reviews \cite{PfenningPhD,FewsterLisbon}.

An absolute quantum energy inequality (QEI) for the renormalized stress-energy
tensor has the general form
\cite{FewsterCovariance1}
\begin{equation}
\int_\mathcal K d\mathcal K \angles{:\T_{\mu\nu}:}_\omega  f^{\mu\nu} \geq
-Q(f^{\mu\nu})\ ,
\end{equation}
where the integral is taken over a region $\mathcal K$ of spacetime and the
sampling tensor $f^{\mu\nu}$ belongs to the class of second rank contravariant
tensor fields, possibly singularly supported on curves or surfaces in $\mathcal K$. Furthermore, $Q$ is a positive, real-valued map on
these tensors, and $\omega$ belongs to a class of states of the theory -- usually the class of Hadamard states if curved spacetime is considered.

Most quantum energy inequalities are obtained by averaging with a sampling
function $f(t)$ along the worldline of an observer,
\begin{equation}
\int_\gamma dt\ f(t) \angles{:\T_{\mu\nu}:(t,t)}_\omega  u^\mu(t) u^\nu(t)
\geq -Q(f)\ ,
\end{equation}
where $u^\mu$ denotes the four-velocity of the observer. Early quantum
inequalities were derived for Lorentzian sampling functions
$f_{t_0}(t)=t_0/\left(\pi(t^2+t_0^2)\right)$. To give an explicit example, for a
free quantized scalar field in Minkowski spacetime it holds
\begin{equation}
\int_{-\infty}^{\infty} dt\ f_{t_0}(t) \angles{:\T_{\mu\nu}:} 
u^\mu u^\nu \geq -\frac{3}{32\pi^2 t_0^4}\ ,
\end{equation}
in the frame of an arbitrary inertial observer  with proper time coordinate $t$
for any state \cite{FordRoman}. In the meantime, the class of admissible sampling
functions has been extended to arbitrary smooth positive functions with
sufficiently nice decay properties.

A simple configuration leading to violation of the AWEC and absolute QEIs is an
observer at rest between uncharged, perfectly conducting plates. Then, the
Casimir effect \cite{Casimir:1948dh} predicts negative vacuum energy for the
quantized electromagnetic field measured by such an observer. In curved
spacetimes a further difficulty is normal ordering of the stress-energy tensor.
This problem is solved by difference QEIs, which are inequalities of the
type
\begin{equation}
\int_\gamma dt\
f(t)\left[\angles{\T_{\mu\nu}(\gamma(t))}_\omega-\angles{\T_{\mu\nu}(\gamma(t))}_{\omega_0}
\right] \gamma^\mu(t) \gamma^\nu(t)\geq -Q(f,\omega_0)\ ,
\end{equation}
where $\omega_0$ is a reference state. Clearly, in the Fock
representation of a Hadamard state $\omega_0$ we would, by Wald's fourth
axiom (see section \ref{HadamardStates}), regain an absolute QEI.
Difference QEIs  have been proved on arbitrary globally hyperbolic spacetimes with general sampling functions on the class of
Hadamard states. See below for a precise statement.

Sampling along timelike curves is not the only possibility to obtain energy
inequalities. One can try sampling in a spacetime region or one may
sample over a spatial region alone:
\begin{equation}
H(\xi,\Sigma) := \int_\Sigma d\mu(\vec x) f(\vec x)
\angles{:\T_{\mu\nu}:} \xi^\mu\xi^\nu \ ,
\end{equation}
where $\xi^\mu$ is a vector field that is orthogonal to $\Sigma$ and $f(\vec x)$
has compact support on $\Sigma$. Only for a restricted class of theories, namely, conformal fields in
two spacetime dimensions, spacelike sampled quantum inequalities have been derived
\cite{Flanagan:1997gn,Vollick:2000pm}. Rather it seems that compactly supported
weighted averages over spacelike surfaces alone are generally unbounded below
for dimensions $n\geq 3$ \cite{Ford:2002kz}. Naively, this kind of objects
could have been hoped to be prototypes of `local Hamiltonians'. 
Interestingly, the integrals involved here resemble the integrals that were
investigated some forty years ago in order to derive symmetry generating
global charge operators from local currents (see
\cite{Voelkel,Orzalesi,Requardt:1976yx}). In particular, these
investigations suggest, that spacelike smearing alone of local currents, does not
yield operators with sensible
properties. In view of these problems with spacelike averaging, we adopt the
viewpoint that the investigation of timelike averaged stress-energy tensor
energy densities is more promising.

Let us make two more remarks. First, we note that there are quantum field
theories which do not satisfy quantum inequalities. For example, the
non-minimally coupled scalar fields violate the energy conditions already on the
classical level and, as expected, their averaged energy density is unbounded
below on the class of Hadamard states. Recently, Fewster and
Osterbrink derived state dependent quantum inequalities for the fields
with coupling $0<\xi\leq\frac{1}{4}$ \cite{FewsterOsterbrink}. Second, the issue
of quantum energy inequalities for interacting fields is not yet resolved,
though, recently, some progress in this direction has been reported by
Fewster and Bostelmann \cite{FewsterBostelmann}.

\subsection{Stability conditions}\label{StabilitityConditions} 
Quantum inequalities seem to be closely related to other stability conditions in
quantum field theory, as has been pointed out by Fewster and Verch
\cite{FewsterVerch,FewsterStability}. On the microscopic level the microlocal
spectrum condition (theorem \ref{MicrolocalSpectrumCondition}) serves as a
suitable stability condition. Now, theorem \ref{WorldlineQuantumInequality}
states that quantum inequality exists for all states that satisfy the microlocal
spectrum condition on arbitrary globally hyperbolic spacetimes. In
\cite{FewsterVerch} a macroscopic stability condition was related to the quantum
inequalities. It was shown that on static spacetimes the existence of
passive states (see below) follows from the existence of quantum weak energy
inequalities. The circle is closed by noting that in \cite{SahlmannVerch} it was
shown that the two-point distributions of passive states of (vector-valued)
quantum fields in a stationary-spacetime satisfy the microlocal spectrum
condition. Now, viewing quantum inequalities as mesoscopic stability conditions
this relates stability conditions on three different scales.

\subsubsection{Passivity} Let us briefly comment on the notion of passivity. A
state on a C*-algebra $(\mathfrak A,\alpha_t)$ is called passive if and
only if
\begin{align}
\omega\left(U^*\frac{\delta}{\i}(U)\right)\geq 0
\end{align}
for any $U\in\mathcal U(\mathfrak A)\cap D(\delta)$, where $\mathcal U_0$
denotes the unit-component of the group $\mathcal U(\mathfrak A)$ of all
unitary elements of $\mathfrak A$ and $D(\delta)$ is the set of all
$A\in\mathfrak A$ for which the the generator
\begin{equation}
\delta(A):=\lim_{t\rightarrow 0}\frac{1}{t}(\alpha_t(A)-A)
\end{equation}

 exists \cite{Pusz:1978hb,BratteliRobinson2}. For example, for bounded operators
 on the Hilbert space  of a quantum mechanical system with Hamiltonian $H$, the
generator amounts to a bounded symmetric derivation $\delta(A)=\i[H,A]$ (see the
monograph \cite{Sakai:1991}). The notion of passivity was introduced in
\cite{Pusz:1978hb} as a precise mathematical formulation of the second law of
 thermodynamics, which says that systems at equilibrium are unable to perform
 mechanical work in cyclic processes. This justifies to call it a macroscopic
 stability condition. It is valid for infinite systems and is closely related to
 the KMS condition (see chapter \ref{ChapterAES}) by the fact that KMS states and
 mixtures of KMS are passive. Under certain technical conditions ensuring that we
 deal with pure phases, the inverse statement, namely, that passive states are
 KMS states or ground states, is also true.
 
 To date, there exists no formulation  of passivity for non-stationary
 spacetimes. A direct implementation would require to replace the (strongly-continuous) group of automorphism $\alpha_t$ by
 a propagator family of automorphisms $\alpha_{t,s}$ (see the related discussion
 in section \ref{SectionQuantization}) and thus to deal with a time-dependent
 family of derivations $\delta_t$. Such dynamical families are not well
 investigated with respect to (non)-equilibrium states -- only two references are
 know to the present author. First, in \cite{Buchholz:2002zc} certain states on
 Robertson-Walker spacetimes, which were obtained by 'transplantation' from
 de~Sitter space, are proved to be locally passive in a certain sense.
 Second, in \cite{Ojima:1986qd} a general framework for the treatment of time-dependent
 non-equilibrium processes was proposed, which makes use of the propagator
 families and their generators. This (perturbative) scheme is equivalent the
 standard thermo field dynamics if and only if the states under consideration are
 equilibrium states and the dynamics is time-independent. This
remark ends our digression on passivity.

\section{A general worldline inequality}
The classical energy-momentum tensor of the minimally coupled massive scalar
field is given by
\begin{align}\label{ClassicalEnergyMomentum}
\T_{\mu\nu}=\nabla_\mu\phi\nabla_\nu\phi-\frac{1}{2}\g_{\mu\nu}\left(\nabla^\sigma\phi\nabla_\sigma\phi+\m^2\phi^2\right)
\end{align}
The quantized and point split version of \eqref{ClassicalEnergyMomentum} is
defined as follows  \cite{Fewster:2000,FewsterSmith}.
For a smooth timelike curve $\gamma(t)$ in $\mathcal M$, where $t$ is the
proper time of the curve, denote by $\mathcal U$ a tubular neighborhood of
$\gamma$. Choose an orthonormal frame $\{e_\alpha^\mu\}_{\alpha=0,1,2,3}$ in
$\mathcal U$ so that $\g^{\mu\nu}=\mathfrak n^{\alpha\beta} e^\mu_\alpha
e^\nu_\beta$ and such that the restriction of $e_0^\mu$ to $\gamma$ equals the
four-velocity of the curve, $e_0^\mu\vert_\gamma=\dot\gamma^\mu(t)$.

Suppose that the two-point distribution of $\omega_0$ obeys the
microlocal spectrum condition. Then one can define a distribution by
\begin{align}
\nonumber\angles{\T_{\mu\nu'}\, v^\mu v^{\nu'}}_{\omega_0}(t,t')
&:=\frac{1}{2} \sum_{\alpha=0}^{3}  \boldsymbol\varphi^* \left[ \left(
e_\alpha^\mu\nabla_\mu \otimes e_\alpha^{\nu'}\nabla_{\nu'}
\right)\omega_0^{(2)}(x,x') \right]  \\
&\phantom{=}+ \frac{1}{2} \m^2 \boldsymbol\varphi^*\left[\left(\mathbb 1\otimes\mathbb 1\right)
\omega_0^{(2)}(x,x')\right] \ . \label{Bidistribution}
\end{align}
Here $\boldsymbol\varphi^*: \mathcal M\times\mathcal
M\rightarrow\mathbb R^2$ denotes the pull-back induced by the map
$\boldsymbol\varphi(t,t')=(\gamma(t),\gamma(t'))$. That
\eqref{Bidistribution} is a well defined distribution on
$\mathbb R^2$ is shown in \cite{Fewster:2000}. Let $\mathfrak A$ be the algebra of observables of the minimally
coupled scalar field of mass $\m\geq 0$ on a globally hyperbolic spacetime $(\mathcal
M,\g_{\mu\nu})$ with dimension $n\geq 2$. Let, furthermore, $\gamma:\mathbb
R\rightarrow \mathcal M$ be a smooth timelike curve. Then the following theorem
holds.
\begin{Theorem}[\cite{Fewster:2000}]\label{WorldlineQuantumInequality}
Let $\omega$ and $\omega_0$ be states on $\mathfrak A(\mathcal M,\g_{\mu\nu})$
with globally Hadamard two-point distributions and define the normal ordered
energy density relative to $\omega_0$ by
\begin{equation}
\angles{:\T:}_\omega:=\angles{\T}_\omega-\angles{\T}_{\omega_0}\ .
\end{equation}
Then $\angles{:\T:}_\omega$ is smooth, and the quantum inequality
\begin{equation}\label{QuantumInequality}
\int_\gamma dt\ f(t)^2\angles{:\T:}_\omega(t,t) \geq
-\int_0^\infty\frac{d\lambda}{\pi}\widetilde{\left[(f\otimes f)\angles{\T}_{\omega_0}\right]}(-\lambda,\lambda)
\end{equation} 
holds for all real-valued $f\in\mathscr D(\mathbb R)$, and the right hand side is
convergent for all such $f$.
\end{Theorem}

\subsection{Stress-energy in Robertson-Walker
spacetimes}\label{AppendixEnergy} As a necessary prerequisite for the construction the almost equilibrium states we
need the explicit form of the left hand side of \eqref{QuantumInequality} in
Robertson-Walker spacetimes; This will be part of our free energy functional to
be defined in chapter \ref{ChapterAES}. We state the result as a lemma.
\begin{Lemma}\label{LemmaEnergy}
Let $\T_{\mu\nu}$ be the stress-energy tensor of real, linear, scalar field in
Robertson-Walker spacetimes and let $\omega,\omega_0$ two homogeneous,
isotropic states. Furthermore, let $f\in\DD(\mathcal
M)$ and $\gamma$ be the worldline of an isotropic observer. Then,
\begin{align}\label{AveragedEnergyRW}
\int \d t\ f(t)^2 \left(\angles{\T}_\omega(t,t)-\angles{\T}_{\omega_0}(t,t) 
\right)&=\int d\mu(\vec k)^\varepsilon \int dt\
f(t)^2\left(\boldsymbol\rho_\k(t) -\boldsymbol\rho_{\k,0}(t) \right)\ ,
\end{align}
where
\begin{align}\label{TheEnergyDensity}
\boldsymbol\rho_\k(t)
&:= b_1(S)  \left(\abs{\dot T_\k(t)}^2 + \omega_\k^2  \abs{T_\k(t)}^2 \right) +
\Re\left\{ b_2(S)   \left(\dot{T}_\k(t)^2 + \omega_\k^2  T_\k(t)^2  
\right)\right\}\ ,
\end{align}
and the coefficients $b_1,b_2$ are defined in \eqref{TheBees}. The energy
density $\boldsymbol\rho_{\k,0}(t)$ is defined analogously. The measures $d\mu(\vec
k)^\varepsilon$ differ by constants from the measures $\d\vec k$ defined in
\ref{Measures}:
\begin{align}
d\mu(\vec k)^{0}&:=\frac{\d\vec k}{(2\pi)^3}\ , &
d\mu(\vec k)^{+}&:=\frac{\d\vec k}{V_{\Sigma^+}}\ , &
d\mu(\vec k)^{-}&:=\frac{\d\vec k}{2\pi^2}\ . 
\end{align}
\end{Lemma}
\proof The stress-energy tensor of the Klein-Gordon field
\eqref{ScalarFieldEquation} is given by equation \eqref{ClassicalEnergyMomentum}.
In order to calculate the left hand side of \eqref{AveragedEnergyRW} we have
to consider the point-split expression
\begin{align}\label{SmearedEnergyDifference}
\nonumber\int dt\ &f(t)^2
\left[\angles{\T_{00}}_\omega(t,t)-\angles{\T_{00}}_{\omega_0}(t,t) \right]\\
&=\int dt\ f(t)^2 \lim_{t'\rightarrow t}\left.\left(\frac{1}{2}
\nabla_0\nabla_0'-\frac{1}{2}\sum_{\mu,\nu=1}^3 \g^{\mu\nu}
\nabla_\mu\nabla_\nu'-\frac{1}{2}\ \m^2
\right)\F(x,x')\right\vert_{\substack{x=\gamma(t) \\ x'=\gamma(t')}}\ ,
\end{align}
where 
\begin{align}
\F(x,x'):=\omega^{(2)}(x,x')-\omega^{(2)}_0(x,x')
\end{align} 
and the two-point distribution $\omega^{(2)}(x,x')$ is given in lemma
\ref{LemmaTwoPoint}. We need the limits $x'\rightarrow x$ of the
involved derivatives of the functional $\F(x,x')$. For the restricted case of
pure states, this calculation has been performed in \cite{Olbermann:2007i}.
See also \cite{ParkerFulling:1974} for similar results. In section
\ref{StatesOfLowEnergy} we will see that the limiting case of pure states
amounts to the values $b_2=0$ and $b_1=\frac{1}{2}$. 

The difference $\F(x,x')$ is a smooth function if $\omega$ and  $\omega_0$ are Hadamard states. In that case, the coincidence
limit $x\rightarrow x'$ is well defined. The derivatives involved in the energy
momentum tensor and which have to be calculated for each of the three cases $\varepsilon=0,1,-1$ are
\begin{align}
&\lim_{x'\rightarrow x} \F(x,x')\ ,&
&\lim_{x'\rightarrow x} \nabla_0\nabla_0' \F(x,x')\ ,&
&\lim_{x'\rightarrow x} \sum_{\mu,\nu=1}^3 \g^{\mu\nu}\nabla_\mu\nabla_\nu'  \F(x,x')\ . 
\end{align}
With the convention that latin indices are summed from $1$ to $3$ we can
simplify the spatial derivative as
\begin{equation}\label{SpatialDerivatives}
\sum_{\mu,\nu=1}^3 \g^{\mu\nu} \nabla_\mu\nabla_\nu' = \h^{ij} \nabla_i
\nabla_j'=\frac{1}{\a^2} \s^{ij} \nabla_i \nabla_j'\ .
\end{equation}
Besides that, we simplify the notation by introducing ellipsis for the second
summand, which always looks like the first except that it belongs to the state
$\omega_0$.

\paragraph{[$\varepsilon=0$]:}
This is, as expected, the simplest case. The modulus $\abs{Y_{\vec k}(\vec
x)}^2=\frac{1}{(2\pi)^3}$ of the spatial solutions is independent of
$\vec x$. Thus, we can calculate directly
\begin{align}
\lim_{x'\rightarrow x} \F(x,x')&=\int\d\vec k\ \abs{Y_{\vec k}(\vec x)}^2
\bigg[
\abs{T_\k(x^0)}^2 \left( b_1 +\frac{1}{2} \right)+T_\k(x^0)^2\cdot b_2\\
&\phantom{=}{\nonumber}+\Bar T_\k(x^0)^2\cdot \bar b_2 +\abs{T_\k(x^0)}^2 
\left( b_1-\frac{1}{2}\right) \bigg] - \ldots\\
&=\frac{1}{(2\pi)^3}\int\d\vec k\left[ 2\abs{T_\k(x^0)}^2\cdot  b_1  + 2\Re\left\{ T_\k(x^0)^2\cdot b_2 \right\} \right] - \ldots\\
&=2\int d\mu(\vec k) \left[
\abs{T_\k(x^0)}^2\cdot  b_1  + \Re\left\{ T_\k(x^0)^2\cdot b_2 \right\} \right] - \ldots\ ,
\end{align}
where we defined the measure $d\mu(\vec k)=\frac{\d\vec k}{(2\pi)^3}$.
Similarly we obtain for the term involving time derivatives
\begin{align}
\lim_{x'\rightarrow x} \nabla_0 \nabla_0' \F(x,x')&=2\int d\mu(\vec k)
\left[ \abs{\dot T_\k(x^0)}^2\cdot  b_1  + \Re\left\{ \dot T_\k(x^0)^2\cdot b_2
\right\} \right] - \ldots\ .
\end{align}
The third object involves spatial derivatives. Is is calculated with the aid
of \eqref{SpatialDerivatives} as
\begin{align}
&\lim_{x'\rightarrow x} \sum_{\mu,\nu=1}^3 \g^{\mu\nu}\nabla_\mu\nabla_\nu' 
\F(x,x')=\frac{1}{\a(x^0)^2}\lim_{x'\rightarrow x}  \s^{ij} \nabla_i \nabla_j' \F(x,x')\\ &=\frac{1}{\a(x^0)^2} \int\d\vec k\ \s^{ij} \nabla_i  Y_{\vec k}(\vec x) \nabla_j'\Bar Y_{\vec k}(\vec x')\left[2\abs{T_\k(x^0)}^2\cdot  b_1  + 2\Re\left\{ T_\k(x^0)^2\cdot b_2 \right\} \right] - \ldots\\
&=\frac{1}{(2\pi)^3}\int\d\vec k\ \frac{E(\k)}{\a(x^0)^2} 
\left[2\abs{T_\k(x^0)}^2\cdot  b_1  + 2\Re\left\{ T_\k(x^0)^2\cdot b_2 \right\} \right] - \ldots\\
&=2\int d\mu(\vec k)  \frac{E(\k)}{\a(x^0)^2} 
\left[\abs{T_\k(x^0)}^2\cdot  b_1  + \Re\left\{ T_\k(x^0)^2\cdot b_2 \right\}
\right] - \ldots\ .
\end{align}
Here we used that $\s^{ij} \nabla_i Y_{\vec k}(\vec x)\nabla_j\Bar Y_{\vec
k}(\vec x)=\frac{1}{(2\pi)^3}\cdot E(\k)$, which follows directly from
\eqref{SpacelikeFlatSolution}.

\paragraph{[$\varepsilon=+1$]:} In this case we have to work a little more. The
states under consideration are assumed to be homogeneous. Thus, the quantities
$\lim_{x'\rightarrow x}\F(x,x')$ and $\lim_{x'\rightarrow x} \nabla_0\nabla_0' \F(x,x')$ 
 cannot depend on the spatial coordinates $\vec x$ and $\vec x'$. By the
 compactness of $\Sigma^+$, we can carry out the limiting procedure, integrate over $\Sigma^+$ using the measure $\d\vec
x$ and divide by the volume $V_{\Sigma^+}$. This gives
\begin{align}
\lim_{x'\rightarrow x} \F(x,x')&=\frac{1}{V_{\Sigma^+}}\int_{\Sigma^+}\d\vec x\lim_{x'\rightarrow x} \F(x,x')\\
&=\frac{1}{V_{\Sigma^+}}\int\d\vec k\int_{\Sigma^+}\d\vec x\ \abs{Y_{\vec
k}(\vec x)}^2 \big[
\abs{T_\k(x^0)}^2  b_1 +T_\k(x^0)^2 b_2 \\
&\phantom{=}{\nonumber}+\Bar T_\k(x^0)^2 \bar b_2
+\abs{T_\k(x^0)}^2 b_1  \big] - \ldots\\
&=\frac{1}{V_{\Sigma^+}}\int\d\vec k\ \delta(\vec k,\vec k) \left[
2\abs{T_\k(x^0)}^2 b_1  + 2 \Re\left\{T_\k(x^0)^2 b_2 \right\} \right] -
\ldots\\ 
&=2 \int d\mu(\vec k) \left[
\abs{T_\k(x^0)}^2   b_1  + \Re\left\{ T_\k(x^0)^2 b_2 \right\}
\right] - \ldots\ ,
\end{align}
where we used \eqref{FourierX} and  defined the measure $d\mu(\vec
k)=\frac{\d\vec k}{V_{\Sigma^+}}$. Again, after taking the time derivatives, we
can perform the same calculation as before:
\begin{align}
\lim_{x'\rightarrow x} \nabla_0 \nabla_0' \F(x,x')
&=2\int d\mu(\vec k) \left[
\abs{\dot T_\k(x^0)}^2  b_1  + \Re\left\{ \dot T_\k(x^0)^2  b_2
\right\} \right] - \ldots\ .
\end{align}
The spatial part gives the preliminary expression
\begin{align}
&\lim_{x'\rightarrow x} \sum_{\mu,\nu=1}^3 \g^{\mu\nu}\nabla_\mu\nabla_\nu'  \F(x,x')= \frac{1}{V_{\Sigma^+}} \int_\Sigma\d\vec x  \frac{1}{\a^2(x^0)}\lim_{x'\rightarrow x}  \s^{ij} \nabla_i \nabla_j' \F(x,x')\\
&=\frac{1}{\a(x^0)^2} \frac{1}{V_{\Sigma^+}} \int_{\Sigma^+}\d\vec x \int\d\vec
k\ \lim_{x'\rightarrow x}  \s^{ij}\  \nabla_i  Y_{\vec k}(\vec x)\ 
\nabla_j'\Bar Y_{\vec k}(\vec x')\\
&\phantom{=}{\nonumber}\times \bigg[ T_\k(x^0) \Bar T_{\k}(x'^0) \left( b_1 + \frac{1}{2} \right)+T_\k(x^0) T_{\k}(x'^0)  b_2 
\\
&\phantom{=}{\nonumber}+\Bar T_\k(x^0) \Bar T_{\k}(x'^0) \bar b_2 +\Bar
T_\k(x^0) T_{\k}(x'^0) \left( b_1-\frac{1}{2}\right) \bigg]  - \ldots\\
&=\frac{1}{\a(x^0)^2}\frac{1}{V_{\Sigma^+}}\int\d\vec k\int_{\Sigma^+}\d\vec x\
\s^{ij}\ \nabla_i Y_{\vec k}(\vec x)\nabla_j\ \Bar Y_{\vec k}(\vec x)\\
&\phantom{=}{\nonumber}\times \bigg[
\abs{T_\k(x^0)}^2  b_1  +T_\k(x^0)^2  b_2 +\Bar T_\k(x^0)^2 \bar b_2
+\abs{T_\k(x^0)}^2  b_1 \bigg] - \ldots\ .
\end{align}
Since $\Sigma^+$ is compact without boundary, we can perform a partial
integration \cite{Taylor1} using
\begin{equation}
\int_{\Sigma^+}\d\vec x\ \s^{ij}\ \nabla_j Y_{\vec k}(\vec x)\  \nabla_i \Bar Y_{\vec k}(\vec x) 
=-\int_{\Sigma^+}\d\vec x\ \Delta Y_{\vec k}(\vec x)\ \Bar Y_{\vec k}(\vec x)\ .
\end{equation} 
Furthermore, since $\Delta_{\s} Y_{\vec k}(\vec x) = -E(\k) Y_{\vec k}(\vec
x)$, it follows that
\begin{align}
&= \frac{1}{V_{\Sigma^+}}\int\d\vec k\int_{\Sigma^+}\d\vec x\ \abs{Y_{\vec
k}(\vec x)}^2 \frac{E(\k)}{\a(x^0)^2} \bigg[ 2\abs{T_\k(x^0)}^2   b_1  + 2
\Re\left\{T_\k(x^0)^2  b_2 \right\} \bigg] - \ldots\\ 
&=\frac{1}{V_{\Sigma^+}}\int\d\vec k\ \delta(\vec k,\vec k) 
\frac{E(\k)}{\a(x^0)^2} \bigg[ 2\abs{T_\k(x^0)}^2   b_1  + 2 \Re\left\{
T_\k(x^0)^2  b_2 \right\} \bigg] - \ldots\\ &=2\int d\mu(\vec k) \frac{E(\k)}{\a(x^0)^2}  \bigg[
\abs{T_\k(x^0)}^2   b_1  + \Re\left\{ T_\k(x^0)^2  b_2 \right\} \bigg] -
\ldots\ .
\end{align}

\paragraph{[$\varepsilon=-1$]:} In this case one uses techniques from the
harmonic analysis of hyperbolic spaces with negative constant curvature, also called
Lobachevskian spaces \cite{GelfandVilenkin5}. In order to find a direct integral
representation of the isometry group $\G^-$, one embeds $\Sigma^-$ as a
three-dimensional hyperboloid with coordinates $\vec x=(x_1,x_2,x_3)$ into the
four-dimensional Minkowski spacetime $(\mathbb M,\mathfrak n_{\mu\nu})$ by means
of the map $\iota:\mathbb R^3\rightarrow\mathbb M,\mathbf x\mapsto
(\sqrt{1+\vec x^2},\vec x)$. One can calculate the metric on $\Sigma^-$ as
the pullback $\g_{\mu\nu}=\iota^*\mathfrak n_{\mu\nu}$, which gives 
\begin{align}
\h_{ij}(\vec x)&=\delta_{ij}-\frac{x^i x^j}{1+\vec x^2}
& &\Leftrightarrow & \h^{ij}(\vec x)&=\delta_{ij}+x_i x_j\ .
\end{align}
Now, define a normalized momentum by $\xi:=\vec k/\k$ and write for each
$\k$ the Fourier transform of a function $h\in\mathscr D(\Sigma^-)$ as
\begin{equation}
\tilde h_\k(\boldsymbol\xi):=\int \d\vec x\ Y_{\k\boldsymbol\xi}(\vec x)h(\vec
x)\ ,
\end{equation}
where $Y_{\vec k}(\vec x)$ are eigenfunctions of the Laplace operator given by
\eqref{SpacelikeSolutionOpen}.
This Fourier transform is a map with values in $\L^2(\mathrm S^2,d\Omega)$,
where $\mathrm S^2$ is the two-sphere embedded in $\mathbb R^3$ and $d\Omega$ denotes
the induced measure on $\mathrm S^2$. The Lorentz transformations $\mathrm g:
f(x) \mapsto f(x \mathrm g)$ on the functions $f(x)$  correspond to operators
\begin{equation}
U_\k(\mathrm g)\tilde h_\k =\left(\widetilde{U(\mathrm g) h}\right)_\k\ .
\end{equation}
Let $\mathrm g_x$ be the Lorentz transformation that maps $(\sqrt{1+\vec x},\vec
x)$ to $(1,0,0,0)$. Then by the unitarity of $U_\k(\mathrm g)$ 
\begin{align}
\abs{Y_\vec k(\vec x)}^2        
&=\frac{1}{(2\pi)^3}\int d\Omega(\boldsymbol\xi) \left(x\cdot \xi\right)^{-2}
\\
&=\frac{1}{(2\pi)^3}\int d\Omega(\boldsymbol\xi) \left(x\cdot
\mathrm g_x^{-1}\xi\right)^{-2}\\
&=\frac{1}{(2\pi)^3}\int d\Omega(\boldsymbol\xi) \left(\mathrm g_x x\cdot
\xi\right)^{-2}\\
&=\frac{1}{(2\pi)^3}\int d\Omega(\boldsymbol\xi) =\frac{4\pi}{(2\pi)^3} \ .
\end{align}
Again, as for $\varepsilon=0$, the modulus of
$Y_\vec k(\vec x)$ is a constant. Using this, we obtain
\begin{align}
\lim_{x'\rightarrow x} \F(x,x')&=\int\d\vec k\ \abs{Y_{\vec k}(\vec x)}^2\\
&\phantom{=}{\nonumber}\times \bigg[
\abs{T_\k(x^0)}^2 \left( b_1 +\frac{1}{2} \right)+T_\k(x^0)^2  b_2\\
&\phantom{=}{\nonumber}+\Bar T_\k(x^0)^2 \bar b_2 +\abs{T_\k(x^0)}^2 
\left( b_1-\frac{1}{2}\right) \bigg] - \ldots\\ 
&=\frac{4\pi}{(2\pi)^3}\int\d\vec k\left[
2\abs{T_\k(x^0)}^2   + 2\Re\left\{ T_\k(x^0)^2  b_2 \right\} \right] - \ldots\\
&=2\int d\mu(\vec k) \left[
\abs{T_\k(x^0)}^2   + \Re\left\{ T_\k(x^0)^2 b_2 \right\} \right] -
\ldots\ ,
\end{align}
where $d\mu(\vec k)=\frac{\d\vec k}{2\pi^2}$. Similarly, one finds
\begin{align}
\lim_{x'\rightarrow x} \nabla_0\nabla_0'\F(x,x')
&=2\int d\mu(\vec k) \left[
\abs{\dot T_\k(x^0)}^2   + \Re\left\{\dot T_\k(x^0)^2  b_2 \right\}
\right] - \ldots\ .
\end{align}
Finally, since $\s^{ij} \nabla_i  Y_{\vec k}(\vec x) \nabla_j'\Bar Y_{\vec
k}(\vec x')=(1+\k^2)\abs{Y_\vec k(\vec x)}=E(\k)\abs{Y_\vec k(\vec x)}$ we
obtain by
\begin{align}
Y_\vec k(x)&=\frac{1}{(2\pi)^{\frac{3}{2}}}\left(x^0-\frac{\vec
x\cdot\boldsymbol\xi}{\k}\right)^{-1+\i\k}
\end{align} 
(see  \eqref{SpacelikeSolutionOpen}) for the spatial derivatives the expression
\begin{align}
&\lim_{x'\rightarrow x} \sum_{\mu,\nu=1}^3 \g^{\mu\nu}\nabla_\mu\nabla_\nu' 
\F(x,x')=\frac{1}{\a^2(x^0)}\lim_{x'\rightarrow x}  \s^{ij} \nabla_i \nabla_j' \F(x,x')\\ 
&=\frac{1}{\a(x^0)^2} \int\d\vec k\ \s^{ij} \nabla_i  Y_{\vec k}(\vec x)
\nabla_j'\Bar Y_{\vec k}(\vec x')\left[2\abs{T_\k(x^0)}^2   + 2\Re\left\{
T_\k(x^0)^2 b_2 \right\} \right] - \ldots\\ &=\frac{1}{(2\pi)^3}\int\d\vec k\ \frac{E(\k)}{\a(x^0)^2} \left[2\abs{T_\k(x^0)}^2  b_1  + 2\Re\left\{ T_\k(x^0)^2  b_2 \right\}
\right] - \ldots\\ &=2\int d\mu(\vec k)  \frac{E(\k)}{\a^2(x^0)} 
\left[\abs{T_\k(x^0)}^2  b_1  + \Re\left\{ T_\k(x^0)^2  b_2 \right\}
\right] - \ldots\ . 
\end{align}
Putting all this together gives the desired result. \qed

\subsubsection{A better parametrization}
For the actual minimization, it will be convenient to parametrize the averaged
stress-energy by the components of the two-point matrix $S$. For this, note that
\begin{align}
\int dt\ f(t)^2
\boldsymbol\rho_{k}(t) =b_1(S) c_1(T_\k,f) + \Re\left\{ b_2(S) c_2(T_\k,f)
\right\}\ ,
\end{align}
where we have defined
\begin{subequations}\label{TheCs}
\begin{align}   
c_1(T_\k,f)&:=\int dt\ f(t)^2 \left(\abs{\dot T_\k(t)}^2 + \omega_\k^2 
\abs{T_\k(t)}^2 \right)\ ,\\
c_2(T_\k,f)&:=\int dt\ f(t)^2 \left(\dot{T}_\k(t)^2 + \omega_\k^2  T_\k(t)^2
\right)\ .
\end{align}
Our definition of the quantities $c_1(T_\k,f)$ and $c_2(T_\k,f)$ differ from the
\end{subequations}
one in \cite{Olbermann:2007ii} by a factor of $2$. Now, inserting $b_1$ and $b_2$
we can write
\begin{align}\label{TheEnergy}
\int dt\ f(t)^2
\boldsymbol\rho_{k}(t) 
&=S_{00}\ d_1(T_\k,f) + \a^6 S_{11}\ d_2(T_\k,f) +\a^3 \RS\ d_3(T_\k,f)\ ,
\end{align}
where 
\begin{subequations}\label{TheDs}
\begin{align}
d_1(T_\k,f)&:= \abs{T_\k(t_0)}^2  c_1(T_\k,f) - \Re\left\{\Bar{T}_\k^2(t_0)
c_2(T_\k,f) \right\}\ ,\\
d_2(T_\k,f)&:= \abs{\dot T_\k(t_0)}^2  c_1(T_\k,f) - \Re\left\{\dot{\Bar
 T}_\k^2(t_0) c_2(T_\k,f)\right\}\ , \\
d_3(T_\k,f)&:= \left(\Bar{T}_\k(t_0){\dot T_\k(t_0) +{T_\k(t_0)}\dot{\Bar
T}_\k(t_0)}\right) c_1(T_\k,f) - \Re\left\{2 \Bar{T}_\k(t_0)\dot{\Bar
T}_\k(t_0) c_2(T_\k,f) \right\}\ .
\end{align}
\end{subequations}

\chapter{Almost equilibrium states}\label{ChapterAES}
In this section we tackle the construction the almost equilibrium states.
First, we explain the basic ideas behind the procedure and the definition of the free
energy functional associated to the states of interest. Then, we calculate the
entropy, which is the part of the free energy, that cannot be inferred from the
quantum energy inequalities discussed in the last chapter. This is followed by
the actual minimization and the definition of the almost equilibrium states. 
In the last section we prove that the almost equilibrium states are
indeed Hadamard states.

\section{KMS states}\label{KMSStates}
Consider a finite quantum system that is described by an algebra of observables
$\mathfrak A$. A general state $\omega$ on $\mathfrak A$ is described by a density matrix $\rho$ and the expectation value
of an observable $A\in\mathfrak A$ is given by
\begin{equation}
\omega(A)=\tr(\rho A)
\end{equation}
Now, a state $\omega$ on $\mathfrak A$ is a thermal equilibrium state at
inverse temperature $\beta\in\mathbb R$ if it is a Gibbs state. A Gibbs state is given by a density
matrix of the form
\begin{align}\label{GibbsState}
\rho_{\beta H}&:=\frac{\e^{-\beta H}}{\tr(\e^{-\beta H})}\ ,
\end{align}
where $H=H^*$ is a self-adjoint operator (the free Hamiltonian) and
$\e^{-\beta H}$ is a trace class operator, which means that $\tr(e^{-\beta
H})\leq\infty$. The trace class property of  $\e^{-\beta H}$  is guaranteed for
finite volume systems by the properties of $H$ in that case \cite{Haag}. A
simple calculation using the cyclic invariance of the trace, $\tr(ABC)=\tr(CAB)$, shows that Gibbs states formally satisfy the relation
\begin{equation}\label{KMSRelation}
\omega(AB)=\omega(B \alpha_{\i\beta}(A))\ ,
\end{equation}
for all $A,B\in\mathfrak A$, where the automorphism
\begin{equation}
\alpha_t(A):=\e^{\i tH} A \e^{-\i tH}
\end{equation}
describes the free time evolution of an observable $A\in\mathfrak A$. Equation
\eqref{KMSRelation} gives the combinatorics inherent to thermal
equilibrium states of finite as well as infinite quantum systems.

Now, consider a state $\omega_\rho$ that is described by a density
matrix $\rho$ and define the entropy of $\omega_\rho$ by the von~Neumann
entropy functional
\begin{align}\label{Entropy}
\mathcal S(\omega_\rho):=-\tr(\rho\ln\rho)\ ,
\end{align}
where we understand that $x\ln x=0$ at $x=0$. If one defines the free energy
of the state $\omega_\rho$ by
\begin{align}\label{FreeEnergyFinite}
\mathcal F(\omega_\rho):=\omega_\rho(H)-\frac{1}{\beta} \mathcal S(\omega_\rho)\
,
\end{align}
then the Gibbs state is the unique state that minimizes the free energy
$\mathcal F(\omega_\rho)$. Equivalently, the Gibbs state maximizes the entropy
at fixed energy \cite{BratteliRobinson2,Wehrl}.

A necessary condition for Gibbs states to be well defined is that $H$ has a
purely discrete spectrum bounded from below (We consider only positive
temperature states: $\beta>0$). For infinite systems, Gibbs states are ill
defined because $\tr(e^{-\beta H})=\infty$. Nonetheless, a generalized
definition of thermal equilibrium states exists for infinite systems, namely,
the notion of KMS states. The above notion of a Gibbs state implies a
characterization in terms of analytic functions $F_{A,B}$ for any pair
$A,B\in\mathfrak A$, that satisfy certain boundary conditions, provided that the
Hamiltonian $H$ is bounded below and $\e^{-\beta H/2}$ is of trace class (see,
e.g., \cite{Emch,Wehrl}). That this characterization survives the
thermodynamical limit and thus can be used for the definition of equilibrium
states in the general case has been shown in \cite{HHW}. Since then, many
applications have shown that KMS states indeed describe thermal equilibrium.

We come to the definition of KMS states. Consider a C*-dynamical system
$(\mathfrak A,\alpha_t)$.
 \begin{Definition}[KMS state] Let
$\beta>0$ denote the inverse temperature. A state $\omega_\beta$ is called a
$\beta$-KMS state if for all $A,B\in\mathfrak A$ there is a function
$F_{A,B}(t):C_\beta\rightarrow \mathbb C$ which is analytic in the strip
$C_\beta:=\{z\in\mathbb C:0<\Im z<\beta\}$, bounded and continuous on its closure
$\overline C_\beta$, and satisfies the boundary conditions
\begin{align}\label{KMSCondition}
F_{A,B}(t)&=\omega_\beta(A\alpha_t(B)), &
F_{A,B}(t+\i\beta)&=\omega_\beta(\alpha_t(B)A)\ .
\end{align}
\end{Definition}
Of course, on a non-stationary spacetime there exists no global
C*-dynamical system. Thus, no global KMS state can exist
in that spacetimes. We remark that a relativistic KMS  condition has
been defined in \cite{BrosBuchholz}.

\section{Almost equilibrium states}
As noted in chapter \ref{ChapterQEI}, it is well known that in
non-stationary spacetimes there is no global generator of time-translations. The
standard Hamiltonian -- the integral of the energy density over the Cauchy
surface -- is not conserved. Even worse, the energy density is not bounded-below
at a spacetime point, and spatially averaged energy densities are ill defined for
dimensions $n>2$. An energy quantity which is unbounded below, poses serious
problems regarding the existence of equilibrium states. However, as we have also
seen, there exists a quantum inequality that gives lower bounds on the
difference of energy densities of Hadamard states averaged along the worldline
of an observer in any globally hyperbolic spacetime. Thus, one can expect to find
ground states with respect to this kind of energy. Indeed, in
\cite{Olbermann:2007ii} states with this property so-called states of low
energy, were constructed (see section \ref{StatesOfLowEnergy}). However, states
of low energy are pure states; consequently, they do not describe
systems in thermal equilibrium. They are merely ground states of a class
of almost equilibrium states, as we will show in this chapter.

Our guiding idea for the construction of the almost equilibrium states is to
define a sensible free energy functional $\mathcal F(\omega)$ on the class of
homogeneous, isotropic, quasi-free states that is based on the quantum
inequality stated in theorem \ref{QuantumInequality}. We write this functional
exclusively in terms of the two-point matrix \eqref{RealS}. Subsequently, we
minimize this functional with respect to the components of \eqref{RealS}.

Suppose that we can define a sensible free energy functional $\mathcal 
F(\omega)$, where we replace the energy expectation value $\omega_\rho(H)$ present in 
\eqref{FreeEnergyFinite} by the worldline averaged expectation value of the
(renormalized) stress-energy tensor,
\begin{align}\label{TimelikeSampledEnergyMomentum}
\omega(H_f):=\int_{\gamma} dt\ f(t)^2 \angles{:\T:}_\omega
(t,t)\ ,
\end{align}
for some real-valued $f\in\mathscr D(\mathbb R)$. Additionally, we assume that
$\gamma$ is the worldline of an isotropic observer. To remind the reader, an isotropic observer
moves along a timelike geodesic with tangent vector orthogonal to the
homogeneous spacelike surfaces. For such an observer, the spacetime looks
spatially isotropic at any instant of time. Note also that $\mathcal F(\omega)$
is actually a free energy density rather than a free energy and, at this stage,
it is unclear how to define the entropy $\mathcal S(\omega)$.

The minimization of $\mathcal F(\omega)$ means the following. We consider the
class of homogeneous, isotropic, quasi-free states. All such states are
completely determined by their two-point distributions \eqref{QuasifreeOmegaTwo}.
Within this class we look for a distinguished state $\omega_{\aes}$ such that for
any other state $\omega$ in the same class the free energy is larger, i.e.,
\begin{align}\label{FreeEnergyDiff}
\mathcal F(\omega)-\mathcal F(\omega_{\aes})\geq 0\ . 
\end{align}
In the rest of this chapter we will show that such a state $\omega_{\aes}$ exists
and is uniquely determined by its two-point distribution. Furthermore, we show
that the two-point distribution of the state $\omega_{\aes}$ satisfies the
Hadamard condition. This latter step is crucial for the whole construction by the
following argument.

The difference of the energy densities involved in the left hand side of
\eqref{FreeEnergyDiff} is
\begin{align}\label{FreeEnergy2}
\begin{split}
\omega(H_f)-\omega_{\aes}(H_f))=\int_{\gamma} dt\ f(t)^2
\left(\angles{\T}_{\omega}(t,t)-\angles{\T}_{\omega_{\aes}}(t,t)\right)\ .
\end{split}
\end{align}
Provided that the states $\omega$ and $\omega_{\aes}$ satisfy the Hadamard
condition and $\dim \mathcal M\geq 2$, the integral on the right hand side of
\eqref{FreeEnergy2} is well defined and bounded from below as we know from
theorem \eqref{QuantumInequality}. At this stage, it is, as a matter of fact,
not known if the state that eventually minimizes this functional is a Hadamard
state. So, the initial ansatz is justified afterwards
by proving that the resulting state $\omega_{\aes}$ is indeed a Hadamard state.

The integral on the right hand side of \eqref{FreeEnergy2} has the
form given in lemma \ref{LemmaEnergy}. It is a difference of integrals of the energy density
$\boldsymbol\rho_\k$ for each mode $\vec k$. Note that this mode decomposition of
the energy is due to the symmetry of the spacelike surfaces in Robertson-Walker
spacetimes and familiar from standard quantum field theory, which, after all, can
be viewed as a theory of infinitely many, independent, harmonic oscillators. We
infer from this observation that, when looking for the minimum of the averaged
energy density, we can confine ourselves to a single mode $\vec k$. Put
differently, we can minimize the integrand instead of the integral to obtain the
overall minimum. In order to utilize this principle for the free energy, we write
\begin{align}\label{FreeEnergyModes}
\mathcal F(\omega)&:=\int d\mu(\vec k)\mathcal F_\vec k\ ,
\end{align}
where we have defined the free energy of a single mode by 
\begin{align}\label{EntropyModes}
\mathcal F_\vec k:=\int_\gamma d t\ f(t)^2
\boldsymbol\rho_\k(t)-\frac{1}{\beta} \mathcal S_\vec k(\omega)\ .
\end{align}

We have yet to define the entropy $\mathcal S_\vec k$. Since each mode $\vec k$
can be considered as an independent quantum system with one degree of freedom, it
is natural to ascribe an entropy given by the von~Neumann entropy functional
\eqref{Entropy} to it. Then, $\mathcal S_\vec k$ is completely characterized by
the density matrix one ascribes to the single-mode system. Since we want our
states to exhibit thermal behavior, we take our modes to be Gibbs equilibrium
states. This means, we consider density  matrices
\begin{align}
\rho_{\beta K}:=\frac{\e^{-\beta K}}{\tr(e^{-\beta K})}\ ,
\end{align}  
where we assume that $K$ is a positive definite, quadratic form that is not
diagonalized from the outset. The latter point is important, since, such a form is determined
by three real parameters, just as the two-point distribution \eqref{RealS} of a
quasi-free state.

The further course of action will be as follows. In lemma \ref{LemmaKMS} we state
the generating functional of KMS states with respect to the evolution generated
by $K$. Then (by equation \eqref{QuadraticM}) we obtain a one-to-one
correspondence between the generator $K$ and the two-point matrix $S$. This
relation is used to express the free energy $\mathcal F_\vec k$ in terms of
$S_{11},S_{22}$, and $\det S$, which then allows to minimize with respect to
these variables. The result of the minimization will be a uniquely determined
state of inverse temperature $\beta$ associated to the sampling function
$f(t)$.

\section{The entropy}
For the entropy of a mode $\vec k$, we use the von~Neumann entropy associated to
the Gibbs state with density matrix $\rho_{\beta K}$:
\begin{align}
\mathcal S_\vec k(\omega_{\rho_{\beta K}})&=-\tr(\rho_{\beta K}\ln\rho_{\beta
K})\ ,
\end{align}
where we have defined a positive definite, real quadratic
form for the position and momentum operators $(q,p)$ by
\begin{align}
K(q,p):=K_{00} q^2 + K_{01} (qp+pq) + K_{11} p^2 =
\begin{pmatrix}
q \\ p   
\end{pmatrix}^\transpose
\cdot
\begin{pmatrix}
K_{00} & K_{01} \\
K_{01} & K_{11}  
\end{pmatrix}
\cdot
\begin{pmatrix}
q \\ p   
\end{pmatrix}\ .
\end{align}
In order to calculate the entropy we use canonical diagonalization of $K$. This
is possible for any positive definite quadratic form on the phase space.

\subsubsection{Canonical diagonalization of $K$}\label{DiagonalizingK}
The canonical diagonalization of a quadratic Hamiltonian is a well
investigated technique (for general results see, e.g.,
\cite{Moshinsky,Bogdanovic}). We consider the simple case of linear
transformation matrices $M$ for the coordinates $q,p$ that are canonical, which
means 
\begin{align}
M^\transpose \mathbb J M&=\mathbb  J\ , &
\mathbb J&=\begin{pmatrix}
0 & 1 \\
-1 &  0
\end{pmatrix}\ .
\end{align}
The eigenvalues of the matrix 
\begin{equation}
\mathbb J K=
\begin{pmatrix}
K_{01} & K_{11} \\
-K_{00} & -K_{01}
\end{pmatrix}
\end{equation}
are $\pm \i\sqrt{\det{K}}$. It turns out that in this model
$\Omega:=2\sqrt{\det K}$ plays the role of the oscillator frequency. The
eigenvectors $\vec v=\vec m_1+\i \vec m_2$ and $\Bar{\vec v}=\vec m_1-\i \vec
m_2$ are
\begin{align}
\vec v&=
y\left[ 
  \begin{pmatrix}
 K_{11} \\
 -K_{01}
\end{pmatrix}
+
\i
\begin{pmatrix}
0 \\
\sqrt{\det{K}}
\end{pmatrix}
\right]
\ ,& 
\Bar{\vec v}=
y\left[ 
  \begin{pmatrix}
 K_{11} \\
 -K_{01}
\end{pmatrix}
-
\i
\begin{pmatrix}
0 \\
\sqrt{\det{K}}
\end{pmatrix}
\right]\ ,
\end{align}
which defines the vectors $\vec m_1,\vec m_2$ needed for the transformation
matrix $M=(\vec m_1,\vec m_2)$:
\begin{equation}
M=y
\begin{pmatrix}
K_{11} & 0\\
-K_{01} & \sqrt{\det{K}}
\end{pmatrix}\ .
\end{equation}
The normalization constant $y$ is fixed by the condition $\vec m_i^\transpose \mathbb J
\vec m_j = \mathbb J_{ij}$, which gives
\begin{align}
y&=\pm\frac{1}{\sqrt{K_{11}\sqrt{\det{K}}}}\ ,
\end{align}
so that 
\begin{equation}
M=\pm\frac{1}{\sqrt{K_{11}\sqrt{\det{K}}}}
\begin{pmatrix}
K_{11} & 0\\
-K_{01} & \sqrt{\det{K}} 
\end{pmatrix}\ .
\end{equation}
This $M$ diagonalizes the quadratic form $K$ by 
\begin{equation}
M^\transpose K M=\begin{pmatrix}
\sqrt{\det{K}}&0 \\
0&\sqrt{\det{K}}
\end{pmatrix}\ .
\end{equation}
The new coordinates are given by 
\begin{align}
Q=\frac{1}{\sqrt{K_{11}\sqrt{\det{K}}}}
\begin{pmatrix}
K_{11}\\
-K_{01}
\end{pmatrix}^\transpose
\begin{pmatrix}
p\\
-q
\end{pmatrix}
=\frac{K_{11}p+K_{01}q}{\sqrt{K_{11}\sqrt{\det{K}}}}\ ,\\
P=-\frac{1}{\sqrt{K_{11}\sqrt{\det{K}}}}
\begin{pmatrix}
0\\
\sqrt{\det{K}}
\end{pmatrix}^\transpose
\begin{pmatrix}
p\\
-q
\end{pmatrix}
=\frac{\sqrt{\det{K}} q}{\sqrt{K_{11}\sqrt{\det{K}}}}\ .
\end{align}
If we define the frequency $\Omega:=2\sqrt{\det K}$ then the diagonalized matrix
reads $\frac{1}{2}  \left(\begin{smallmatrix}
\Omega&0 \\
0&\Omega
\end{smallmatrix} \right)
$. However there is still some freedom left in the choice of the coordinates. We
can perform a simultaneous transformation $q':=y Q$ and
$p':=y^{-1} P$ in order to gain a more suitable form of the
matrix. Choosing $y=\sqrt{\Omega}$ leads to 
\begin{equation}
K=\frac{1}{2}\begin{pmatrix}
\Omega^2&0 \\
0&1
\end{pmatrix}\ .
\end{equation}

We conclude that choosing appropriate coordinates $(q',p')$
is is possible to diagonalize the 'Hamiltonian' matrix $K$ and write it in the
standard harmonic oscillator form
\begin{align}
K(q',p')&=\frac{1}{2}(\Omega^2  q'^2+ p'^2)\ .
\end{align}
Therefore, for all basis-independent quantities, we
can use standard results from the the theory of the one dimensional, harmonic
oscillator with frequency $\Omega$. For example the eigenvalues of the system are
\begin{equation}
K_n=\left(n+\frac{1}{2} \right)2\sqrt{\det{K}}\ ,
\end{equation}
and the partition function is (see, e.g., \cite{Reif})
\begin{equation}
Z:=\tr(e^{-\beta K})= \frac{e^{-\beta \sqrt{\det{K}}}}{1-e^{-2\beta
\sqrt{\det{K}}}}=\frac{1}{2\sinh\left(\beta\sqrt{\det{K}}\right)}\ .
\end{equation}
From the partition function $Z$ one obtains the entropy as
\begin{align}\label{EntropyInK}
\mathcal S_\vec k(\omega_{\rho_{\beta K}})&=-\ln\left( 1-
e^{-2\beta\sqrt{\det{K}}}\right)+2\beta\sqrt{\det{K}}\frac{e^{-2\beta\sqrt{\det{K}}}}{1-e^{-2\beta\sqrt{\det{K}}}}\ .
\end{align}

\subsection{The generator of KMS states}\label{AppendixKMS}
Now we know the entropy $\mathcal S_\vec k(\omega_{\rho_{\beta K}})$ of the modes
in terms of the Hamiltonian matrix $K$. Note, that the expression
\eqref{EntropyInK} involves only the the determinant of $K$. Next, as we want
to characterize states by their two-point distribution, we need the correspondence between the matrices
$K$ and $S$. For this, we characterize the KMS states on a Weyl algebra with
respect to the time-evolution generated by $K$. The result, which is stated in
the following lemma, generalizes a result of \cite{NT}.
\begin{Lemma}\label{LemmaKMS}
Let $\mathfrak A$ be the Weyl algebra generated by the exponentials of the
position and momentum operators and let the time evolution be
generated by a positive Hermitian form $K$, with $\det K:=\operatorname{det}
K\neq 0$ on the phase space elements $\vec z\in\mathbb R^2$. Then, the generator
of a KMS state associated to the inverse temperature $\beta>0$ on $\mathfrak A$ is given by
\begin{align}
\omega(W(\vec z))&=\e^{-\frac{1}{4} \sqrt{\det K}\ \vec z^\transpose
K^{-1} \vec z\ \coth\left(\sqrt{\det K}\beta\right)}.
\end{align}
\end{Lemma}
\proof To prove this, we consider the exponentiated one-dimensional Heisenberg
*-algebra generated by $q$ and $p$, which are subject to the commutation
relations
\begin{gather}\label{ToyModelCommutationRelations}
[q,p]=\i \ ,\\  [q,q]=0=[p,p]\ ,
\end{gather}
and invariant under involution, $q^*=q$, $p^*=p$. The Weyl operators are given by
\begin{equation}
W(\vec z):=\e^{-\i(z_1 q +z_2 p)}\ ,
\end{equation}
where $\vec z=(z_1,z_2)\in\mathbb R^2$. 
We define a symplectic form $\sigma$ by
\begin{align}
\sigma(\vec z,\vec z')&={z_1}{z_2'}-{z_2}{z_1'}\ .
\end{align}
The Weyl operators $W(z)$, which generate the algebra $\mathfrak W$, satisfy the
Weyl relations
\begin{align}
W(\vec z)^*&=W(-\vec z)\ ,\\
W(\vec z)W(\vec z')&=\e^{-\frac{\i}{2}\sigma(\vec z,\vec z')}W(\vec z+\vec z')\
.
\end{align}

A state $\omega$ is, as usual, defined as a positive, linear, normalized
functional on $\mathfrak W$. By the GNS theorem we have a
representation $\pi_\omega$ and a representation space $\mathcal H_\omega$ with vectors
$\ket{\vec z}=W(\vec z)\ket{0}$ and $\ket{0}$ being the cyclic vector of
$\mathcal H_\omega$. A state $\omega$ on $\mathfrak W$ is determined by its
action on the Weyl operators
\begin{equation}
f(\vec z):=\omega(W(\vec z))=\sproduct{0}{W(\vec z)0}\ .
\end{equation}
The scalar product of two arbitrary vectors is given by
\begin{align}
\sproduct{\vec z'}{\vec z}
&=\sproduct{W(\vec z')0}{W(\vec z)0}
=\e^{\frac{\i}{2}\sigma(\vec z',\vec z)}f(\vec z-\vec z')\ .
\end{align}
The time automorphisms $\alpha_t$ over $\mathfrak W$ correspond to symplectic
transformation matrices $B_t$ on the phase space. The KMS condition
\eqref{KMSCondition} together with the Weyl relations yields
\begin{align}
\e^{-\frac{\i}{2}\sigma(\vec z,\vec z')}f(\vec z+\vec
z')&=\e^{-\frac{\i}{2}\sigma(\vec z',B_{\i\beta}\vec z)}f(B_{\i\beta}\vec
z+\vec z')\ .
\end{align}
If $\vec z'=-\vec z$ this simplifies to (Note that $f(0)=1$):
\begin{align}
1&=\e^{\frac{\i}{2}\sigma(\vec z,B_{\i\beta}\vec z)}f((B_{\i\beta}-\mathbb
1)\vec z)\\ f((B_{\i\beta}-\mathbb 1)\vec z)&=\e^{-\frac{\i}{2}\sigma(\vec
z,B_{\i\beta}\vec z)}\ .
\end{align}
Since for invertible $(B_{\i\beta}-\mathbb 1)$ we have
\begin{align}
\e^{-\frac{\i}{2}\sigma(\vec z,B_{\i\beta}\vec z)}
&=\e^{-\frac{\i}{2}\left(\sigma\left(\frac{1}{B_{\i\beta}-\mathbb
1}(B_{\i\beta}-\mathbb 1)\vec z,(B_{\i\beta}-\mathbb 1)\vec z\right)\right)}\ ,
\end{align}
a transformation $(B_{\i\beta}-\mathbb 1)\vec z \rightarrow \vec z$ gives
\begin{align}
f(\vec z)=\e^{-\frac{\i}{2}\sigma\left(\frac{1}{B_{\i\beta}-\mathbb 1}\vec
z,\vec z\right)}\ .
\end{align}

Next, we tackle the relation between the matrices $K$ and $S$. The time evolution generated by $K$ is given by
\begin{equation}
\alpha_t(W(z))=\e^{-\i Kt} \e^{\i(z_1 q + z_2 p)} \e^{\i
Kt}=\e^{\i(z_1(t)q+z_2(t)p)}\ .
\end{equation}
The time evolution of $\vec z(t)$ is given by the Heisenberg equation $\dot{\vec
z}=\i[K,\vec z]$. The commutators $[K,q]$ and $[K,p]$ can be calculated by \eqref{ToyModelCommutationRelations} and the identity
$[A,BC]=[A,B]C+B[A,C]$. The Heisenberg equation gives the system of
differential equations
\begin{align}
\begin{pmatrix}
\dot z_1 \\ \dot z_2 
\end{pmatrix}
&=
\begin{pmatrix}
2K_{01} & -2K_{00} \\
2K_{11} & -2K_{01}  
\end{pmatrix}
\cdot
\begin{pmatrix}
z_1 \\ z_2 
\end{pmatrix}\ .
\end{align}
We introduce the frequency 
\begin{equation}
\Omega:=2\sqrt{\det K}\ ,
\end{equation}
where $\det{K}:=K_{00} K_{11}-K_{01}^2$ is the determinant of the matrix $K$.
Furthermore, we set
\begin{align}
\kappa_{01}&:=\frac{K_{01}}{\sqrt{\det{K}}}\ , &
\kappa_{00}&:=\frac{K_{00}}{\sqrt{\det{K}}}\ , &
\kappa_{11}&:=\frac{K_{11}}{\sqrt{\det{K}}}\ .
\end{align}
Then, the time evolution can be written as
\begin{align}
\vec z(t)&=B_t \vec z(0)\ ,
\end{align}
where 
\begin{align}
B_t&:=
\begin{pmatrix}
  \cos(\Omega t)+\kappa_{01} \sin(\Omega t) 
& -\kappa_{00} \sin(\Omega t) 
\\
  \kappa_{11} \sin(\Omega t) 
& \cos(\Omega t) - \kappa_{01} \sin(\Omega t) 
\end{pmatrix}\ .
\end{align}
The matrix $B_t$ generalizes the usual harmonic oscillator
evolution matrix. To implement the KMS condition we complexify
time, $t\rightarrow \i\beta$, and since $\sin(\i\beta)=\i\sinh(\beta)$ and
$\cos(\i\beta)=\cosh(\beta)$, we obtain
\begin{align}
B_{\i\beta}&=
\begin{pmatrix}
  \cosh(\Omega\beta)+\i\ \kappa_{01} \sinh(\Omega\beta) 
& - \i\ \kappa_{00} \sinh(\Omega\beta) 
\\
  \i\ \kappa_{11} \sinh(\Omega\beta) 
& \cosh(\Omega\beta) - \i\ \kappa_{01} \sinh(\Omega\beta) 
\end{pmatrix}\ .
\end{align}
The matrix $(B_{\i\beta}-\mathbb 1)$ is invertible for $\det{K},\beta\neq 0$
(note that $\coth\left(\frac{x}{2}\right)=\frac{\sinh(x)}{\cosh(x)-1}$)
\begin{align}
(B_{\i\beta}-\mathbb 1)^{-1}
&=-\frac{1}{2}
\begin{pmatrix}
  1- \i\ \kappa_{01}\coth\left(\frac{\Omega\beta}{2}\right) 
& \i\ \kappa_{00}\coth\left(\frac{\Omega\beta}{2}\right) 
\\
  - \i\ \kappa_{11}\coth\left(\frac{\Omega\beta}{2}\right) 
& 1+\i\ \kappa_{01}\coth\left(\frac{\Omega\beta}{2}\right) 
\end{pmatrix}\ ,
\end{align}
and thus we find
\begin{align}
\sigma\left(\frac{1}{B_{\i\beta}-\mathbb 1}\vec z,\vec z\right)
&=-\frac{\i}{2}\frac{1}{\sqrt{\det K}}
\begin{pmatrix}z_1 \\ z_2\end{pmatrix}^\transpose
\begin{pmatrix}
K_{11} & -K_{01} \\
-K_{01} &  K_{00} 
\end{pmatrix}
\begin{pmatrix}z_1 \\ z_2\end{pmatrix}
\coth\left(\sqrt{\det K}\beta\right)\ .
\end{align}
So we have found the generator 
\begin{align}
f(\vec z)
&=\e^{-\frac{1}{4} \sqrt{\det K}\ \vec z^\transpose K^{-1} \vec z\
\coth\left(\sqrt{\det K}\beta\right)}\ .
\end{align}
This proves lemma \ref{LemmaKMS}. \qed

Defining the quadratic form $S(\vec z,\vec z): =
\frac{1}{2}(\vec z^\transpose  S \vec z)$ and recalling that the generator of
a quasi-free state is given by $f(\vec z)=\e^{-\frac{1}{2}S(\vec z,\vec z)}$ we
infer that
\begin{align}\label{QuadraticM}
\begin{pmatrix}
S_{00} 
& \RS   \\
 \RS  
& S_{11} 
\end{pmatrix}
=\frac{1}{2\sqrt{\det K}}
\coth\left(\beta\sqrt{\det{K}}\right)
\begin{pmatrix}
K_{11} 
& - K_{01}  \\
- K_{01} 
& K_{00} 
\end{pmatrix}\ .
\end{align}
Inverting  \eqref{QuadraticM} gives
\begin{align}\label{KforS}
\begin{pmatrix}
K_{00} & K_{01}\\
K_{01} & K_{11}
\end{pmatrix}
&=\frac{1}{\beta\sqrt{\det{S}}}\arcoth(2\sqrt{\det{S}})
\begin{pmatrix}
S_{11}   & - \RS \\
- \RS  & S_{00} 
\end{pmatrix}\ .
\end{align}
The determinants of $K$ and $S$ are related by
\begin{align}
\sqrt{\det K}&=\frac{1}{\beta}\arcoth(2\sqrt{\det S})\ ,&  
\sqrt{\det S}&=\frac{1}{2}\coth(\beta\sqrt{\det K})\ .
\end{align}
Thus, we can write the entropy per mode as
\begin{align}
\mathcal S_\vec k(\omega_{\rho_{\beta K}})&=-\ln\left( 1-
\e^{-2\arcoth(2\sqrt{\det{S}})}\right)+2\arcoth(2\sqrt{\det{S}})\frac{\e^{-2\arcoth(2\sqrt{\det{S}})}}{1-\e^{-2\arcoth(2\sqrt{\det{S}})}}\\
&= 2\sqrt{\det{S}} \arcoth(2\sqrt{\det{S}}) +
\frac{1}{2}\ln\left(\frac{4\det{S}-1}{4} \right) \label{TheEntropy}\ ,
\end{align}
where, for the second line, we used that
$\arcoth(y)=\frac{1}{2}\ln\left(\frac{y+1}{y-1}\right)$ and thus $\e^{-2\arcoth(4y)}=\frac{4y+1}{4y-1}$.

\section{Minimization of the free energy}
Using the energy density \eqref{TheEnergy} and the entropy formula
\eqref{TheEntropy} we can write the free energy functional $\mathcal F_\vec k$
exclusively in terms of the two-point matrix \eqref{RealS}.
\begin{align}\label{FreeEnergyFunctional}
\mathcal F_\vec k&:=
S_{00} d_1(T_\k,f) + \a^6 S_{11} d_2(T_\k,f) \pm \a^3  \sqrt{S_{00}S_{11}-\det{S}} d_3(T_\k,f)\\
&\phantom{=}{\nonumber} -\frac{1}{\beta}\left[
 2\sqrt{\det{S}} \arcoth(2\sqrt{\det{S}}) +
 \frac{1}{2}\ln\left(\frac{4\det{S}-1}{4} \right)\right]\ ,
\end{align}
where we used $\det{S}:=S_{00}S_{11}-(\RS)^2$ to eliminate
$\RS$ in favour of the determinant $\det S$. 
The remaining task is a simple exercise, namely, the minimization of this
expression. A necessary condition for the minimization of $\mathcal F_\vec k$ is
that we have a critical point, i.e., the gradient of the function vanishes:
\begin{align}
\nabla_{(S_{00},S_{11},\det{S})}\mathcal F_\vec k &\stackrel{!}{=}0\ .
\end{align}
This amounts to the equations
\begin{align}
d_1(T_\k,f)+\frac{\a(t_0)^3
d_3(T_\k,f)}{2}\frac{S_{11}}{\sqrt{S_{00}S_{11}-\det{S}}}&=0 \label{eq1}\ ,\\
d_2(T_\k,f)+\frac{d_3(T_\k,f)}{2\a(t_0)^3}\frac{S_{00}}{\sqrt{S_{00}S_{11}-\det{S}}}&=0
\label{eq2}\ ,\\
-\frac{1}{\beta\sqrt{\det{S}}}\arcoth(2\sqrt{\det{S}})\pm\frac{\a(t_0)^3
d_3(T_\k,f)}{2} \frac{1}{\sqrt{S_{00}S_{11}-\det{S}}}&=0 \label{eq3}\ ,
\end{align}
where the coefficients $d_i(T,f)$ are given by \eqref{TheDs}. Multiplying
\eqref{eq1} by  \eqref{eq2} we obtain
\begin{align}
\sqrt{S_{00}S_{11}-\det S}&=\frac{\abs{d_3}}{\sqrt{4d_1d_2-d_3^2}}\sqrt{\det S}\ .
\end{align}
Using this, and assuming $\beta>0$, the critical point is obtained from
\eqref{eq3} as
\begin{align}\label{TheThirdEquation}
\sqrt{\det{S}}&=\frac{1}{2}\coth\left(\frac{\beta}{2} \a(t)^3  \sqrt{4 d_1 d_2-
d_3^2}\right)\ ,
\end{align}
where we omitted the combination of signs from \eqref{eq3} and $d_3/\abs{d_3}$
giving no solution for $\beta> 0$. So, given a sampling function $f(t)$ there is
for every $\beta>0$ a unique solution to equation \eqref{TheThirdEquation} and
thus a unique minimum of the free energy functional $\mathcal
F(\omega_{\rho_{\beta K}})$ given in \eqref{FreeEnergyFunctional}.
We state this result as a theorem.
\begin{Theorem}
Consider the Weyl algebra $\mathfrak A$ of the minimally coupled Klein-Gordon
field in a Robertson-Walker spacetime. Define the free energy of a
homogeneous, isotropic, quasi-free state $\omega$ on $\mathfrak A$ by
\begin{align}
\mathcal F(\omega)&:=\int d\mu(\vec k)\mathcal F_\vec k\ ,
\end{align}
where
\begin{align}
\mathcal F_\vec k:=\int_\gamma d t\ f(t)^2
\boldsymbol\rho_\k(t)-\frac{1}{\beta} \mathcal S_\vec k(\omega)\ .
\end{align}
and $\boldsymbol\rho_\k$ is the energy density of a mode. Averaging takes
place along a timelike curve $\gamma$ with a sampling function $f\in\mathscr
D(\mathbb R)$, and each mode is assigned the von Neumann entropy $\mathcal
S_\vec k(\omega)$. Then, for every $\beta>0$ and every function $f$, there is a
unique state $\omega_{\aes}$ that minimizes $\mathcal F(\omega)$. This state is determined by
\begin{align}\label{AESDefining}
\sqrt{\det{S}}&=\frac{1}{2}\coth\left(\frac{\beta}{2} \a(t)^3  \sqrt{4
d_1(T_\k,f) d_2(T_\k,f)- d_3^2(T_\k,f)}\right)
\end{align}
supplemented by equations \eqref{eq1}, and \eqref{eq2}, where $T_\k$ is an
arbitrary solution to \eqref{KGRWTimeDependentPart} satisfying \eqref{Wronskian}. We refer to
$\omega_{\aes}$ as an almost equilibrium state of inverse temperature $\beta$
associated to $f$.
\end{Theorem}
The almost equilibrium states are parametrized by arbitrary solutions $T_\k(t)$
of the time-dependent part of the  Klein-Gordon equation. As we will see in the
next section, they resemble equilibrium states in Minkowski space if one uses the
states of low energy for the parametrization.

\section{Hadamard property}\label{HadamardProperty}
In this section, we prove that the two-point distribution of the almost
equilibrium states satisfies the Hadamard condition. This is of twofold
importance. First, it is a necessary condition to make our ansatz meaningful, as
otherwise the energy difference involved in the free energy \eqref{FreeEnergy2}
is not bounded below and our formal result remains formal. Second, it
proves that the almost equilibrium states can be considered as physical states
in the sense we described in chapter \ref{ChapterQFTCST}.

We are in the fortunate situation that we can base our proof on previous related
work \cite{Junker:1996,JunkerSchrohe,Olbermann:2007ii,Olbermann:2007i}. It was
shown in \cite{Olbermann:2007ii} that the states of low energy are Hadamard
states. The proof used the fact that for large momenta $\k$ and large iteration
order $n$ the difference of the two-point distributions of the states of low
energy and the adiabatic vacuum states converges to zero. Thus, the two-point
distributions differ only by a smooth function. Since adiabatic vacuum states of
infinite order are Hadamard states \cite{Junker:1996,Junker:2002}, this proves
the Hadamard property of the states of low energy. Along the same line of
reasoning, we show that the two-point distributions of the almost equilibrium
states and of the states of low energy also differ only by a smooth function. The
calculations are simplified a great deal, due to the fact that the states of low
energy are the natural ground states associated to the almost equilibrium states.

\subsubsection{Parametrization by states of low energy}
\label{StatesOfLowEnergy} Let us first introduce the states of low energy \cite{Olbermann:2007ii}. Consider the Weyl algebra
$\mathfrak{A}$ of the free Klein-Gordon field over a Robertson-Walker spacetime
$(\mathcal M,\g_{\mu\nu})$ and $f(t)\in\mathscr{D}(\mathbb{R})$. In the set of
homogeneous, isotropic, pure quasi-free states on $\mathfrak{A}$, there is a state
$\omega_\sole$ for which the averaged energy density
\begin{align}
\omega(H_f)=\int_\gamma dt\ f(t)^2 \angles{\ren\T}_\omega(t)
\end{align} 
is minimal. The averaging is understood along the path of an isotropic observer.
This state is given by the two-point distribution
\begin{align}  
\omega_\sole^{(2)}(x,x')=\int \d\vec k\ Y_{\vec k}(\vec x)\Bar{Y}_{\vec k}(\vec x')\,\Bar{L}_\k(t)L_\k(t')
\end{align}
with
\begin{align}
L_\k(t)=\lambda T_\k(t)+\mu \Bar{T}_\k(t),
\end{align}
where $T_\k$ is an arbitrary solution of the differential equation
\eqref{KGRWTimeDependentPart} fulfilling the continuity condition
\eqref{Wronskian} and $\lambda,\mu$ are given by the equations
\begin{align}
\lambda&=\exp\left(i\alpha\right)\sqrt{\frac{c_1}{2\sqrt{c_1^2-\abs{c_2}^2}}+\frac{1}{2}}\ , &
\mu&=\sqrt{\frac{c_1}{2\sqrt{c_1^2-\abs{c_2}^2}}-\frac{1}{2}}\ . 
\end{align}
The coefficients $c_1=c_1(T_\k,f)$ and $c_2=c_2(T_\k,f)$, up to a factor of $2$, are
defined in equation \eqref{TheCs}. Setting $T_\k=L_\k$, it follows that a state given by $L_\k$ is a state of low  energy if and only if it satisfies $c_2(L_\k,f)=0$.

For the proof of the Hadamard property, and presumably for most other purposes,
the expression \eqref{AESDefining} defining almost equilibrium states is
simplified considerably by using the states of low energy for the
parametrization. So, we plug in a state of low energy $L_\k(t)$ into
\eqref{TheThirdEquation}. Note that the squared modulus of \eqref{Wronskian}, is
given by the useful relation
\begin{equation}\label{AUsefulRelation}
4  \abs{T}^2 \abs{\dot T}^2  - \left(\Bar T \dot T + T \dot{\Bar T} \right)^2
=\abs{\Bar T \dot T - T \dot{\Bar T}}^2 = \frac{1}{\a^6} \ ,
\end{equation}
which turns up several times in the calculations. Plugging in the states of low
energy  into the expression $4 d_1 d_2 - d_3$ gives, due to $c_2(L_\k(t),f)=0$ in
equation  \eqref{TheDs},
\begin{align}
4 d_1 d_2 - d_3^2 
&= 4 \abs{L_\k(t_0)}^2 \abs{\dot L_\k(t_0)}^2  c_1^2(L_\k,f)-\left({\Bar L_\k(t_0)}\dot{L}_\k(t_0)+{L}_\k(t_0)\dot{\Bar L}_\k(t_0)\right)^2 c_1^2(L_\k,f)\\
&=\frac{1}{\a(t_0)^6} c_1^2(L_\k,f)\ .
\end{align}
Using this, equation \eqref{TheThirdEquation} becomes
\begin{align}
\sqrt{\det{S}}&=\frac{1}{2}\coth\left(\frac{\beta}{2}\ c_1(L_\k,f) \right)\ .
\end{align}
Solving for $S_{00}$ and $S_{11}$ at the minimum of the free energy gives
\begin{align}\label{SolvedForTheS}
S_{11}&= \frac{2\abs{d_1}}{\a^3\sqrt{4 d_1 d_2 - d_3^2}}\sqrt{\det{S}}\ ,\\
S_{00}&= \frac{2\a^3 \abs{d_2}}{\sqrt{4 d_1 d_2 - d_3^2}}\sqrt{\det{S}}\ ,
\end{align}
where $d_1$ and $d_2$ have the same sign. Noting that $d_1$ and
$d_2$ are always positive for states of low energy, this reduces to
\begin{align}
S_{11}&=2\abs{L_\k(t_0)}^2 \sqrt{\det{S}}\ ,\\  
S_{00}&=2\a(t_0)^6 \abs{\dot L_\k(t_0)}^2 \sqrt{\det{S}}
\end{align}
for the almost equilibrium states. 

Now, we calculate the coefficients $b_1$ and $b_2$. We can use the expressions
for $S_{00}$ and $S_{11}$ to find an expression of $\RS$ from $\det S=S_{00}S_{11}-(\RS)^2$:
\begin{align}
\RS &=\pm\sqrt{4\a(t_0)^6 \abs{L_\k(t_0)}^2 \abs{\dot L_\k(t_0)}^2
-1}\cdot \sqrt{\det S}\ .
\end{align}
On the other hand we know that with the normalization \eqref{Wronskian} we have
equation \eqref{AUsefulRelation} giving
\begin{align}
\a(t_0)^3 \left({\Bar L_\k(t_0)}\dot{L}_\k(t_0)+{L}_\k(t_0)\dot{\Bar
L}_\k(t_0)\right) &=\pm\sqrt{4\a(t_0)^6 \abs{L_\k(t_0)}^2 \abs{\dot
L_\k(t_0)}^2 - 1}\ .
\end{align}
Each of the last two equations has two possible signs. If the signs adjust such that their product is negative we can write
\begin{align}
b_1&=S_{00} \abs{L_\k(t_0)}^2+\a(t_0)^6 S_{11}\abs{\dot L_\k(t_0)}^2 +
\left(\Bar L_\k(t_0){\dot L_\k(t_0)}+{L}_\k(t_0)\dot{\Bar L}_\k(t_0)\right)
\a(t_0)^3\RS \\ &=4\a(t_0)^6 \abs{L_\k(t_0)}^2 \abs{\dot L_\k(t_0)}^2  \sqrt{\det{S}} - \left( 4\a(t_0)^6 \abs{L_\k(t_0)}^2 \abs{\dot L_\k(t_0)}^2 - 1\right)
\sqrt{\det S} \\ &=\sqrt{\det S}\ .
\end{align}
Assuming the same adjustment of signs as before, we have an alternative
expression for $\RS $, namely
\begin{align}\label{AlternativeS01}
  \RS &=- \a(t_0)^3 \left({\Bar
  L_\k(t_0)}\dot{L}_\k(t_0)+{L}_\k(t_0)\dot{\Bar L}_\k(t_0)\right) \sqrt{\det
  S}\ ,
\end{align}
which inserted into $b_2$ gives
\begin{align}
b_2&=-S_{00} \Bar{L}_\k^2(t_0)-\a(t_0)^6S_{11}\dot{\Bar L}^2_\k(t_0) -
2\Bar{L}_\k(t_0)\dot{\Bar L}_\k(t_0) \a(t_0)^3\RS \\ &=2\a(t_0)^6
\sqrt{\det S} \bigg[ \abs{\dot L_\k(t_0)}^2 \Bar{L}_\k^2(t_0)- \abs{L_\k(t_0)}^2 \dot{\Bar L}^2_\k(t_0)\\
&\phantom{=}+ \Bar{L}_\k(t_0)\dot{\Bar
L}_\k(t_0) \left({\Bar L_\k(t_0)}\dot{L}_\k(t_0)+{L}_\k(t_0)\dot{\Bar
L}_\k(t_0)\right) \bigg]\\ &=0\ .
\end{align}

\subsection{Almost equilibrium states are Hadamard states}
Now we are prepared to prove that our construction indeed yields Hadamard
states.
\begin{Proposition}
The two-point distribution of an almost equilibrium state $\omega_\aes$
satisfies the Hadamard condition.
\end{Proposition}
\proof The strategy for the proof is the following. We show that that the
difference of the two-point distributions, i.e.,
$(\omega_{\aes}^{(2)}-\omega^{(2)}_\sole)(x,x')$, is a smooth function and thus
the singular parts of $\omega_{\aes}$ and $\omega_{\sole}$ coincide. As
indicated in section \ref{HadamardStates} an implicit infrared cutoff for the relevant
integrals can be assumed, i.e., we have only to care for the large $\k$
behaviour. (For the closed case $\varepsilon=+1$ no such assumption is needed.)
Since the states of low energy $\omega_\sole$ are known to have a Hadamard
singularity structure, this proves the same for the almost equilibrium  states
$\omega_\aes$.

We have shown that $b_1=\sqrt{\det S}$ and $b_2=0$. Hence,
\begin{align}\label{HadamardDifference}
\left(\omega_{\aes}^{(2)}-\omega^{(2)}_\sole\right)(x,x')
&= \int\d\vec k\   
Y_{\vec k}(\vec x) \Bar{Y}_{\vec k}(\vec x') \frac{L_\k(x^0) \Bar L_{\k}(x'^0) + \Bar L_\k(x^0) L_{\k}(x'^0)}{\e^{\beta c_1(L_\k,f)}-1}\ .
\end{align}
In \cite{LR} the following estimates for the growth of $Y_{\vec k}(\vec x)$ and
$T_\k(t)$ and their derivatives for large $\k$ are given ($\alpha$ is a
multi-index, $j\in\mathbb N$):
\begin{align}
\abs{\D_\vec x^\alpha Y_\vec k(\vec
x)}=O(k^{\abs\alpha +2})\ ,
\end{align}
and 
\begin{align} 
\D_t^{j} L_\k(t)=O(k^{j-\frac{1}{2}})\ .
\end{align}
Using this, we can then estimate the growth of the
following expression
\begin{align}
&\sup\left\lvert \D_{(x,x')}^{\nu}\left[  
  Y_{\vec k}(\vec x) \Bar{Y}_{\vec k}(\vec x') \left(L_\k(t) \Bar L_{\k}(t') + \Bar L_\k(t) L_{\k}(t') \right) \right]\right\rvert \\
&=\sup\left\lvert   
\D_{\vec x}^{\abs\sigma}  Y_{\vec k}(\vec x)\ \D_{\vec x'}^{\abs{\sigma'}}\Bar{Y}_{\vec k}(\vec x') \left(\D_{t}^{j} L_\k(t)\ \D_{t'}^{j'} \Bar L_{\k}(t') + \D_{t}^{j} \Bar L_\k(t)\ \D_{t'}^{j'} L_{\k}(t') \right) \right\rvert \\
&\leq\left\lvert  \D_{\vec x}^{\abs\sigma}  Y_{\vec k}(\vec x) \right\rvert  \left\lvert \D_{\vec x'}^{\abs{\sigma'}}\Bar{Y}_{\vec k}(\vec x') \right\rvert 
\left( \left\lvert \D_{t}^{j} L_\k(t) \right\rvert  \left\lvert \D_{t'}^{j'} \Bar L_{\k}(t')  \right\rvert
+ \left\lvert \D_{t}^{j} \Bar L_\k(t) \right\rvert  \left\lvert \D_{t'}^{j'} L_{\k}(t') \right\rvert  \right)  \\
&=O(k^{3+\abs\nu})\ , 
\end{align}
where  $\abs\nu=\abs\alpha + \abs{\alpha'}$ and $\alpha=\abs \sigma + j $,
$\alpha'=\abs{\sigma'} + j' $. We also need an estimate for the growth of
$c_1(\k):=c_1(T_\k,f)$. In \cite{Olbermann:2007i} it has been shown that the
$\k$-dependence of $c_1(T_\k,f)$ for large $\k$ is such that there exist
constants $a_1, a_2>0$ with 
\begin{equation}
a_1 (1+\k) \leq c_1(\k) \leq  a_2 (1+\k)\ .
\end{equation}
Thus, the factor $\frac{1}{\e^{\beta c_1(L_\k,f)}-1}$, which vanishes faster
than any polynomial, makes the integral
\begin{align}
&\D_{(x,x')}^\alpha(\omega^{(2)}-\omega^{(2)}_\sole)(x,x')\\
&=\int\d\vec k\ \D_{(x,x')}^\alpha \left\{  
  Y_{\vec k}(\vec x) \Bar{Y}_{\vec k}(\vec x') \frac{L_\k(t) \Bar
  L_{\k}(t') + \Bar L_\k(t) L_{\k}(t')}{\e^{\beta c_1(L_\k,f)}-1} \right\}
\end{align}
converge absolutely. This proves that the two-point difference
\eqref{HadamardDifference} is a smooth function which in turn proves the
Hadamard property of $\omega^{(2)}_\aes(x,x')$. \qed

\chapter{Summary and Outlook}
The present thesis is the successful accomplishment of a natural task suggested
by recent results in quantum field theory in curved spacetimes. In
\cite{Olbermann:2007ii} ground states, so-called states of low energy, for
the worldline averaged, renormalized stress-energy tensor of the Klein-Gordon
field in Robertson-Walker spacetimes were constructed. They are obtained by
minimizing the averaged energy density that an isotropic observer measures in a
pure, homogeneous, isotropic, quasi-free state. The states of low energy depend
on the sampling function used in the averaging procedure.
 
In this thesis we constructed, in the same setting, a family of states that we
consider as almost equilibrium states to the inverse temperature $\beta$.
The almost equilibrium states are obtained by a suitable application of the
principle of minimal free energy -- a cornerstone of statistical mechanics. More
precisely, we defined a free energy functional on the homogeneous, isotropic,
quasi-free states of the Klein-Gordon field, where the 'inner energy' was
constructed from the worldline averaged stress-energy tensor since that quantity
is known to be lower bounded on the class of Hadamard states by a quantum energy
inequality \cite{Fewster:2000}. We showed how this energy can be written in terms
of the same parameters that determine the two-point distribution of a
homogeneous, isotropic, quasi-free state. Viewing each mode of the quantum field
as a quantum system with one degree of freedom, we assigned to the modes the
usual von~Neumann entropy. An essential point of the whole construction was the
determination of the entropy of such a system in terms of the two-point
distribution. This was accomplished by calculating the generator of a KMS state
on the Weyl algebra of a system with a single degree of freedom. By subsequent
minimization of the free energy, we obtained an explicit expression for the two-point distribution of a family of
states, determined, again, by the sampling function of the averaging procedure
and the inverse temperature $\beta$. Finally, we showed that the two-point
distributions of the almost equilibrium states satisfy the Hadamard condition.
The latter step is a vital part of the construction as otherwise the utilized quantum
inequality would not guarantee the existence of a finite lower bound for the
averaged stress-energy.

The formulation of quantum field theory in curved spacetimes disentangles the
construction of the quantum algebra from the construction of the quantum states.
It is well known how to construct the algebra of a linear quantum field in a
globally hyperbolic spacetime. The construction of physically meaningful states, however,
is less straightforward. With decreasing symmetry properties of the underlying
spacetime it becomes increasingly complicated to single out such states or
rather classes of them.  Regarding timelike symmetries, the step from
stationary to non-stationary spacetimes is the biggest one. It brings along unitarily
inequivalent representations of the algebra of observables and makes the concept
of a conserved Hamiltonian that generates global time evolution useless. In view
of this situation, it is important to watch for other notions of energy that may
be more meaningful. We advocate that such a notion is provided by the averaged
energy densities used in the quantum energy inequalities.

The almost equilibrium states defined here are, to date, the sole example of
explicitly constructed (global) Hadamard states with thermal properties on a
non-stationary spacetime. For quantum fields in a non-stationary spacetime it is
not reasonable to expect the existence of true equilibrium states with definite
temperature $1/\beta$ since the influence of tidal forces will always destroy
such property. We claim that the almost equilibrium states are reasonably defined
approximations to equilibrium, and that they provide an interesting starting
point for future investigations. Due to the setting, namely, quantum fields in
cosmological spacetimes, they might provide a class of states useful in the
inflationary scenario and the analysis of the cosmic microwave background.

Like the states of low energy, the two-point distribution of an almost
equilibrium state involves a sampling function $f(t)\in\mathscr D(\mathbb R)$.
The function $f(t)$ describes the measurement of energy by an isotropic
observer. This function introduces some freedom in the construction which may be used to
design states with desired properties. This might be useful with respect
to models of the early universe.

Admittedly, our construction of almost equilibrium states makes heavy use of the
mode decomposition, which is available in Robertson-Walker spacetimes because of
the underlying symmetries. In general, there is no satisfactory mode
decomposition on an arbitrary non-stationary spacetime, and thus there is no
direct route for a generalization of our method.   

Quantum energy inequalities, which play a major role in our construction, are
among a family of interconnected criteria for dynamical stability of quantum
systems \cite{FewsterVerch}. In \cite{SchlemmerVerch}, a link between local
thermal equilibrium states of a linear scalar field on a curved spacetime and
quantum energy inequalities was alluded. It was proved that the existence
of a linear scalar quantum field fulfilling some local thermal condition, in the sense
of \cite{BOR}, implies a quantum energy inequality for these states. However, no
states are known that fulfill the hypothesis of the paper. Due to our
construction, which, starts with a quantum inequality rather than to derive one,
it might turn out that the almost equilibrium states provide examples for local
thermal equilibrium states (see below).

Among the mentioned stability criteria for quantum systems there is passivity,
which has not been generalized to non-stationary spacetimes so far. It is known
that KMS states and mixtures of KMS states are passive. It might be worth
considering, whether almost equilibrium states (and mixtures of almost
equilibrium states) are 'almost passive' in a sense to be specified. Such a
notion of passivity should be able to measure the amount of mechanical work that can be performed,
owing to the present tidal forces, by a system in a cyclic process in a
non-stationary spacetime.

\subsubsection{Local thermal equilibrium states}\label{ChapterLTE}
The compact support of the sampling function $f(t)$ introduces intrinsically a
kind of localization in time and, by taking the associated closed double cone, a
localization in spacetime into the almost equilibrium states. An open question
for future investigations might be how almost equilibrium states are connected to
other notions of local thermal equilibrium. The most interesting states would be
the local thermal equilibrium (LTE) states in the sense of Buchholz,
Ojima, and Roos in \cite{BOR}. This approach gives a precise
meaning to the saying that by a local measurement one cannot distinguish between local
and global equilibrium states. This abstract idea is implemented in field theory
by the following procedure. First, one chooses a set of global thermal
equilibrium states, e.g., mixtures of KMS states in different inertial systems.
Then, for a point $x$, one chooses a certain set $\borS_x$ of reference
observables. A state $\omega$ is called $\borS_x$-thermal if its expectations
values for all observables in $\borS_x$ coincide with the expectation values
for some global thermal equilibrium state $\omega_x$. Thus, as long as one
considers observables from $\borS_x$ it is impossible to distinguish between
the local state $\omega$ and the global thermal equilibrium state $\omega_x$. Of
course, nothing is said or assumed about the expectation values of other
observables. Thus, in general, $\omega$ does not coincide with the global
state $\omega_x$. The spaces $\borS_x$ are chosen as the linear spaces generated by the
balanced derivatives of the Wick-square $:\phi^2:(x)$. Balanced derivatives are
defined in \cite{BOR} as
\begin{equation}\label{BalacedDerivative}
\eth_{\boldsymbol\mu} :\phi^2:(x) = \lim_{\zeta\rightarrow 0}\left[
\phi(x+\zeta)\phi(x-\zeta)-\omega_0(\phi(x+\zeta)\phi(x-\zeta))\mathbb 1
\right]\ ,
\end{equation}
where $\eth_{\boldsymbol\mu}=\partial_{\zeta^{\mu_1}}
\dots\partial_{\zeta^{\mu_n}}$ with the multi-index
$\boldsymbol\mu:=(\mu_1,\dots,\mu_n)\in\mathbb N^n$ and $\omega_0$ is the vacuum
state. Here, one takes the limit along spacelike directions $\zeta\in\mathbb M$
in Minkowski spacetime, so that $\phi(x+\zeta)\phi(x-\zeta)$ is well defined as
a quadratic form \cite{Bostelmann}.
Now, with this approach one calculates for the Wick-square, which corresponds to
the balanced derivative of zero-th order, of the massless field in Minkowski
space
\begin{equation}
\omega^{\beta e}(:\phi:(x))=\frac{1}{12\beta^2}\ .
\end{equation}
independently of $x$, where $e$ is an orthonormal tetrad defining the Lorentz
frame of the KMS state (see the relativistic KMS condition \cite{BrosBuchholz}).
This is interpreted as an indication that the the Wick-square serves as a scalar thermometer. 

It is natural to ask for the Wick-square of an almost equilibrium state and if
there is a similar interpretation as thermometer. Hereby, one has to resolve an
ambiguity in the definition the Wick-square since one has to choose a ground
state. By construction, it is natural to choose the states of low energy for this
purpose. But, by the method of locally covariant quantum fields \cite{BFV}, one
could also choose the Minkowski vacuum. The latter procedure should be viable by
the generalization of LTE states for curved spacetimes that is proposed
\cite{SchlemmerVerch}.

To the very end, let us make some speculative remarks regarding the
former procedure. We use use the states of low energy for the parametrization and as the ground state. For simplicity, we
consider flat Robertson-Walker spacetimes. By some formal manipulations, one
easily, obtains the expression
\begin{align}
\lim_{\zeta\rightarrow 0}\left(\omega_{\aes}^{(2)}-\omega^{(2)}_{\sole}\right)(x+\zeta,x-\zeta)
&=\frac{1}{(2\pi)^{3}}\int\d\vec k\ \frac{2\abs{L_\k(x^0)}^2}{\e^{\beta
c_1(L_k,f)}-1}\ .
\end{align}
Note that this expression bears a strong resemblance to its Minkowski spacetime
counterpart. In that case by equation \eqref{StaticSolution} we have
$2\abs{L_\k(x^0)}^2=\frac{1}{\omega_\k}$ and $c_1(L_k,f)=\omega_\k$ provided that
$\int f(t)^2 dt=1$. We learn from this that $c_1(T_\k,f)$ possibly plays the
role of a generalized frequency. In order to proceed into the direction of LTE states
on needs estimates on $c_1(L_k,f)$. Such estimates may, e.g, lead to the
interpretation that the almost equilibrium states are a mixture of LTE states
with a range of temperatures $a(f)<\beta<b(f)$.

\appendix
\backmatter
\cleardoublepage

\cleardoublepage
\setlength{\parindent}{0em}
\setlength{\parskip}{0.5em}
\pagestyle{empty}
\vspace*{\stretch{1}}
\section*{Acknowledgements} I am deeply grateful to my supervisor Prof. Klaus
Fredenhagen for initiating and caring for this project; I thank him
for the offered trust and the kind hospitality in his working group.

I am also grateful to Nicola Pinamonti for many useful discussions on the
subject and critical reading of the manuscript. It helped a lot.

Furthermore, I have to thank Prof. Hans-J\"urgen Seifert for his amiableness,
which made life much easier in the past years.

And, of course, I thank all friends and colleagues, past and
present, at the AQFT group, the II. Institute for Theoretical Physics, DESY, and
elsewhere.

Finally, I thank my parents for still supporting me, as always, in every
respect, as much as they can, in whatever I do. We have come long
way in one generation. The honor belongs to you.

Financial support of the Emmy-Noether Projekt ``Hawking-Strahlung Schwarzer
L\"ocher'' and of the Graduiertenkolleg ``Zuk\"unftige Entwicklungen in der
Teilchenphysik'' is gratefully acknowledged. 
\vspace*{\stretch{2}}

\end{document}